\documentclass[11pt]{article}
\pdfoutput=1

\usepackage{jheppub}
\usepackage{url,comment}
\usepackage{times}
\usepackage{latexsym}
\usepackage{graphicx, graphics, hyperref, amsmath, amssymb, slashed, xcolor, bbm,bm,amsthm, array}
 \usepackage{subfigure}
 \usepackage{listings} 
 \usepackage{trimclip}
\usepackage{multirow}
\usepackage{rotating}
\usepackage{afterpage}

\newcommand{\nc}{\newcommand}

\nc{\beq}{\begin{equation}}
\nc{\eeq}{\end{equation}}
\nc{\barray}{\begin{eqnarray}}
\nc{\earray}{\end{eqnarray}}
\nc{\barrayn}{\begin{eqnarray*}}
\nc{\earrayn}{\end{eqnarray*}}
\nc{\bcenter}{\begin{center}}
\nc{\ecenter}{\end{center}}
\nc{\mc}{\mathcal}
\nc{\er}[1]{(\ref{eq:#1})}
\nc{\onehalf}{\frac{1}{2}} 
\nc{\partialbar}{\bar{\partial}}
\nc{\psit}{\widetilde{\psi}}
\nc{\Tr}{\mbox{Tr}}
\nc{\hc}{\mbox{H.c.}}
\nc{\ev}{\;\mathrm{eV}}
\nc{\mev}{\;\mathrm{MeV}}
\nc{\gev}{\;\mathrm{GeV}}
\nc{\kev}{\;\mathrm{keV}}
\nc{\tev}{\;\mathrm{TeV}}

\def\chii0{\chi_i^0}
\def\chij0{\chi_j^0}

\newcommand{\gsim}{\lower.7ex\hbox{$\;\stackrel{\textstyle>}{\sim}\;$}}
\newcommand{\lsim}{\lower.7ex\hbox{$\;\stackrel{\textstyle<}{\sim}\;$}}
\nc{\ttbar}{t\bar t}

\newcommand{\fref}[1]{Fig.~\ref{f.#1}}
\newcommand{\eref}[1]{Eq.~(\ref{e.#1})}

\newcommand{\aref}[1]{Appendix~\ref{a.#1}}
\newcommand{\sref}[1]{Section~\ref{s.#1}}
\newcommand{\ssref}[1]{Section~\ref{ss.#1}}

\newcommand{\cref}[1]{Chapter~\ref{c.#1}}
\newcommand{\tref}[1]{Table~\ref{t.#1}}

\def\mbarsubscript{\mathrm{halo}}

\graphicspath{{plots/}}

\title{Direct Detection of Mirror Matter in Twin Higgs Models}

\author[f]{Zackaria Chacko,}
\author[f, i,n]{David Curtin,}
\author[f,l]{Michael Geller,}
\author[f,y]{Yuhsin Tsai}

\affiliation[f]{Maryland Center for Fundamental Physics, Department of Physics,
University of Maryland, College Park, MD 20742-4111 USA}
\affiliation[i]{Department of Physics, University of Toronto, Toronto, ON M5S 1A7, Canada}
\affiliation[n]{Perimeter Institute for Theoretical Physics, Waterloo, Ontario N2L 2Y5, Canada}
\affiliation[l]{School of Physics and Astronomy, Tel Aviv University, Tel Aviv 6997801, Israel}
\affiliation[y]{Department of Physics, University of Notre Dame, IN 46556, USA}

\emailAdd{zchacko@umd.edu}
\emailAdd{dcurtin@physics.utoronto.ca}
\emailAdd{micgeller@tauex.tau.ac.il}
\emailAdd{ytsai3@nd.edu}

\abstract{
 We explore the possibility of discovering the mirror baryons and 
electrons of the Mirror Twin Higgs model in direct detection 
experiments, in a scenario in which these particles constitute a 
subcomponent of the observed DM. We consider a framework in which the 
mirror fermions are sub-nano-charged, as a consequence of kinetic mixing 
between the photon and its mirror counterpart. We consider both nuclear 
recoil and electron recoil experiments. The event rates depend on the 
fraction of mirror DM that is ionized, and also on its distribution in 
the galaxy. Since mirror DM is dissipative, at the location of the Earth 
it may be in the form of a halo or may have collapsed into a disk, 
depending on the cooling rate. For a given mirror DM abundance we 
determine the expected event rates in direct detection experiments for 
the limiting cases of an ionized halo, an ionized disk, an atomic halo 
and an atomic disk. We find that by taking advantage of the 
complementarity of the different experiments, it may be possible to 
establish not just the multi-component nature of mirror dark matter, but 
also its distribution in the galaxy. In addition, a study of the recoil 
energies may be able to determine the masses and charges of the 
constituents of the mirror sector. By showing that the mass and charge 
of mirror helium are integer multiples of those of mirror hydrogen, 
these experiments have the potential to distinguish the mirror nature of 
the theory. We also carefully consider mirror plasma screening effects, 
showing that the capture of mirror dark matter particles in the Earth 
has at most a modest effect on direct detection signals.
 }

\begin{document}

\begin{flushright}
\small{.}
\end{flushright}

\maketitle

\section{Introduction}\label{s.introduction}

The Mirror Twin Higgs (MTH) 
framework~\cite{Chacko:2005pe,Barbieri:2005ri,Chacko:2005vw} offers a 
simple and distinctive solution to the little hierarchy problem of the 
Standard Model (SM). In this class of theories the spectrum of light 
states includes a complete mirror (``twin") copy of the SM, with the same 
particle content and gauge groups. A discrete $\mathbb{Z}_2$ interchange 
symmetry relates the particles and interactions in the mirror sector to 
those in the SM. The Higgs emerges as the pseudo-Nambu Goldstone boson 
of an approximate global symmetry which is spontaneously broken. 
Quadratically divergent corrections to the Higgs mass are cancelled by a 
combination of the global symmetry and the discrete $\mathbb{Z}_2$ 
symmetry that relates the SM and twin sectors. MTH models stabilize the 
Higgs mass against radiative corrections up to scales of order 5-10 TeV, 
above which an ultraviolet completion~\cite{Falkowski:2006qq,Chang:2006ra,Craig:2013fga,Katz:2016wtw,Badziak:2017syq,Badziak:2017kjk,Badziak:2017wxn,Geller:2014kta,Barbieri:2015lqa,Low:2015nqa} is required.

In the MTH construction, there are no new light states charged under the 
SM gauge groups. Therefore, this class of models is free from the strong 
constraints on top partner searches at the Large Hadron Collider 
(LHC)~\cite{Aaboud:2017ayj, Aaboud:2017nfd, Aaboud:2017ejf, 
Aaboud:2016tnv, Sirunyan:2017pjw, Sirunyan:2017cwe, 
Khachatryan:2017rhw}. The only coupling between the SM and mirror 
sectors that is required by the construction is a Higgs portal 
interaction between the SM Higgs and its twin partner. Then, after 
electroweak symmetry breaking in the two sectors, the SM Higgs boson 
mixes with its twin counterpart. As a result of this mixing, the Higgs 
acquires couplings to twin fermions and gauge bosons. This allows twin 
particles to be produced at collider experiments through the Higgs 
portal. The twin particles are invisible and give rise to missing 
energy signals at colliders. However, the event rate is low, and missing 
energy searches at the LHC have only limited sensitivity to twin 
particle production.

The tightest collider constraints on MTH models are from precision Higgs 
measurements at the LHC. As a result of the mixing between the SM Higgs 
and its twin counterpart, the couplings of the Higgs to SM particles are 
suppressed. In addition, the Higgs acquires couplings to mirror fermions 
and gauge bosons, and can decay into them. Both these effects contribute 
to a reduction in the number of Higgs events at the LHC as compared to 
the SM prediction~\cite{1992MPLA....7.2567F}. The fact that the number 
of Higgs events observed at the LHC is consistent with the expectation 
from the SM can be used to place constraints on MTH models.  In order to 
satisfy this constraint, we require a mild hierarchy between the scale 
of electroweak symmetry breaking in the twin sector, denoted by $\hat{v}$, 
and the corresponding scale in the SM sector $v$. The constraint is 
satisfied provided $\hat{v}/v \gtrsim 3$~\cite{Burdman:2014zta}. We can 
realize this hierarchy by introducing a soft explicit breaking of the 
discrete $\mathbb{Z}_2$ symmetry that relates the two sectors, albeit at 
the expense of mild tuning. 
(This $\mathbb{Z}_2$ breaking can also occur spontaneously, with a possible reduction of tuning, see e.g.~\cite{Harnik:2016koz,Beauchesne:2015lva,Yu:2016bku,Yu:2016swa,Batell:2019ptb}.) 
A phenomenologically important consequence 
of the difference in the scales of electroweak symmetry breaking is that 
the elementary fermions and gauge bosons in the twin sector are heavier 
by a factor of $\hat{v}/v$ than their SM counterparts.

The MTH framework is severely constrained by cosmology. The Higgs portal 
interaction keeps the SM and twin sectors in thermal equilibrium until 
temperatures of order a GeV~\cite{Barbieri:2005ri}. Below this 
temperature, even though the two sectors are decoupled, mirror states 
continue to contribute almost half of the total energy density in the 
universe. This results in a large contribution to the energy density in 
dark radiation during the CMB epoch, $\Delta N_{eff}=5.7$ 
\cite{Chacko:2016hvu,Craig:2016lyx}. An effect of this size is excluded 
by the current bounds, which require $\Delta 
N_{eff}\lsim0.25\,(2\sigma)$ from Planck 2018+lensing+BAO, or $\Delta 
N_{eff}\lsim0.49\,(2\sigma)$~\cite{Aghanim:2018eyx} if including the 
$H_0$ measurement from Ref.~\cite{2018ApJ...855..136R}. This problem can 
be solved if there is an additional source of breaking of the discrete 
$\mathbb{Z}_2$ symmetry. This would allow the number of degrees of 
freedom in the twin sector at the time when the two sectors decouple to 
be much less than in the SM, leading to a suppression of $\Delta 
N_{eff}$ 
\cite{Barbieri:2005ri,Chang:2006ra,Farina:2015uea,Barbieri:2016zxn,Csaki:2017spo,Barbieri:2017opf,Bishara:2018sgl,Liu:2019ixm,Harigaya:2019shz}. 
The same result can be achieved by making the mirror sector vector-like 
\cite{Craig:2016kue}. The most radical proposal in this regard is the 
Fraternal Twin Higgs (FTH) construction~\cite{Craig:2015pha}, in which 
the first two generations of twin fermions, which do not play a 
significant role in the solution of the little hierarchy problem, are 
simply removed from the theory. The FTH framework leads to distinctive 
collider signatures at the LHC involving displaced 
vertices~\cite{Curtin:2015fna,Csaki:2015fba,Kilic:2018sew,Alipour-fard:2018mre}.

An alternative approach to address the problem, which does not require 
additional breaking of the $\mathbb{Z}_2$ symmetry that relates the two 
sectors, is to introduce into the theory an asymmetric reheating process 
that preferentially heats up the SM 
sector~\cite{Berezhiani:1995yi,Berezhiani:1995am,Adshead:2016xxj}. The 
asymmetric reheating should occur at temperatures below 1 GeV, after the 
two sectors have decoupled, but before Big Bang nucleosynthesis (BBN). 
This reduces the fraction of energy density contained in the twin 
sector, allowing the bounds on $\Delta N_{eff}$ to be satisfied 
\cite{Chacko:2016hvu,Craig:2016lyx}. Although the mirror sector 
contribution to $\Delta N_{eff}$ is suppressed in this scenario, it is 
expected to be large enough to be observed in future CMB experiments. 
Both the MTH and FTH frameworks contain several promising dark matter (DM) 
candidates~\cite{Craig:2015xla,Garcia:2015loa,Garcia:2015toa,Farina:2015uea,Farina:2016ndq,Prilepina:2016rlq,Hochberg:2018vdo,Cheng:2018vaj,Terning:2019hgj,Koren:2019iuv,Badziak:2019zys,Feng:2020urb}.

 If there is a baryon asymmetry in the twin sector, the mirror baryons 
and electrons will constitute a subcomponent of the DM in 
the universe. If the discrete $\mathbb{Z}_2$ twin symmetry is only 
softly broken, so that the masses of mirror particles are fixed by the 
ratio of electroweak VEVs in the two sectors, $\hat{v}/v$, the mirror 
baryons are primarily composed of mirror hydrogen and helium. The 
relative abundances of these two species is determined by the dynamics 
of BBN in the twin sector. Together with the mirror photons and 
neutrinos, the mirror baryons and electrons can give rise to highly 
distinctive signals in large scale structure and in the cosmic microwave 
background~\cite{Chacko:2018vss}. Baryon acoustic oscillations in the 
mirror sector prior to recombination lead to a suppression of structure 
on large scales. Current limits on the size of this effect bound the 
mirror contribution to DM in the MTH framework to be less than 
$\mathcal{O}(10\%)$.

Apart from the Higgs portal coupling, the gauge symmetries of the MTH 
construction allow only one other renormalizable interaction that 
connects the SM and twin sectors, a kinetic mixing between hypercharge 
and its twin counterpart,
 \begin{equation}
\frac{\epsilon}{2 {\rm cos} \; \theta_W} B_{\mu \nu} B^{\prime \mu \nu} \;.
 \end{equation}
 If this operator is present, the twin fermions acquire a charge under 
electromagnetism proportional to $\epsilon$. Avoiding thermalization of 
the hidden and visible sector after asymmetric reheating constrains any 
such mixing to be very small, less than or of order 
$10^{-9}$~\cite{Vogel:2013raa}. However, even such small values of the 
mixing are radiatively stable in the minimal MTH construction, since 
this mixing is not generated through 3-loop order~\cite{Chacko:2005pe}. 
If there is a baryon asymmetry in the twin sector, the mirror fermions 
will therefore constitute sub-nano-charged DM and can scatter off 
ordinary matter through processes involving the exchange of the photon. 
Interestingly, it has recently been shown that higher-order loop 
diagrams involving gravitons could generate a kinetic mixing between the 
visible and hidden sectors~\cite{Gherghetta:2019coi}. This contribution, 
which primarily arises from energies of order the Planck scale, can give 
rise to mixings of the order $\epsilon \sim 10^{-13}$. Tantalizingly, 
this tiny mixing is compatible with the asymmetric reheating mechanism 
while providing an achievable sensitivity goal for direct detection 
experiments.

 In this paper we explore the possibility of discovering the twin baryons 
and electrons of the MTH scenario in current and next-generation direct 
detection experiments. We consider a framework in which mirror matter is 
sub-nano-charged, as a consequence of kinetic mixing between the hypercharge 
gauge boson of the SM and its massless mirror counterpart. For 
concreteness, we assume that the discrete $\mathbb{Z}_2$ twin symmetry 
is only softly broken, so that the masses of mirror particles are fixed 
by the ratio of electroweak VEVs in the two sectors, $\hat{v}/v$. The 
constraint on this scenario from $\Delta N_{eff}$ is assumed to be 
satisfied as a consequence of late-time asymmetric reheating. Since 
mirror matter, like visible matter, is dissipative, some fraction of the 
twin DM in the galaxy may have collapsed into a disk. The direct 
detection signal then depends in part on whether a twin disk is present 
and, if so, the fraction of mirror matter it contains, its alignment 
relative to the visible disk, and whether it extends out to the location 
of the Earth. The size of the signal also depends on whether the mirror 
matter in the galaxy is in ionic form or has condensed into atoms.

In order to understand the distribution of mirror matter in the galaxy, 
and whether it is in the form of ions or atoms, it is necessary to track 
how this subcomponent of DM evolved in time as the Milky Way was 
forming. When halo formation begins at redshifts of $\mathcal{O}(10)$, 
the shock wave induced by the in-falling twin atoms heats up and 
reionizes the mirror sector. The mirror sector can dissipate its energy 
through the emission of twin photons in processes involving the 
scattering of twin particles. The timescale of this cooling process 
depends on the abundance of mirror particles and is longer than in the 
SM. We find that this timescale can 
nevertheless be shorter than the age of the universe for sufficiently large abundances of mirror matter. This indicates 
that some fraction of the the mirror halo may have collapsed into a 
disk.

In our analysis we consider experiments based on both nuclear recoil (NR) and 
electron recoil (ER) signals. For a given mirror DM abundance we determine 
the expected event rates in direct detection experiments for the 
limiting cases of an ionized halo, an ionized disk, an atomic halo and 
an atomic disk. We are careful to account for the effects of mirror 
matter capture in the Earth. We find that in most of the relevant 
parameter space, its effect on the direct detection signals we consider 
is negligible or at most modest. By taking advantage of the complementarity of the different 
experiments, we find that it may be possible to establish not just the 
multi-component nature of mirror DM, but also its distribution in the 
galaxy. In addition, a study of the recoil energies may be able to 
determine the masses of the mirror DM constituents. By establishing that 
the masses and charges of mirror hydrogen and helium are integer 
multiples of each other, these experiments may be able to diagnose the 
mirror nature of the theory. 
There is also an important complementarity between direct detection experiments and astrophysical probes of mirror matter. The reach of the former is best for mirror baryons arranged in a halo. On the other hand, dark disk scenarios can be probed very sensitively via white dwarf cooling bounds~\cite{Curtin:2020tkm}, and are also more likely to lead to the formation of mirror stars, which can be detected in optical and $X$-ray observations~\cite{Curtin:2019lhm,Curtin:2019ngc}, microlensing surveys~\cite{Winch:2020cju}, and gravitational wave observations of mirror neutron star mergers~\cite{Hippert:2021fch}.

Although our focus is on MTH models with softly broken twin symmetry, 
the direct detection signatures we study are also features of the more 
general class of models in which mirror baryons and electrons constitute 
some or all of the observed DM. Earlier work on the distribution of 
mirror DM in the galaxy may be found, for example, 
in~\cite{Mohapatra:1996yy,Mohapatra:2000qx,Roux:2020wkp}. Direct 
detection of mirror matter has been considered, for example, 
in~\cite{Foot:2003iv,An:2010kc,Foot:2010hu,Addazi:2015cua,Clarke:2016eac}. 
Reviews of mirror models and mirror DM, with many additional references, 
may be found 
in~\cite{Berezhiani:2003xm,Okun:2006eb,Ciarcelluti:2010zz,Foot:2014mia}. 
In detail, however, the direct detection signals depend sensitively on 
the masses of the mirror particles and their distribution in the galaxy. 
From this perspective, our paper represents a detailed study of the 
direct detection signals of generalized mirror-like models in the region of parameter 
space motivated by the little hierarchy problem.

The outline of this paper is as follows. In the next section we give a 
quick review of the parameter space of the MTH model. In 
Sec.~\ref{s.distribution} we study the distribution of mirror particles 
in the Milky Way, based on an estimate of the rate of twin particle 
cooling after the shock wave heating process. In 
Sec.~\ref{s.directdetection}, we estimate the signal rates in direct 
detection experiments, considering both nuclear and electron recoils. 
Our conclusions are in Sec.~\ref{s.conclusion}. Mirror matter capture in 
the Earth and its effects on direct detection are carefully analyzed in 
\aref{capture}.

\section{Parameters of the Model}\label{s.parameters}

 Our focus is on the direct detection signals of MTH models in which the 
mirror nuclei and electrons constitute a subcomponent of DM. We restrict 
our analysis to the case when the Yukawa couplings respect the discrete 
$Z_2$ symmetry that relates the two sectors. Then the elementary 
fermions in the twin sector are heavier than their visible counterparts 
by a factor of $\hat{v}/v$, the ratio of electroweak symmetry breaking 
scales in the two sectors. The energy density in twin radiation is 
assumed to be diluted by late time asymmetric reheating after the two 
sectors have decoupled, allowing the current CMB and BBN constraints on 
dark radiation to be satisfied.\footnote{We assume the masses of both 
visible and hidden sector neutrinos can be neglected.}
Then, in this framework, the direct detection signals depend on four 
parameters,
 \begin{equation}
 \label{eq:rall}
\epsilon, \qquad  \hat{v}/v, \qquad 
r_\mathrm{all}  = \Omega_{\text{all\ mirror\ baryons}}/\Omega_{\text{DM}},\qquad \hat 
Y_p(^4\mathrm{\hat He})=\frac{\rho_{^4\mathrm{\hat He}}}{\rho_{\mathrm{\hat H}}+\rho_{^4\mathrm{\hat He}}}.
 \end{equation}
 Here $\epsilon$ parametrizes the kinetic mixing between the hypercharge 
gauge bosons in the two sectors, while $r_\mathrm{all}$ denotes the 
\emph{total} asymmetric mirror baryon density relative to the total DM 
density today. Just as in the case of the SM, the contributions of the 
twin sector to the matter density are almost entirely from mirror 
hydrogen and helium,
 \begin{equation}
r_\mathrm{all} = r_{\hat{\text{H}}} + r_{\hat{\text{\text{H}}}\text{e}}.
 \end{equation}
 The parameter  $\hat Y_p(^4\mathrm{\hat He})$ represents the 
mass fraction contributed by twin helium.

The masses of particles in the twin sector depend on the ratio 
$\hat{v}/v$. While Higgs coupling measurements at the LHC constrain 
$\hat{v}/v \gtrsim 3$, the requirement that the Higgs mass be only 
modestly tuned limits $\hat{v}/v \lesssim 5$. The mass of the twin 
electron is simply $\hat{v}/v$ times the corresponding value in the SM. 
Since the quark masses are also $\hat{v}/v$ times larger than in the SM, 
the different running of the mirror QCD gauge coupling leads to a larger 
confinement scale in the mirror sector than in the SM by about 30-50\% 
in the range $\hat{v}/v = 3$-$5$. This makes twin baryons heavier than 
SM baryons by about $30-50\%$~\cite{Chacko:2018vss}.

 Since mirror particles constitute an acoustic subcomponent of DM, they
lead to a suppression of large scale structure on scales that enter the
horizon prior to recombination in the twin sector. This can be used to
place limits on the contribution of mirror matter to the observed
density of DM, $r_\mathrm{all} \lesssim 10\%$~\cite{Chacko:2018vss}.

The relative fractions of mirror hydrogen and helium in the early 
universe are determined by the dynamics of BBN in the twin sector. This 
in turn depends on the masses of the mirror baryons and also on the 
energy density in mirror radiation at the time of BBN. 
In~\cite{Chacko:2018vss}, the Boltzmann equations for the number 
changing process $\hat{n}\,\hat{\nu}\leftrightarrow\hat{p}\,\hat{e}$ 
were solved for the MTH model, and the timescale for mirror deuterium 
formation was determined. It was found that $\hat Y_p(^4\mathrm{\hat 
He})\approx 75\%$ for $\hat{v}/v$ in the range we consider and realistic 
values of $\Delta N_{eff}$. However, in our analysis we also consider 
the cases in which the twin baryons are composed entirely of mirror 
hydrogen or mirror helium, corresponding to $\hat Y_p(^4\mathrm{\hat 
He}) = 0$ and $\hat Y_p(^4\mathrm{\hat He})= 1$. These provide some 
insight into the direct detection signals of MTH models in which the 
Yukawa couplings of the light quarks exhibit hard breaking of the 
discrete $Z_2$ symmetry, so that the spectrum of mirror nuclei is 
composed of only a single species, either hydrogen or helium.

\section{Mirror Baryon Distribution in the Milky Way}\label{s.distribution}

Structure formation reaches the regime of nonlinear halo formation at 
redshifts $z \sim \mathcal{O}(10)$~\cite{2010gfe..book.....M}. At these 
redshifts, DM and the SM particles, which include nuclei, electrons, and 
photons, undergo complicated collective dynamics that gives rise to the 
structure of the Milky Way and the other galaxies that we observe today. 
Collisionless DM particles clump under the action of gravity, eventually 
giving rise to cold DM (CDM) distributions such as the NFW or Burkert profiles 
(see e.g.~\cite{Nesti:2013uwa}). The SM baryons, which are initially 
bound in atoms following recombination, fall into the overdense regions 
and collide with each other, leading to the formation of a shock wave 
that expands outwards to heat the baryonic medium. The maximum 
temperature of the baryons is dictated by the the virial theorem and the 
available gravitational energy, which is dominated by the CDM 
halo. The immediate aftermath of shock heating is a fully ionized baryon 
distribution that is in hydrostatic equilibrium. This distribution 
initially satisfies the adiabatic equation of state, but quickly evolves 
to reach thermal equilibrium. Subsequently, processes such as 
bremsstrahlung and ionization cooling lower the temperature of the 
baryons, leading to a loss of pressure support and the eventual onset of 
catastrophic collapse. If the halo has sufficient angular momentum and a 
quiet merger history, this collapse eventually gives rise to a disk such 
as the one in our own Milky Way galaxy. Even after the disk has formed, 
a significant fraction of the baryonic gas remains outside the disk at 
large distances from the galactic core~\cite{2014ApJ...792....8W}.

If mirror baryons make up a small $\lesssim \mathcal{O}(10\%)$ fraction 
of DM, we expect that, during halo formation, they will undergo broadly 
similar dynamics to SM baryons. Mirror atoms will also fall into 
overdense regions, undergo shock heating and ionization, and reach 
adiabatic equilibrium before eventually settling into thermal 
equilibrium. They then cool, potentially leading to collapse and the 
formation of a mirror disk. This kind of dynamics for a DM component has 
been studied in the past, primarily in the context of exact mirror DM 
models that are perfect hidden sector copies of the SM, though there 
have been some early studies of how cosmology and galactic evolution 
might have proceeded for mirror matter with a larger mirror Higgs vev 
than in the SM~\cite{Mohapatra:1996yy}. More recently, this scenario has 
also been explored in the more general context of dissipative DM models 
that could form a so-called `dark disk'~\cite{Fan:2013yva, Fan:2013tia, 
Kramer:2016dqu, Kramer:2016dew}, though without the presence of `dark 
nuclear physics' which, as we describe below, can significantly 
complicate galactic evolution in mirror models.

In spite of this general understanding, it is not possible to make 
precise predictions about the distribution of mirror DM in the 
galaxy in the MTH scenario. Even in the case of ordinary baryonic 
matter, collapse and disk formation is a highly nonlinear process that 
depends sensitively on various radiative and mechanical feedback 
mechanisms including star formation, stellar winds, and heating from 
supernovae, as well as the influence of the central supermassive black 
hole. Even for visible matter, modelling these processes requires 
detailed $N$-body simulations incorporating magnetohydrodynamics and 
stellar 
feedback~\cite{10.1093/mnras/stu1738,10.1093/mnras/sty674,10.1093/mnras/sty1690,10.1093/mnras/stx1160,10.1093/mnras/sty3336, 
Vogelsberger:2014kha,Vogelsberger:2014dza,Genel:2014lma,Sijacki:2014yfa, 
Schaye:2014tpa,Schaller:2015vsa,Schaller:2014uwa}. While the formation 
of our Milky Way disk is beginning to be better understood, the 
simulations are not yet fine-grained enough to make direct contact with 
the astrophysics of individual stars, let alone fundamental physics 
parameters. Instead, these simulations rely on large-scale 
parameterizations of processes like star formation and heating from 
supernovae explosions to reproduce the known Milky Way structure. Given 
that the particle spectrum and dynamics of the MTH are different enough 
from the SM that detailed analogies break down, it is not feasible to 
robustly predict the distribution of mirror baryons in the galaxy.

We therefore take a more modest approach. In this section, we compute 
the mirror baryon distributions resulting from hydrostatic equilibrium 
in the gravitational background of the CDM halo (neglecting 
the effects of halo angular momentum). These distributions can be 
interpreted as the rough starting point for the nonlinear processes of 
collapse. By examining the cooling rates arising from various processes 
and comparing to the predictions for SM-like baryons computed under the 
same assumptions, we can obtain some insight into how the mirror baryon 
distribution might be expected to evolve in our Milky Way. The most 
important question is whether the mirror matter collapses to form its 
own disk. We find that in a large part of the parameter range, because 
of the large uncertainties, the answer is ambiguous. For this reason, in 
our study of direct detection in Section~\ref{s.directdetection}, we 
consider both halo and disk distributions of mirror matter.

\subsection{Initial Mirror Baryon Distribution}

In this subsection we compute the initial mirror matter distribution in 
our Milky Way prior to the onset of cooling, assuming a standard NFW or 
Burkert CDM distribution for the primary DM component and 
hydrostatic equilibrium. This will allow us to estimate the cooling 
timescale in Section~\ref{s.darkbaryoncooling}. We pay particular 
attention to how the distribution of mirror baryons compares to that of 
SM baryons computed under the same assumptions. This will give us some 
insight into how the cooling timescale of mirror baryons, and 
consequently their current distribution, might be expected to differ 
from that of the SM baryons in our Milky Way.

\subsubsection{CDM profile}

We assume that the mirror particles contribute only a small component of 
the total energy density in DM, which is dominated by standard 
CDM with distribution $\rho_\mathrm{CDM}(r)$. Then the contribution of 
mirror particles to the gravitational potential can be neglected. To 
examine the sensitivity of our results on the CDM distribution, we 
consider two possibilities, an NFW profile and a Burkert profile,
 \begin{eqnarray}
\rho_\mathrm{NFW}(r) &=& \frac{\rho_H}{\frac{r}{R_H}\left(1+\frac{r}{R_H}\right)^2} \ , \\ \nonumber \\
\rho_\mathrm{BUR}(r) &=& \frac{\rho_H}{\left(1+\frac{r}{R_H}\right)\left(1+\frac{r^2}{R_H^2}\right)} \ . 
 \end{eqnarray}
 The benchmark parameters we assume are based on the studies 
in~\cite{Nesti:2013uwa}, and are summarized in 
\tref{CDMprofileparameters}. To facilitate comparison, we assume common 
values for $R_\odot$, the distance of the Sun from the Milky Way center, 
and the local CDM density $\rho_\odot = \rho(R_\odot)$. The profiles are 
then completely fixed by specifying the remaining parameter 
$R_H$.\footnote{We have explored the effect of the uncertainties on the 
fitted parameters of a given CDM profile on our results and found them 
to be negligible compared to the difference between these two profiles.} 
For $R > R_\mathrm{\odot}$, the NFW and Burkert profiles are very 
similar, but the NFW profile predicts much higher densities closer to 
the core. These profiles are shown as black curves in 
\fref{mirrorprofiles} (top).

Let $M_\mathrm{CDM}(R)$ represent the total mass of DM enclosed 
within radius $R$. We define the total size of the halo by the virial 
radius $R_{vir}$. This is determined from the standard overdensity 
criterion, $R_{vir} \equiv R_{200}$, as
 \begin{equation}
M_\mathrm{CDM}(R_{200}) =
\Delta  \cdot \rho_\mathrm{crit}  \cdot \frac{4 \pi}{3} R_{200}^3 \ , 
 \end{equation} 
 where $\Delta = 200$ and $\rho_\mathrm{crit} \approx 4.8 \times 10^{-6} 
\gev/\mathrm{cm}^3$ is the critical density.

The virial theorem allows us to relate the average kinetic and potential energy of CDM particles, 
 $\frac{1}{2} |U_\mathrm{tot}| = \mathrm{KE}_\mathrm{tot} = \frac{1}{2} 
M(R_{vir}) v_0^2$. 
We can then determine the average 
velocity-squared of the particles that constitute the primary component 
of DM,
 \begin{equation}
 \label{e.Tvir}
v_0^2 =  \frac{\int_0^{R_{vir}}{G M_\mathrm{CDM}(r) \,\rho_\mathrm{CDM}(r) \ r \ dr}}{\int_0^{R_{vir}}{\rho_\mathrm{CDM}(r) \ r^2  \ dr}} \ .
 \end{equation}
 If mirror matter were collisionless, the mirror atoms would exhibit the 
same distribution as the primary DM component, with the same 
root-mean-square velocity. On timescales short compared to the cooling 
timescale, the only effect of collisions is to redistribute the energy 
of the mirror particles amongst themselves, leaving their total energy 
unchanged. In what follows, we use this fact, together with the condition 
of hydrostatic balance, to determine the distribution of mirror matter 
in adiabatic and thermal equilibrium prior to the onset of cooling.

\begin{table}
\begin{center}
\begin{tabular}{|l||l|l|l|l|l|}
\hline
Profile &
$R_{\odot}$ 
& 
$\rho(\odot)$ 
&
$R_H$ 
&
$R_{vir}$ 
&
$T_{vir}/\bar m_\mbarsubscript$
\\
\hline \hline
NFW & 8 & 0.5  & 16 & 235 & $1.35 \times 10^{-7}$ \\
\hline
Burkert & 8 & 0.5  & 9 & 209 & $1.21 \times 10^{-7}$\\
\hline
\end{tabular}
\end{center}
\caption{
Parameters of CDM distributions, which dominate the gravitational potential of the Milky Way. 
All distances in kpc, $\rho$ in GeV/cm$^3$. 
$\bar m_\mbarsubscript$ is the halo-averaged mass of a virialized sub-population of the halo, such as the MTH mirror baryons. 
}
\label{t.CDMprofileparameters}
\end{table}

\subsubsection{Mirror Matter in Hydrostatic Equilibrium}
\label{s.HSEQ}

The mirror baryons, like the SM baryons, are shock heated and ionized as 
they fall into the collapsing CDM halo. During this process, and 
immediately afterwards, the mirror baryons remain well mixed. Matter is 
churned around in the profile at the convection timescale,
 \begin{equation}
t_{convection}(r)\sim\sqrt{\frac{1}{G\,\rho_{\rm CDM}(r)}}\sim 10^8\,{\rm yrs} \ \ \ \mbox{ at $r = R_\odot$}.
 \end{equation}
 See also Figs.~\ref{f.timescalesNFW} and~\ref{f.timescalesBUR}. This is 
comparable to the time scale when the non-linear halo formation sets in at $z=\mathcal{O}(10)$, so 
convection quickly establishes the mirror baryons in an adiabatic 
distribution. In this configuration, the mirror matter forms a dark 
plasma that is pressure supported while it remains hot enough to stay 
ionized. However, as we shall see, the distribution quickly evolves to 
become isothermal, on a timescale dictated by the diffusion timescale in 
the adiabatic profile.

In computing the mirror halo profiles, we will assume that mirror 
hydrogen and helium are fully ionized. The fraction of partially or 
fully recombined mirror baryons is far too small to affect the 
calculation of mirror density and temperature profiles. It does however 
have an important impact on the various cooling mechanisms considered in 
\sref{cooling}. We will therefore self-consistently determine the actual 
degree of ionization consistent with the computed mirror baryon profiles 
in \sref{ionization}.
We define the average mass of the mirror matter particles 
at any location in the halo as
 \begin{equation}
\bar{m}(\vec{r}) \equiv \frac{\sum_i n_i(\vec{r}) m_i}{\sum_i n_i(\vec{r})} \ ,
 \end{equation}
 where $i$ runs over mirror hydrogen and helium, and also over the 
electrons.  Approximating $m_{\mathrm{\hat He}} = 4 m_\mathrm{\hat H}$, 
the resulting \emph{local} average mass of mirror particles in the limit 
of full ionization is given by,
 \begin{equation}
\bar{m}(\vec{r}) \ \  = \ \  m_\mathrm{\hat H}  \ \frac{4}{8 - 5 \hat Y(\vec{r})} \; ,
\label{relatehatm&Y}
 \end{equation}
 where $\hat{Y}(\vec{r})$ is the local mirror helium mass fraction. 

Initially, the mirror baryons and electrons are well-mixed, so that 
$\bar{m}$ has the same value $\bar{m}_\mbarsubscript$ everywhere in the 
halo. The value of $\bar{m}_\mbarsubscript$, the average mirror particle 
mass over the whole halo, can be obtained from Eq.~(\ref{relatehatm&Y}) 
by assuming that the total mirror helium mass fraction follows the 
cosmological average $\hat Y_p(^4\mathrm{\hat He})$,
 \begin{equation}
\bar{m}_\mbarsubscript \ \  = \ \  m_\mathrm{\hat H}  \ \frac{4}{8 - 5 \hat Y_p(^4\mathrm{\hat{H}e})} \; .
 \end{equation}

We now determine the distribution of mirror particles in the adiabatic 
configuration. We start from the observation that in both the adiabatic 
and isothermal configurations the mirror baryons satisfy the condition 
of hydrostatic equilibrium in the gravitational background of the CDM 
distribution,
 \begin{equation}
\label{hydrostatic}
\frac{dP}{dr} = - \frac{G M_\mathrm{CDM}(r) \rho(r)}{r^2} \ .
 \end{equation}
 Here $P(r)$ and $\rho(r)$ are the local pressure and density of the 
mirror particles, which are related by their equation of state. The 
density $\rho(r)$ can be determined from the local number densities of 
the mirror particles.

The next step is to determine the average energy of a mirror particle 
prior to cooling. Initially, mirror DM and the primary 
component of DM are well-mixed. Since the primary component of 
DM dominates the gravitational potential of the halo, a small 
density of mirror matter would arrange itself in exactly the same 
distribution as the primary component if the mirror particles were 
collisionless. In this scenario, the average energy of a mirror particle 
would be $-(3/2)T_\mathrm{vir}$, with $T_{vir}/\bar m_\mbarsubscript \sim 10^{-7}$, see 
\tref{CDMprofileparameters}. The dominant effect of collisions between 
the mirror particles is just to redistribute their energy amongst 
themselves, so that their total energy is conserved, up to the small 
fraction of energy that is used to ionize the mirror atoms. The cooling 
timescale is long compared to the timescales for convection or 
diffusion. This means that for mirror particles in an adiabatic or 
isothermal halo, the average energy per particle (the sum of kinetic and 
potential energy) is still given by $-(3/2)T_\mathrm{vir}$ minus the 
average ionization energy. We now use this fact, together with the 
condition of hydrostatic equilibrium, Eq~(\ref{hydrostatic}), to 
determine the distribution of mirror particles in an adiabatic halo 
prior to cooling.

Initially the mirror baryons are uniformly mixed, so that $\hat{Y}(r)$ 
is equal to the cosmic value $\hat Y_p(^4\mathrm{\hat He})$ and $\bar{m}$ 
constant throughout and equal to $\bar{m}_\mbarsubscript$. In the adiabatic regime, the 
mirror baryons obey the equation of state $P = A \rho^\gamma$ with 
$\gamma = {5}/{3}$ for a monoatomic gas, where $A$ is a constant 
independent of position. The total number density $n(r)$ is related to 
the local density $\rho(r)$ as $n(r) = \rho(r)/\bar m_\mbarsubscript$. The ideal gas law 
$\bar m_\mbarsubscript P(r) = \rho(r) T(r)$ then yields,
 \begin{eqnarray}
\rho(r) &=& \left( \frac{ A \bar m_\mbarsubscript}{T(r)}\right)^{\frac{1}{1-\gamma}} \;,
 \end{eqnarray}
 which relates the density profile to the temperature. Hydrostatic 
equilibrium then allows us to relate the temperature at an arbitrary 
point in the halo to the temperature at the center,
 \begin{equation}
T(r) = T(0) - G \bar m_\mbarsubscript \left( \frac{\gamma-1}{\gamma}\right) \int_0^r \frac{M_\mathrm{CDM}(\tilde r)}{\tilde r^2} d \tilde r \;.
 \end{equation}
 The temperature at the center of the distribution $T(0)$ is obtained by 
requiring that the average total energy of a mirror particle in the halo is 
$-(3/2)T_\mathrm{vir}$, up to small corrections arising from the 
ionization energies of the atoms.

 The adiabatic distribution quickly becomes isothermal, on timescales 
dictated by the diffusion process in the adiabatic profile. We can 
estimate the diffusion timescale by focusing on scattering between 
ionized $\hat{X} = \hat{\mathrm{H}}^+, \hat{\mathrm{He}}^{+,++}$, which 
can transfer heat more efficiently between ions than $\hat{X}\hat{e}$ 
scattering. The scattering cross section between $\hat{X}$'s can be 
estimated as $\sigma\sim \alpha^2/T^2$. For diffusion with a free 
streaming length $\lambda_{FS}\sim (\rho_{\rm 
mirror}\,\sigma/\bar{m}_\mbarsubscript)^{-1}$, the number of scatterings involved in 
moving in a random walk across a distance $L$ can be estimated as $N\sim 
(L/\lambda_{FS})^2$. Since the time it takes to undergo $N$ scatterings 
is of order $t\sim N\lambda_{\rm FS}/v_{\hat{X}}$, the characteristic 
timescale for diffusing through a distance $L$ is given by,
 \begin{eqnarray}
t_{\rm diffusion}(r,L)\sim \frac{\rho_{\rm mirror}\,\sigma}{\bar{m}_\mbarsubscript\,v_{\hat{X}}}\,L^2 \ .
 \end{eqnarray}
 The diffusion timescale in the adiabatic profile depends only modestly 
on $r$ and is in the range $\sim 10^6 - 10^8$ years for $L \sim 10$ kpc. 
It follows that the adiabatic gas quickly reaches thermal equilibrium 
and arranges itself in an isothermal distribution. We will therefore use 
the isothermal mirror halo distributions in our discussion of cooling in 
the next section.

We now determine the distribution of mirror matter in an isothermal halo 
prior to the onset of cooling. The isothermal halo is at a constant 
temperature $T = T_\mathrm{iso}$, but $\hat{Y}(r)$ and $\bar{m}(r)$ are 
now spatially dependent since constituents with different atomic weights 
settle at different distances from the center. 

For an isothermal distribution, the hydrostatic equilibrium condition Eqn.~(\ref{hydrostatic}) applies separately for different gas components $X$. 
Charge separation due to the different masses of mirror nuclei and electrons occurs on scales of the Debye length, which is negligible on galactic scales. It is therefore an excellent approximation to define the two dominant components of the mirror baryon distribution to be 
$\hat{X} = \{{\rm \hat{H}^+ + \hat e ^-} ,  {\rm \hat{H}e^{++}} + 2 \hat e ^-\}$, 
which allows us to only consider gravitational forces when solving for hydrostatic equilibrium. 
Applying the ideal gas law for each of these two components gives a partial pressure $P_X$ that is $(1+ Q_{\hat X})$ times higher than the pressure for neutral mirror hydrogen or helium atoms, where 
$Q_{\hat{X}} = 1$ (2) for  ${\rm \hat{H}^+}$ (${\rm \hat{H}e^{++}}$)~\cite{polsstellarstructure}. 
The  solution to Eqn.~(\ref{hydrostatic}) for each component is therefore
\begin{equation}
 n_{\hat{X}}(r) = n_{\hat{X}} (0) \exp\left[
- \frac{G m_{\hat{X}}}{T_\mathrm{iso}} \frac{1}{1 + Q_{\hat{X}}}
\int_0^r  \frac{M_\mathrm{CDM}(\tilde r)}{\tilde r^2} d\tilde r
\right] \;.
 \end{equation} 
 The number densities 
$n_X(0)$ at the center of the halo and the temperature $T_\mathrm{iso}$ 
are determined from $r_{\rm all}$ and $\hat{Y}$, and the condition that the 
average energy of mirror particles in the halo is $-(3/2)T_\mathrm{vir}$ 
minus the energy used to ionize the mirror atoms. The isothermal 
distribution has a greater density of matter near the center of the halo 
than the corresponding CDM distribution. The corresponding 
reduction in the gravitational potential energy results in an increase 
in temperature, so that $T_\mathrm{iso}$ is a factor of about $2$ larger 
than the the virial temperature $T_\mathrm{vir}$.

\subsubsection{Results}
\label{s.mirrorhaloresults}

\begin{table}
\begin{center}
\hspace*{-10mm}
\begin{tabular}{|m{29mm}||l|l|l|l|l|l|l|}
\hline
& & & & & & &
\\ 
Scenario (NFW)&  
$\displaystyle\frac{T_\mathrm{iso}}{\mathrm{eV}}$ & 
$\displaystyle\frac{v_\odot}{\mathrm{km}/\mathrm{s}}$ & \hspace*{-2mm} $(\chi_{\mathrm{H}^+}, \chi_{\mathrm{He}^{+}}, \chi_{\mathrm{He}^{++}})$ \hspace*{-3mm}& 
$\displaystyle\frac{\rho_\mathrm{CDM}}{\mathrm{GeV}/\mathrm{cm}^3}$ & 
$\displaystyle\frac{\rho_\mathrm{mirror}}{\mathrm{GeV}/\mathrm{cm}^3}$ & 
$\frac{1}{r_\mathrm{all}} \frac{\rho_\mathrm{mirror}}{\rho_\mathrm{CDM}} $ & $ \hat Y_\odot$
\\
& & & & & & & 
\\
\hline \hline
\hspace*{-4mm}
\begin{tabular}{l}SM Baryons \\ $r_\mathrm{all} = 0.01$ \end{tabular}
& 173 & 197 & $(1, 1.7\times 10^{-7}, 1)$ &  0.5 & 0.12 & 1.29 & $0.96$
\\ 
\hline \hline
\hspace*{-4mm}
\begin{tabular}{l}$\frac{\hat{v}}{v} = 3, r_\mathrm{all} = 0.01,$ \\ $\hat Y_p(^4\mathrm{\hat He}) = 0.75$ \end{tabular}
& 309 & 230 & $(1, 4\times 10^{-7}, 1)$ &  0.5 & 0.004 & 0.77 & $0.992$
\\
\hline 
\hspace*{-4mm}
\begin{tabular}{l}$\frac{\hat{v}}{v} = 3, r_\mathrm{all} = 0.01,$ \\ $\hat Y_p(^4\mathrm{\hat He}) = 0$ \end{tabular}
& 142 & 253 & $(1, -, -)$ &  0.5 & 0.0019 & 0.39 & 0
\\
\hline
\hspace*{-4mm}
\begin{tabular}{l}$\frac{\hat{v}}{v} = 3, r_\mathrm{all} = 0.01,$ \\ $\hat Y_p(^4\mathrm{\hat He}) = 1$ \end{tabular}
& 369 & 249 & $(-, 3\times 10^{-7}, 1)$ &  0.5 & 0.0021 & 0.42 & 1
\\
\hline
\end{tabular}

\vspace{10mm}

\hspace*{-10mm}
\begin{tabular}{|m{29mm}||l|l|l|l|l|l|l|}
\hline
& & & & & & &
\\ 
Scenario (Burkert)&  
$\displaystyle\frac{T_\mathrm{iso}}{\mathrm{eV}}$ & 
$\displaystyle\frac{v_\odot}{\mathrm{km}/\mathrm{s}}$ & \hspace*{-2mm} $(\chi_{\mathrm{H}^+}, \chi_{\mathrm{He}^{+}}, \chi_{\mathrm{He}^{++}})$ \hspace*{-3mm}& 
$\displaystyle\frac{\rho_\mathrm{CDM}}{\mathrm{GeV}/\mathrm{cm}^3}$ & 
$\displaystyle\frac{\rho_\mathrm{mirror}}{\mathrm{GeV}/\mathrm{cm}^3}$ & 
$\frac{1}{r_\mathrm{all}} \frac{\rho_\mathrm{mirror}}{\rho_\mathrm{CDM}} $ & $ \hat Y_\odot$

\\
& & & & & & & 
\\
\hline \hline
\hspace*{-4mm}
\begin{tabular}{l}SM Baryons \\ $r_\mathrm{all} = 0.01$ \end{tabular}
& 145 & 181 & $(1, 2\times 10^{-7}, 1)$ &  0.5 & 0.12 & 1.22 & $0.95$
\\ 
\hline \hline
\hspace*{-4mm}
\begin{tabular}{l}$\frac{\hat{v}}{v} = 3, r_\mathrm{all} = 0.01,$ \\ $\hat Y_p(^4\mathrm{\hat He}) = 0.75$ \end{tabular}
& 263 & 212 & $(1, 5\times 10^{-7}, 1)$ &  0.5 & 0.0039 & 0.78 & $0.99$\\
\hline 
\hspace*{-4mm}
\begin{tabular}{l}$\frac{\hat{v}}{v} = 3, r_\mathrm{all} = 0.01,$ \\ $\hat Y_p(^4\mathrm{\hat He}) = 0$ \end{tabular}
& 119 & 231 & $(1, -, -)$ &  0.5 & 0.0021 & 0.42 & 0
\\
\hline 
\hspace*{-4mm}
\begin{tabular}{l}$\frac{\hat{v}}{v} = 3, r_\mathrm{all} = 0.01,$ \\ $\hat Y_p(^4\mathrm{\hat He}) = 1$ \end{tabular}
& 311 & 229 & $(-, 4\times 10^{-7}, 1)$ &  0.5 & 0.0023 & 0.46 & 1
\\
\hline
\end{tabular}

\end{center}
\caption{
Isothermal halo parameters at $r = R_\odot = 8 \ \mathrm{kpc}$ for some benchmark mirror baryon scenarios. $v_\odot$ is the mean velocity of mirror baryon constituents following a thermal distribution of temperature $T_\mathrm{iso}$.  Temperature and ionization is constant throughout the halo. 
For each $\hat Y_p(^4\mathrm{\hat He})$ separately, the local mirror helium fraction $\hat Y_\odot$, local mean velocity $v_\odot$, and the local $\frac{1}{r_\mathrm{all}} \frac{\rho_\mathrm{mirror}}{\rho_\mathrm{CDM}} $  are almost independent of $\hat{v}/v$ and $r_\mathrm{all}$, see text for discussion. 
As the halo cools, $\hat Y_\odot$ will be reduced as mirror helium sinks to the bottom of the gravity well. 
The SM comparison scenario  corresponds to evaluating the mirror baryon profile for $\hat{v}/v = 1, r_\mathrm{all} = 0.2, \hat Y_p(^4\mathrm{\hat He}) = 0.25$. This is presented to compare mirror halo parameters to a SM-like halo evaluated under the same assumptions, but does not represent the actual SM baryon distribution today.
}
\label{t.mirrorprofilesummary}
\end{table}

\begin{figure}
\begin{center}
\begin{tabular}{cc}
NFW & Burkert
\\
\\
\includegraphics[width=0.45\textwidth]{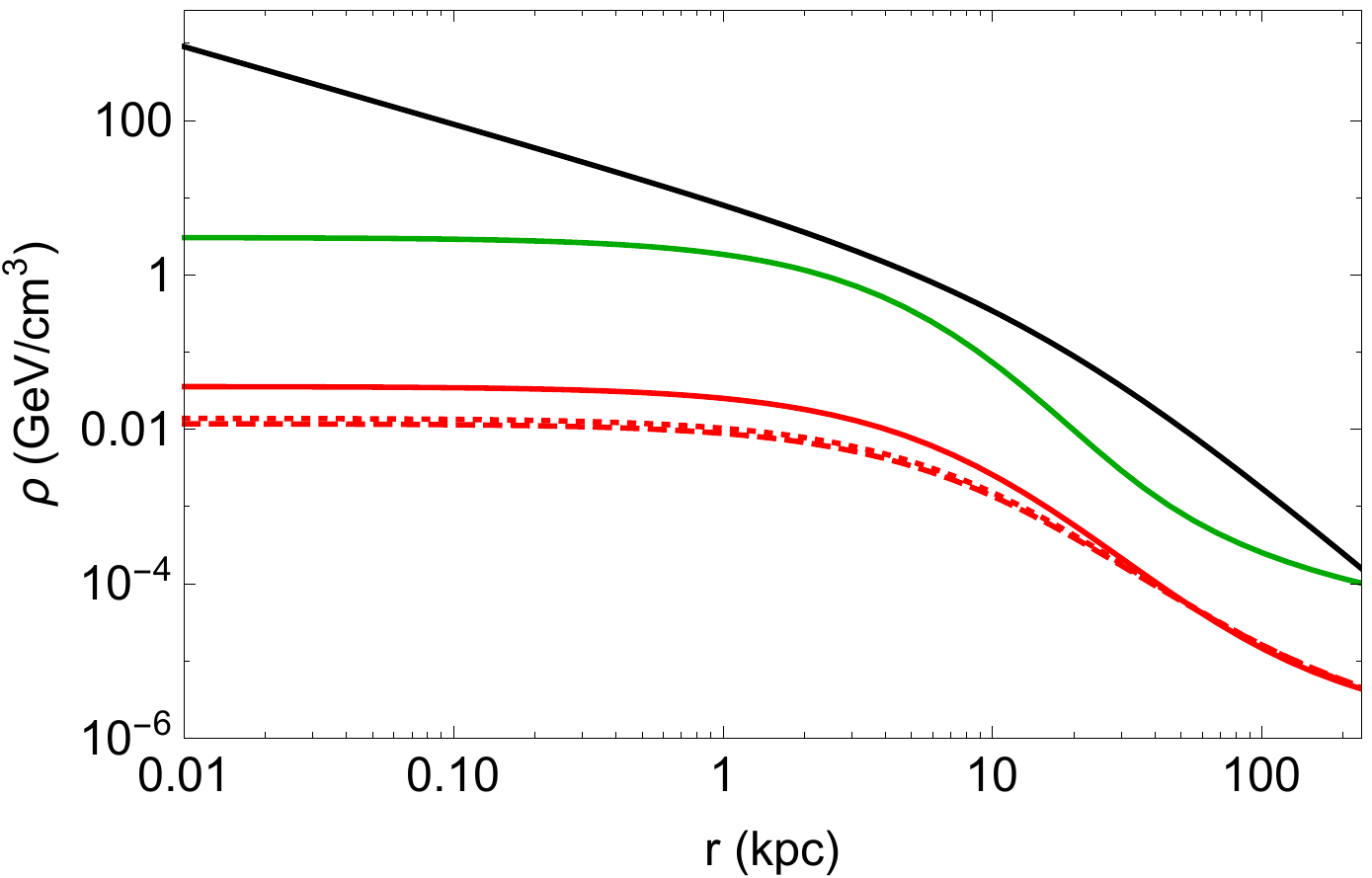} 
&
\includegraphics[width=0.45\textwidth]{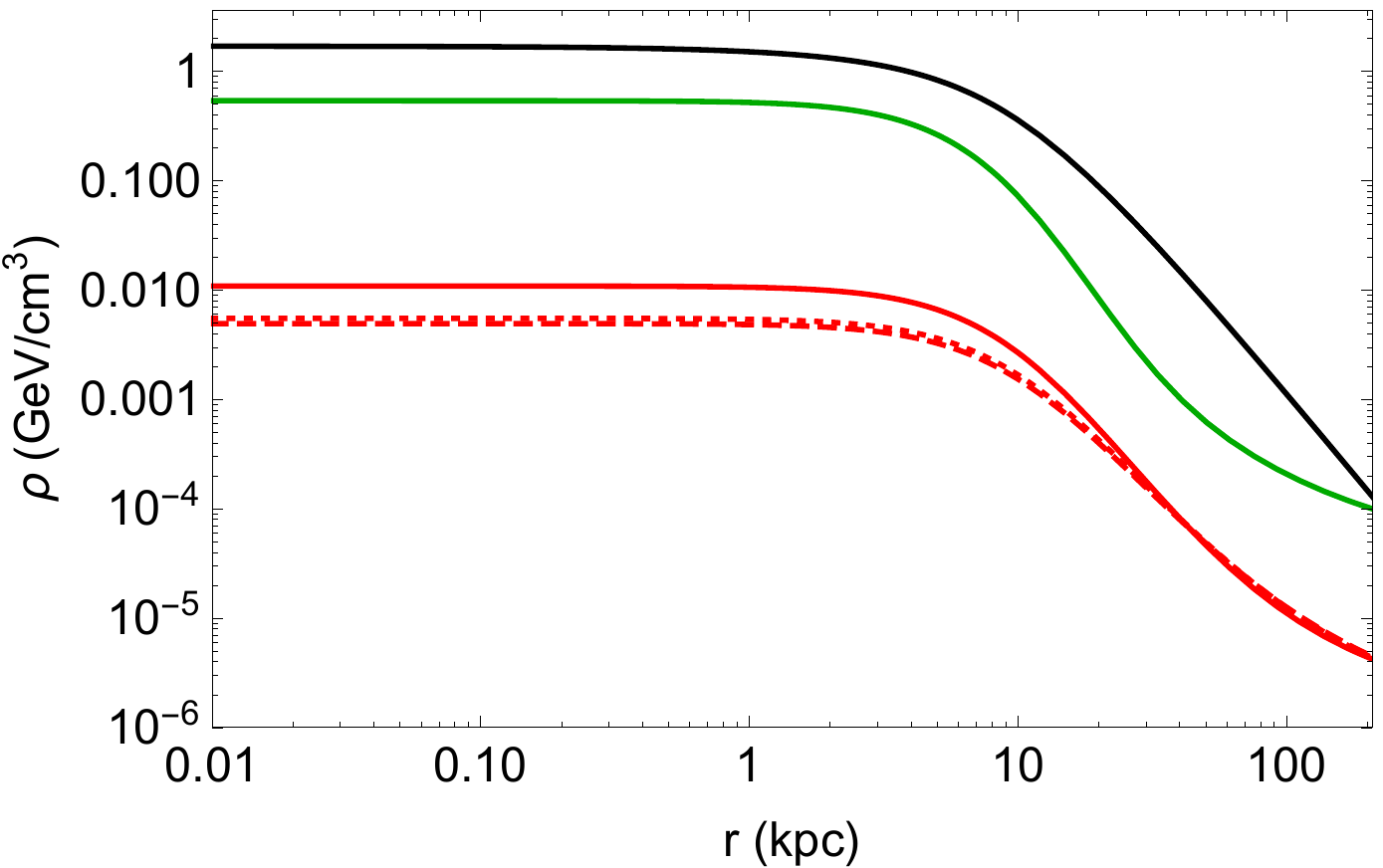} 
\\
\includegraphics[width=0.45\textwidth]{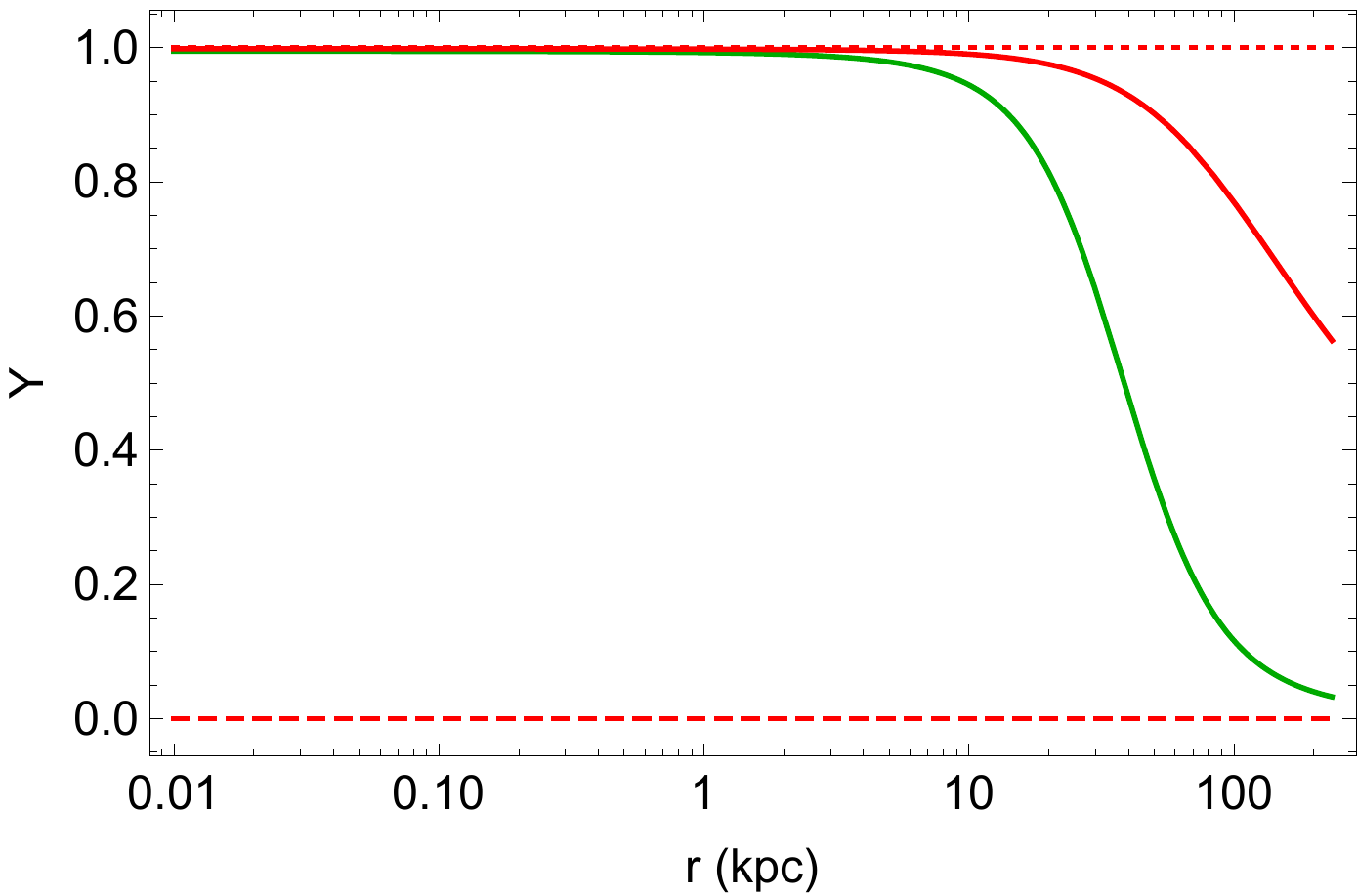} 
&
\includegraphics[width=0.45\textwidth]{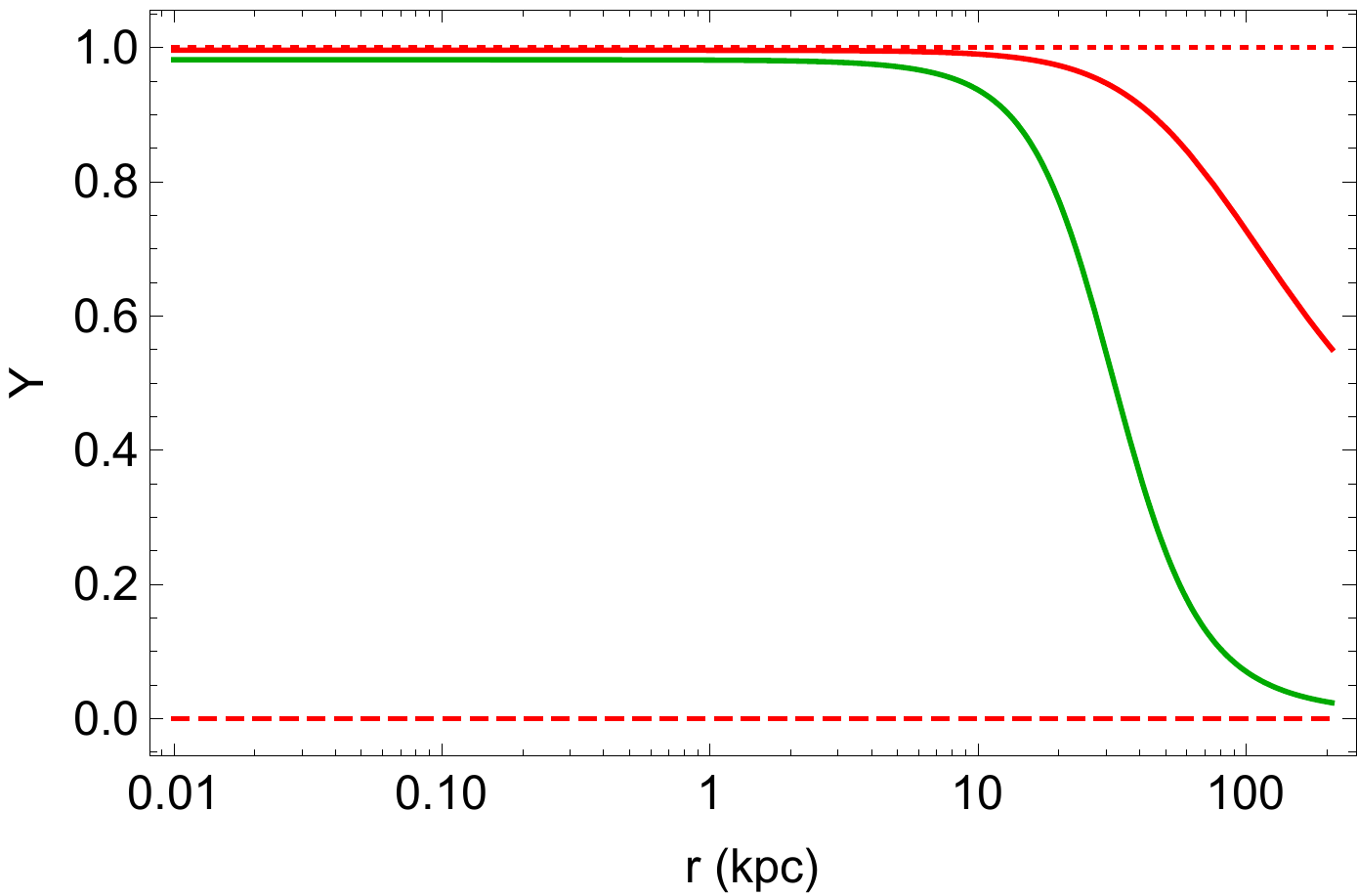} 
\\
\includegraphics[width=0.45\textwidth]{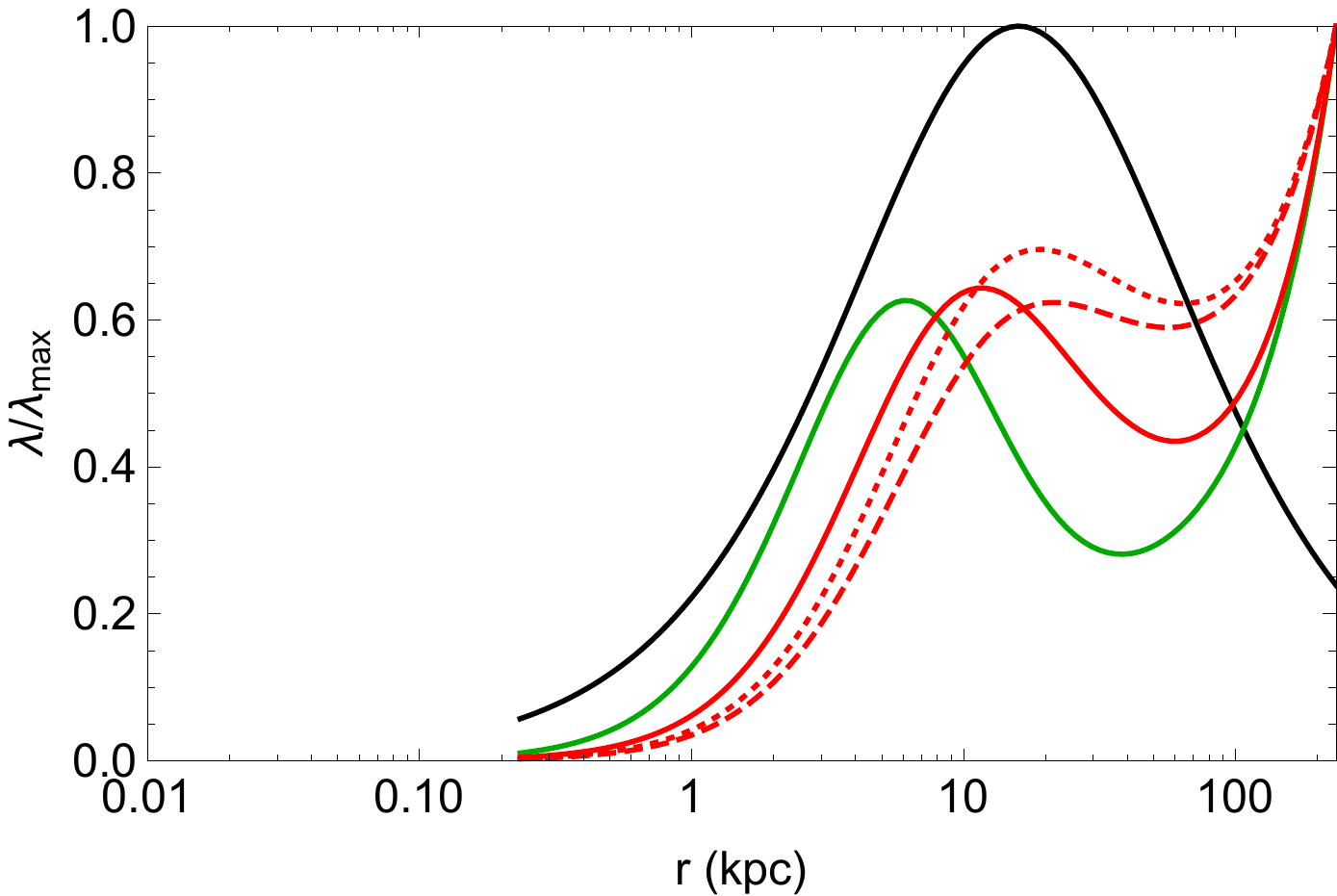} 
&
\includegraphics[width=0.45\textwidth]{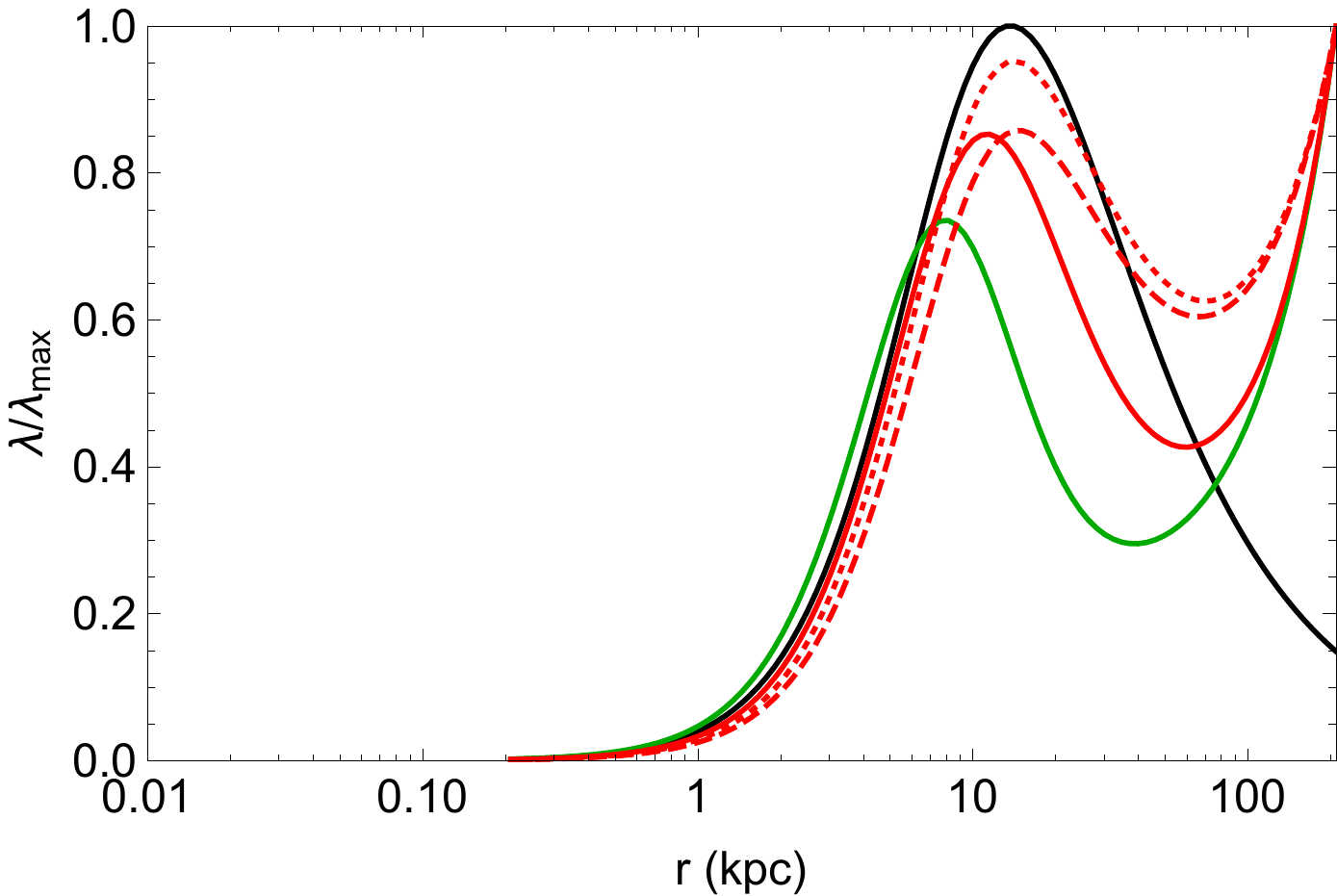} 
\end{tabular}

\includegraphics[width=0.9\textwidth]{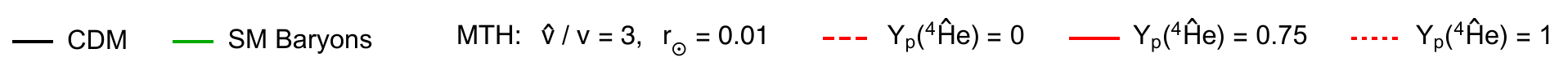}
\end{center}
\caption{
\emph{Top:} Density profile for the mirror halo benchmarks in \tref{mirrorprofilesummary}, evaluated in the background of the NFW or Burkert CDM Profile. 
\emph{Middle:} Mirror helium mass fraction $\hat Y(r)$ as a function of $r$. 
\emph{Bottom:} mass density per spherical shell $\lambda = 4 \pi r^2 \rho(r)$, normalized to its maximum value for each profile. This shows the distance where most of the constituent atoms are situated. 
The SM comparison scenario  corresponds to evaluating the mirror baryon profile for $\hat{v}/v = 1, r_\mathrm{all} = 0.2, Y = 0.25$. This is presented to compare mirror halo parameters to a SM-like halo evaluated under the same assumptions, but does not represent the actual SM baryon distribution today.
}
\label{f.mirrorprofiles}
\end{figure}

As we have seen, the gas of mirror particles quickly evolves to reach 
thermal equilibrium, and arranges itself in an isothermal distribution. 
We will therefore use the isothermal mirror halo distributions as the 
basis for our discussion of cooling. In general, the form of the 
isothermal distribution depends on $r_\mathrm{all}$, $\hat{v}/v$ and 
$\hat Y_p(^4\mathrm{\hat He})$. However, the dependence of the mirror baryon 
profile on $r_\mathrm{all}$ is simple; the density scales proportionally 
with $r_\mathrm{all}$ as long as the mirror sector contribution to the 
gravitational potential can be neglected. Similarly, the effect of 
$\hat{v}/v$ on the distribution is very minor for values in the range of 
interest. However, the dependence on the mirror helium fraction 
$\hat  Y_p(^4\mathrm{\hat He})$ is non-trivial, and we therefore focus on illustrating this. To this end, we show the profiles for three different MTH benchmark points. For each benchmark point the values of $\hat{v}/v = 3$ and $r_\mathrm{all} = 0.01$ are held fixed, but we consider three different values of the helium fraction, $\hat Y_p(^4\mathrm{\hat He})= 0.75$ (close to the BBN prediction for the asymmetrically reheated MTH 
scenario~\cite{Chacko:2018vss}), and the limiting cases of 
$\hat Y_p(^4\mathrm{\hat He})= 0$ and $\hat Y_p(^4\mathrm{\hat He}) = 1$ (pure 
mirror hydrogen and helium). To serve as a basis for comparison, we also 
compute a profile representative of SM baryons, corresponding to the MTH 
parameters $\hat Y_p(^4\mathrm{\hat He}) = 0.25$, $r_\mathrm{all} = 0.2$ and $\hat{v}/v = 1$.

\tref{mirrorprofilesummary} summarizes the isothermal mirror baryon halo 
parameters at the location of our Sun, while spatial distributions are 
shown in \fref{mirrorprofiles}. We define
\begin{eqnarray}
\hat Y_\odot &=& \hat Y(R_\odot)\\
v_\odot &=& v_0(R_\odot)
\end{eqnarray}
to be the local values of the mirror helium fraction and mirror baryon velocity dispersion respectively. Temperature and ionization are constant 
throughout the halo, and both mirror hydrogen and helium are almost 
completely ionized. The local mean velocity $v_\odot$ is comparable to, 
although somewhat higher than, the canonical CDM halo expectation of 
$\sim 220$ km/s. The local density of mirror baryons differs from 
$r_\mathrm{all} \cdot \rho_\mathrm{CDM}$ by an $\mathcal{O}(1)$ factor. 
For $\hat Y_p(^4\mathrm{\hat He}) \neq 0, 1$, the local mirror helium 
fraction $\hat Y_\odot$ is highly temperature dependent. A reduction in the 
halo temperature by a factor of two could reduce $\hat Y_\odot$ to almost 
zero as the region of helium-dominance, see \fref{mirrorprofiles} 
(middle), retreats towards the center of the gravity well.  Therefore, 
any variation in the halo results in a large change in the local value 
of $\hat Y_\odot$.

The four chosen benchmarks are representative of the behavior in the MTH 
framework. For each $\hat Y_p(^4\mathrm{\hat He})$ separately, the local 
values of $\hat Y_\odot$, $v_\odot$ and 
$({1}/{r_\mathrm{all}})({\rho_\mathrm{mirror}}/{\rho_\mathrm{CDM}})$, as 
well as the halo temperature $T_\mathrm{iso}$, are relatively 
insensitive to the values of $\hat{v}/v$ and $r_\mathrm{all}$. 
\footnote{Small differences arise due to the increased ionization 
energies corresponding to higher values of $\hat{v}/v$, resulting in 
slightly lower temperatures for higher values of the mirror Higgs vev. 
However, this effect is too minor to affect our discussion.} 
On the other hand, the dependence of the profiles on $\hat 
Y_p(^4\mathrm{\hat He})$ is nontrivial. In particular, the SM-like 
value, $\hat Y_p(^4\mathrm{\hat He}) = 0.25$, is close to optimal for 
concentrating mirror baryons near the galactic center, with larger or 
smaller values leading to puffier profiles. As we shall see, this 
interesting coincidence has important implications for the cooling 
rates.

\subsubsection{Mirror Baryon Ionization Fractions}
\label{s.ionization}

Before the onset of cooling, the mirror halo is sufficently hot that the
mirror baryons are very close to fully ionized.\footnote{If this were
not the case, the dependence of the mirror plasma heat capacity on
ionization would have to be taken into account.} However, the small
fraction of partially or fully recombined mirror helium and hydrogen is
important for non-bremsstrahlung cooling processes. We therefore now
discuss how to determine the degree of ionization for the computed
mirror baryon profiles.

In general, we wish to determine the local ionization fractions assuming 
some local temperature and number densities of mirror hydrogen and 
helium. If the conditions of detailed balance are satisfied, as in the 
early universe, these ionization fractions can be obtained from Saha's 
equation~\cite{2010gfe..book.....M}. However, mirror halos are usually 
\emph{optically thin}, meaning that photons emitted from bremsstrahlung 
or atomic cooling processes escape the galaxy. In such a scenario, the 
ionization fractions are determined, not from Saha's equation, but from 
the ratios of ionization and recombination rates.

 The local ionization fractions are defined as ratios of number densities
for individual atom species,
 \begin{eqnarray}
\chi_{\mathrm{\hat{H}}^+}(r) &=& \frac{n_{\mathrm{\hat{H}}^+}}{n_{\mathrm{\hat{H}}^0} + n_{\mathrm{\hat{H}}^+}}
\nonumber \\
\chi_{\mathrm{\hat{H}e}^+}(r) &=& \frac{n_{\mathrm{\hat{H}e}^+}}{n_{\mathrm{\hat{H}e}^0} + n_{\mathrm{\hat{H}e}^+} + n_{\mathrm{\hat{H}e}^{++}}}
\nonumber \\
\chi_{\mathrm{\hat{H}e}^{++}}(r) &=& \frac{n_{\mathrm{\hat{H}e}^{++}}}{n_{\mathrm{\hat{H}e}^0} + n_{\mathrm{\hat{H}e}^+} + n_{\mathrm{\hat{H}e}^{++}}}
\;.
 \end{eqnarray}
 Solving the equation $d n_{\mathrm{\hat{H}}^+}/dt = 0$, assuming no
photoionization and neglecting double-ionization and
double-recombination processes yields,
 \begin{equation}
\chi_{\mathrm{\hat{H}}^+}
=
\frac{
\langle \sigma_{\mathrm{ion}(\mathrm{\hat{H}}^0)} v\rangle
}
{
\langle \sigma_{\mathrm{ion}(\mathrm{\hat{H}}^0)} v\rangle
+ \langle \sigma_{\mathrm{rec}(\mathrm{\hat{H}}^+)} v\rangle 
} \;,
\end{equation}
 where $\sigma_{\mathrm{ion}(\mathrm{\hat{H}}^0)}$ and
$\sigma_{\mathrm{rec}(\mathrm{\hat{H}}^+)} $ are the relevant thermally averaged ionization and
recombination cross sections. Similarly, solving $d n_{\mathrm{\hat{H}e}^+}/dt
= d n_{\mathrm{\hat{H}e}^{++}}/dt = 0$ for helium leads to,
 \begin{eqnarray}
\nonumber
\chi_{\mathrm{\hat{H}e}^{\{+,++\}}}
&=&
\frac{
\langle \sigma_{\mathrm{ion}(\mathrm{\hat{H}e}^0)} v\rangle
\langle
\sigma_{
\{
\mathrm{rec}(\mathrm{\hat{H}e}^{++}), \ \mathrm{ion}(\mathrm{\hat{H}e}^+)
\}
}
v\rangle
}
{
\langle \sigma_{\mathrm{ion}(\mathrm{\hat{H}e}^0)} v\rangle
\langle \sigma_{\mathrm{ion}(\mathrm{\hat{H}e}^{+})} v\rangle
+
\langle \sigma_{\mathrm{ion}(\mathrm{\hat{H}e}^0)} v\rangle
\langle \sigma_{\mathrm{rec}(\mathrm{\hat{H}e}^{++})} v\rangle
+
\langle \sigma_{\mathrm{rec}(\mathrm{\hat{H}e}^{++})} v\rangle
\langle \sigma_{\mathrm{rec}(\mathrm{\hat{H}e}^{+})} v\rangle
} \;.
\\
 \end{eqnarray}
 The averaged cross sections $\langle \sigma v \rangle$ are computed
assuming a thermal distribution of the initial state electrons and
neglecting the motion of the relatively slow atoms.
Ref.~\cite{Rosenberg:2017qia} recently summarized the various
ionization, recombination and cooling cross sections and rates for
dissipative DM with one nucleus-like and one electron-like
constituent. These expressions can be directly applied to mirror
hydrogen, with the mirror Rydberg energy given by $\mathrm{Ryd} = (13.54
\ \mathrm{eV}) \times (\hat{v}/v)$. For mirror helium, it is a reasonable
approximation to treat the participating electron as if it were bound in
a hydrogen-like atom with some effective charge $Z_\mathrm{eff}$. We
therefore employ the same expressions for the cross sections as for the
hydrogen-like atom, but with the substitution $\mathrm{Ryd} = (24.48 \
\mathrm{eV}) \times (\hat{v}/v)$ for $\mathrm{He}^0$ and $\mathrm{Ryd} =
(54.17\ \mathrm{eV}) \times (\hat{v}/v)$ for $\mathrm{He}^+$.
For the isothermal profile in the optically thin regime, ionization is trivially constant throughout the halo, since the thermally averaged cross sections do not depend on density.

We discuss the optical depth of the mirror halo in \sref{opticaldepth}
to verify that Rydberg energy photons escape and hence detailed balance
does not apply. This justifies our use of ionization and recombination
cross section ratios to determine the ionization fractions $\chi_i$.
These cross sections 
also give ionization and recombination
timescales for the different mirror atomic species, shown as magenta and
cyan lines in Figs.~\ref{f.timescalesNFW} and~\ref{f.timescalesBUR}. The
high degree of ionization in the halo is reflected by the very short
ionization timescale compared to the other timescales in the system
(including the recombination timescale) in regions where most of the
mirror matter is concentrated.

\subsection{Mirror Baryon Cooling}
\label{s.darkbaryoncooling}

We now discuss various cooling mechanisms that can cause the mirror halo 
to loose pressure support and collapse, potentially leading to disk 
formation. This would have a dramatic effect on the prospects for direct 
and indirect detection of the twin subcomponent of DM~\cite{Randall:2014aa,Randall:2014kta,Foot:2013nea,Agrawal:2017pnb}. 
Cooling occurs through the emission of mirror photons produced in the 
scattering of mirror particles. In \sref{cooling} we discuss the most 
important cooling processes, which include bremsstrahlung and various 
atomic processes such as ionization, recombination and collisional 
excitation. If the cooling timescale is shorter than the age of the 
universe, which we take to be 14 Gyr, there is a possibility that the 
twin particles have condensed into a disk. 
In \sref{opticaldepth} we discuss the optical depth of the mirror halo, and verify that the  photons produced in these cooling processes escape from the galaxy.

The cooling timescale $t_{cool}$ that we evaluate is to be compared to 
the other two relevant timescales; the current age of the universe 
$t_{universe}$ and the dynamical or convection timescale set by the CDM 
halo, $t_{convection}$, which is of order $10^8 - 10^9$ years. If 
$t_{cool} \gg t_{universe}$, the mirror baryon halo has not yet had time 
to cool significantly since the formation of the Milky Way, and is 
likely to still be close to its original halo distribution. At the other 
extreme, if $t_{cool} \ll t_{convection}$ within some radius $r < 
r_{collapse}$, the halo will lose pressure support inside that radius 
and start to undergo catastrophic collapse, which may result in the 
formation of a disk. The size of such a disk would be expected to be of 
the same order as $r_{collapse}$, but as we discuss in 
\sref{mirrorbaryonstoday}, it is very difficult to extrapolate this 
result directly to the mirror baryon distribution today. Finally, if 
$t_{convection} \ll t_{cool} \ll t_{universe}$, the outcome is even more 
uncertain. The halo initially cools gradually without loss of pressure 
support. As the temperature drops, the cooling timescale decreases and 
the halo may eventually reach the aforementioned regime where $t_{cool} 
\ll t_{convection}$ within some radius. Whether this happens depends in 
part on how efficiently the halo maintains an isothermal profile during 
the cooling process, since cooling occurs predominantly via photon 
emission from the inner regions. These complications make quantitative 
predictions about the mirror baryon distribution today 
challenging. Nevertheless, our analysis will still reveal important 
quantitative and qualitative information that serves to illuminate the 
range of possibilities we must consider for direct detection.

\subsubsection{Cooling Timescales}
\label{s.cooling}

We first consider the cooling through the emission of massless mirror 
photons that are produced through Compton scattering of mirror electrons 
off the background mirror CMB photons, 
$\hat{e}\hat{\gamma}\to\hat{e}\hat{\gamma}$, or through bremsstrahlung 
emission, $\hat{e}\hat{X}_i\to\hat{e}\hat{X}_i\hat{\gamma}$, where 
$\hat{X}_i=(\hat{\text{H}}^+,\,\hat{\text{H}}\text{e}^+,\, 
\hat{\text{H}}\text{e}^{++})$. Both processes lead to energy loss of the 
mirror electron. If the mirror photon escapes the halo without being 
reabsorbed, and if $\hat{e}$ and $\hat{X}_i$ remain in thermal 
equilibrium, then $\hat{X}_i$ and $\hat{e}$ cool adiabatically together.

We first determine if the equilibrium condition is satisfied. The time 
for $\hat{X}_i$ and $\hat e$ to reach thermal equilibrium, $t_{eq}^i$, has been estimated as~\cite{1941ApJ....93..369S, Fan:2013yva},
 \begin{eqnarray}
t_{eq}^i&\approx& \frac{m_{\hat{X}_i}\,m_{\hat{e}}}{2\sqrt{3\pi}\hat{\alpha}_{em}^2} \left(\frac{3\,T_\mathrm{iso}}{m_{\hat{e}}}\right)^{\frac{3}{2}} \left[n_{\hat{e}}\log \left(1+ \frac{v_{\hat{e}}^4 m_{\hat{e}}^2}{\hat{\alpha}_{em}^2 n_{\hat{e}}^{2/3}}\right)\right]^{-1}\,, \\
&\approx& 6\times 10^7~\text{yr}~\left(\frac{m_{\hat{X}_i}}{1\text{ GeV}}\right)\left(\frac{T_\mathrm{iso}}{100\,\text{eV}}\right)^{\frac{3}{2}}\sqrt{\frac{m_e}{m_{\hat{e}}}} \left(\frac{10^{-3}\text{cm}^{-3}}{n_{\hat{e}}}\right)\left[\log \left(1+ \frac{9\,T_\mathrm{iso}^2}{\hat{\alpha}_{em}^2 n_{\hat{e}}^{2/3}}\right)\right]^{-1}.\nonumber
 \end{eqnarray}
We find that for the parameter range of 
interest, the characteristic equilibration times for all $\hat X_i$ are 
much less than the age of the universe. As we shall see, the time scale 
of equilibration is also much shorter than the characteristic cooling 
timescales.

We can obtain an expression for the ratio of the rate of energy loss per 
unit volume through Compton cooling to the initial energy density in 
mirror baryons and electrons \cite{1979rpa..book.....R},
 \begin{eqnarray}\label{eq:ComptonCool}
-\frac{dU/dt_{Compton}}{\frac{3}{2}T_\mathrm{iso} \left(\sum n_{\hat{X}_i}+n_{\hat{e}}\right)} 
&\approx& \frac{64\pi^2 \hat{\alpha}_{em}^2}{135} \frac{n_{\hat{e}}}{\sum n_{\hat{X}_i}+n_{\hat{e}}}\frac{\hat{T}^4(z)}{m_{\hat{e}}^3}
\\
\nonumber
&\approx& \frac{1}{1.6 \times 10^{13}~\text{yr}}
\left(\frac{\hat{T}(z)}{4~\text{K}}\right)^4 \left(\frac{m_e}{m_{\hat{e}}}\right)^3.
 \end{eqnarray}
 Here $\hat{T}(z)$ denotes the temperature of the mirror CMB photons at 
redshift $z$. The summation is over the various ions, 
$\hat{X}=(\hat{\text{H}}^+$, $\hat{\text{H}}\text{e}^+$, 
$\hat{\text{H}}\text{e}^{2+})$. The ratio of number densities is 
determined in terms of $\hat Y_p(^4\mathrm{\hat He})$ in the limit that the 
mirror halo is fully ionized. We find from this that for the relevant 
range of redshifts, $z \lesssim \mathcal{O}(10)$, the time scale of 
Compton cooling is greater than the age of the universe, and we do not 
consider this process further.

Cooling via bremsstrahlung emission is much more efficient. The 
corresponding time scale is given by~\cite{Rosenberg:2017qia}\begin{eqnarray}\label{eq:bremcool}
t_{brem}(r) &\approx&
\frac{3^{5/2}}{2^{9/2}\sqrt{\pi}}  \ 
 \frac{\sum n_{\hat{X}_i}+n_{\hat{e}}}{n_{\hat{e}}\,\sum Z_i^2\,n_{\hat{X}_i}} \ \frac{m_{\hat{e}}^{3/2}T_\mathrm{iso}^{1/2}}{\hat{\alpha}_{em}^3}
 \end{eqnarray} 
 (see also \cite{1979rpa..book.....R, Straumann:1984xf}). The relevant 
number densities are obtained from the profiles computed in the previous 
section. We find that in the parameter range of interest, the timescale 
associated with cooling through bremsstrahlung emission can be less than 
the age of the universe in the dense inner regions of the galaxy.

In addition to energy loss via bremsstrahlung, mirror electrons in the 
halo can also lose energy through atomic processes that involve 
$\mathcal{O}(\mathrm{Ryd})$ energies, such as ionization\footnote{The 
collisional ionization comes from a free electron impact that ionizes a 
formerly bound electron, taking energy from the free electron. The 
temperature of the particles is therefore lower after the ionization.} 
($\hat e^- + \hat X_i \to \hat X_i^+ + 2 \hat e^-$), recombination 
($\hat e^- + \hat X_i^+ \to \hat X_i + \hat \gamma$), and collisional 
excitation ($\hat e^- + \hat X_i \to \hat e^- + \hat X_i^* \to \hat e^- 
+ \hat X_i + \hat \gamma$). For all these processes, recent estimates of 
the relevant cross sections and energy loss rates can be found 
in~\cite{Rosenberg:2017qia}, and we adapt them for use with singly- or 
doubly-ionized helium as described in \sref{ionization}. We find that in 
the inner regions of the galaxy, the cooling timescale from atomic 
processes can be less than the age of the universe. In principle 
molecular cooling processes can also play a 
role~\cite{Ghalsasi:2017jna}, but given the high degree of ionization in 
the mirror halo, we neglect them in our simple analysis.

Based on this discussion, we can make some observations about the 
dependence of the cooling rate on the parameters of the MTH.
 \begin{itemize}
 \item 
  We find that the dependence of $t_\mathrm{cool}$ on the
electroweak VEV in the twin sector is quite modest in the $\hat{v}/v \sim 1-5$ range of interest. 
Near the SM value of 1, bremsstrahlung cooling dominates. 
For MTH-like values of $3-5$, the larger mirror electron mass and ionization energies give rise to a small but important neutral atom population, which makes collisional cooling processes dominant. 
This compensating behavior explains the insensitivity of cooling time scales on the (mirror-) Higgs vev, and at our level of precision $\hat{v}/v$ is not important to our discussion. 
 \item The cooling time scale scales inversely with
the number density of mirror particles, $t_\mathrm{cool} \sim
r_\mathrm{all}^{-1}$, so that lower mirror baryon densities are
associated with slower cooling. This is because the cooling processes
arise from the collisions of mirror particles, and their rates go down
if the number density of mirror particles is reduced. 
 \item Cooling depends nontrivially on the mirror helium fraction 
$\hat Y_p(^4\mathrm{\hat He})$, with the SM-like value of $\hat Y_p(^4\mathrm{\hat 
He})$ being near-optimal for cooling, and other values near 0 or 1 
cooling much less efficiently. This can be traced back to our finding 
that SM-like values of $\hat Y_p(^4\mathrm{\hat He})$ lead to the most 
tightly packed profiles.
 \end{itemize}
 Since the SM particles are much more abundant than their 
mirror counterparts, and because $\hat Y_p(^4\mathrm{\hat He})$ is near the 
optimum value for cooling, we can immediately conclude that mirror 
baryons cool much slower than the SM baryons. 
However, determining how the
the cooling rate compares to the convection timescale requires explicit calculation for each profile. We discuss these results in \sref{mirrorhaloresultsfinal}.

\begin{figure}
\begin{center}
\begin{tabular}{cc}
\includegraphics[width=0.45\textwidth]{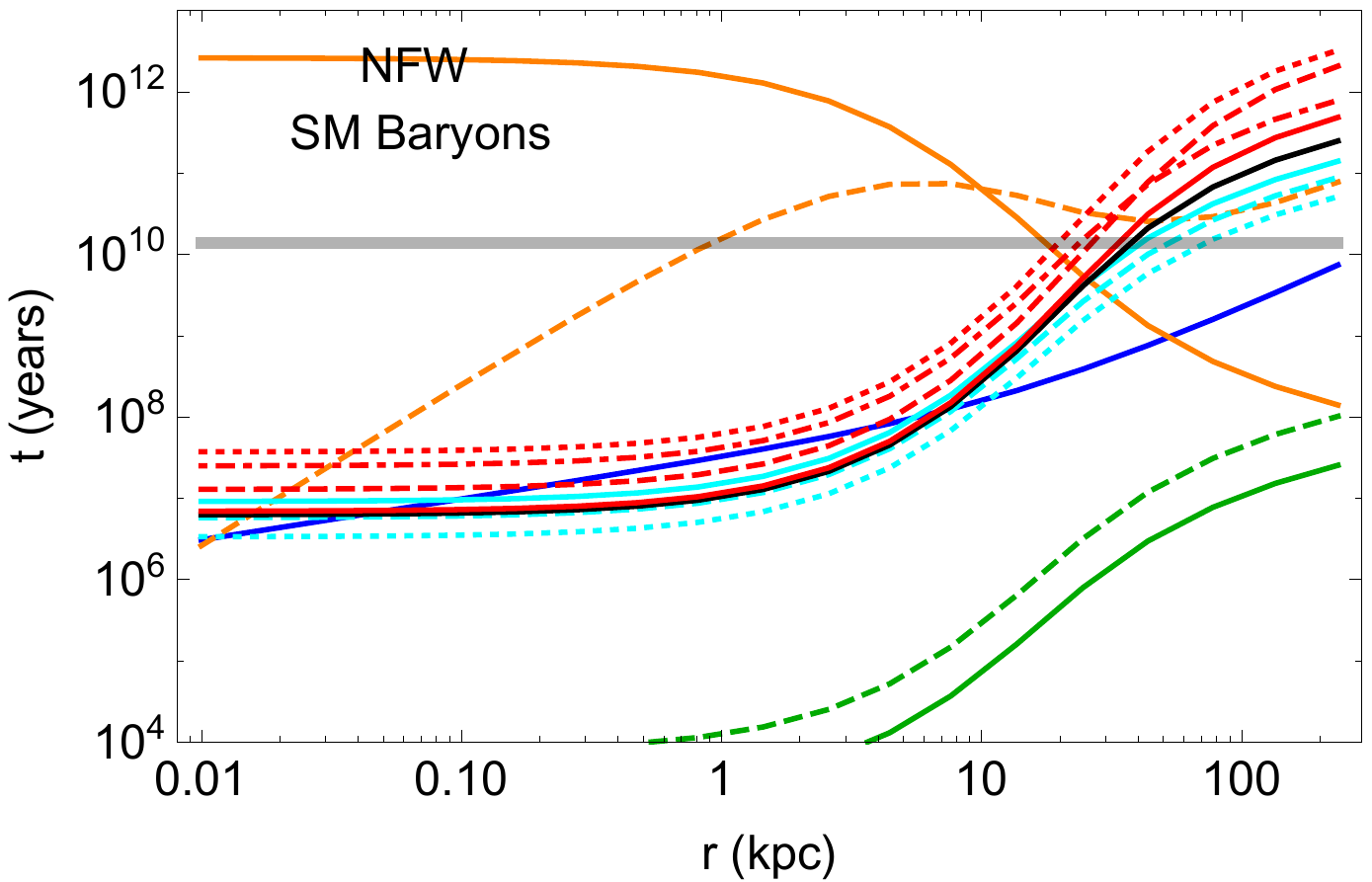} 
&
\includegraphics[width=0.45\textwidth]{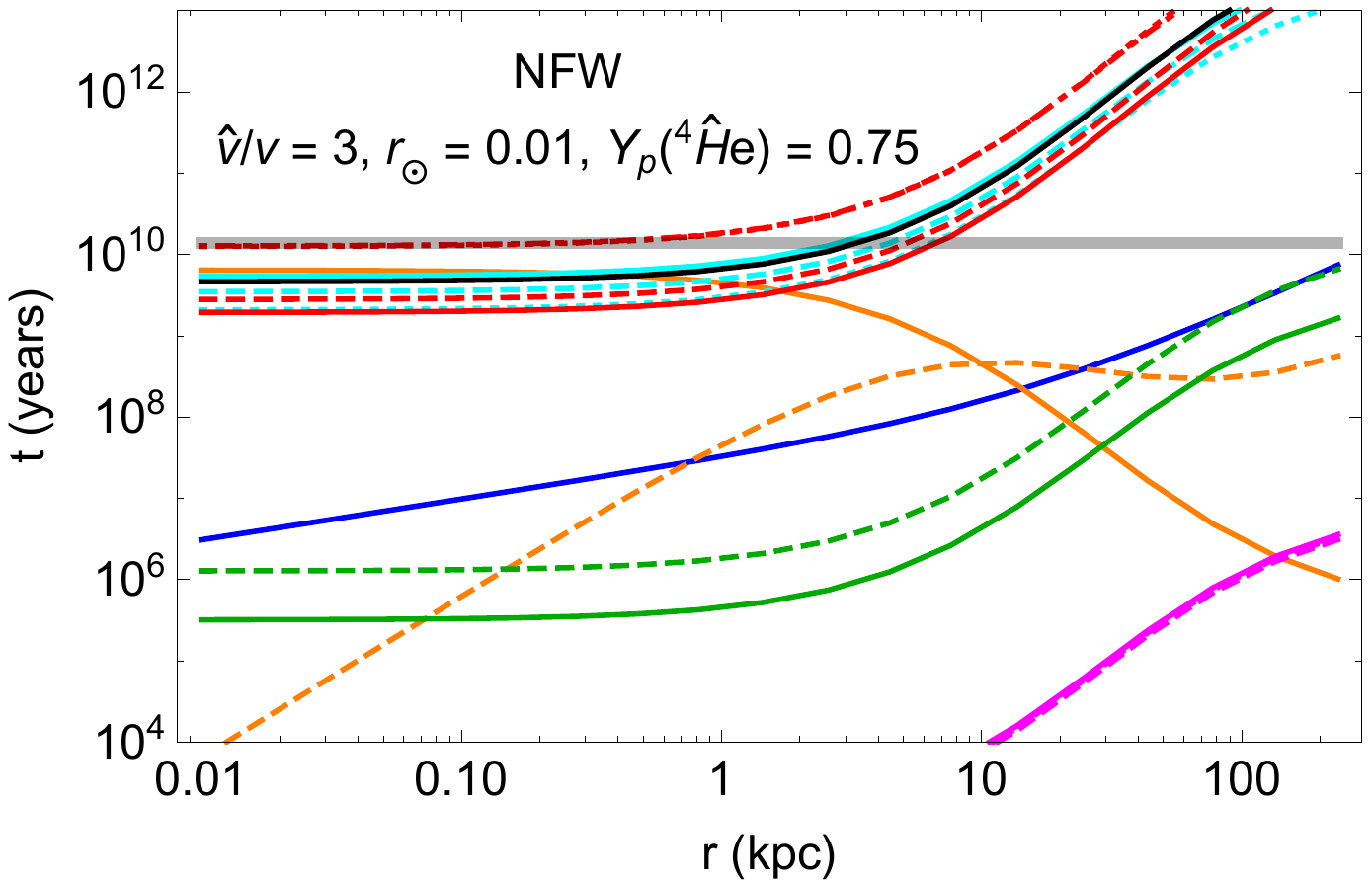} 
\\
\includegraphics[width=0.45\textwidth]{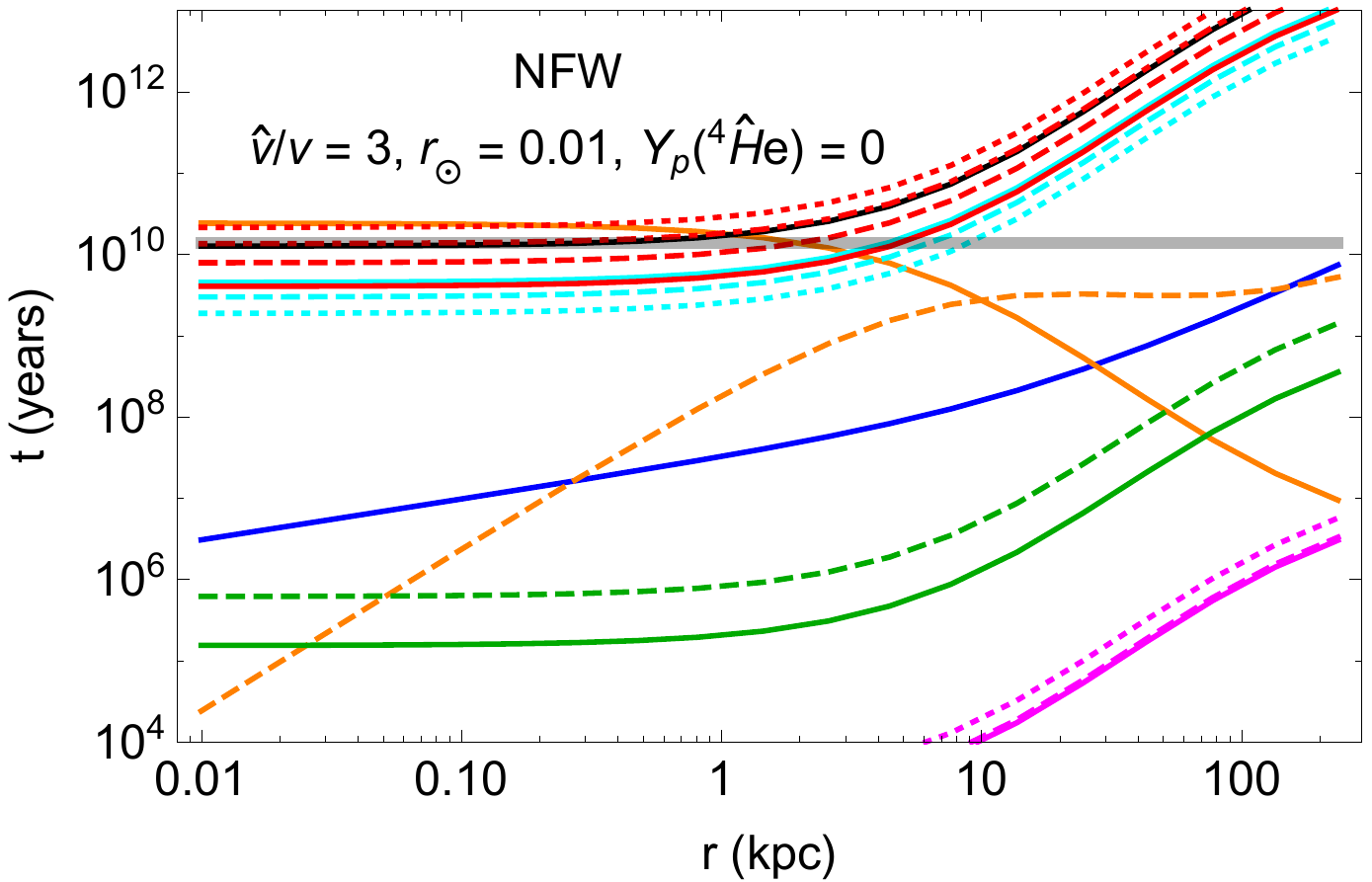} 
&
\includegraphics[width=0.45\textwidth]{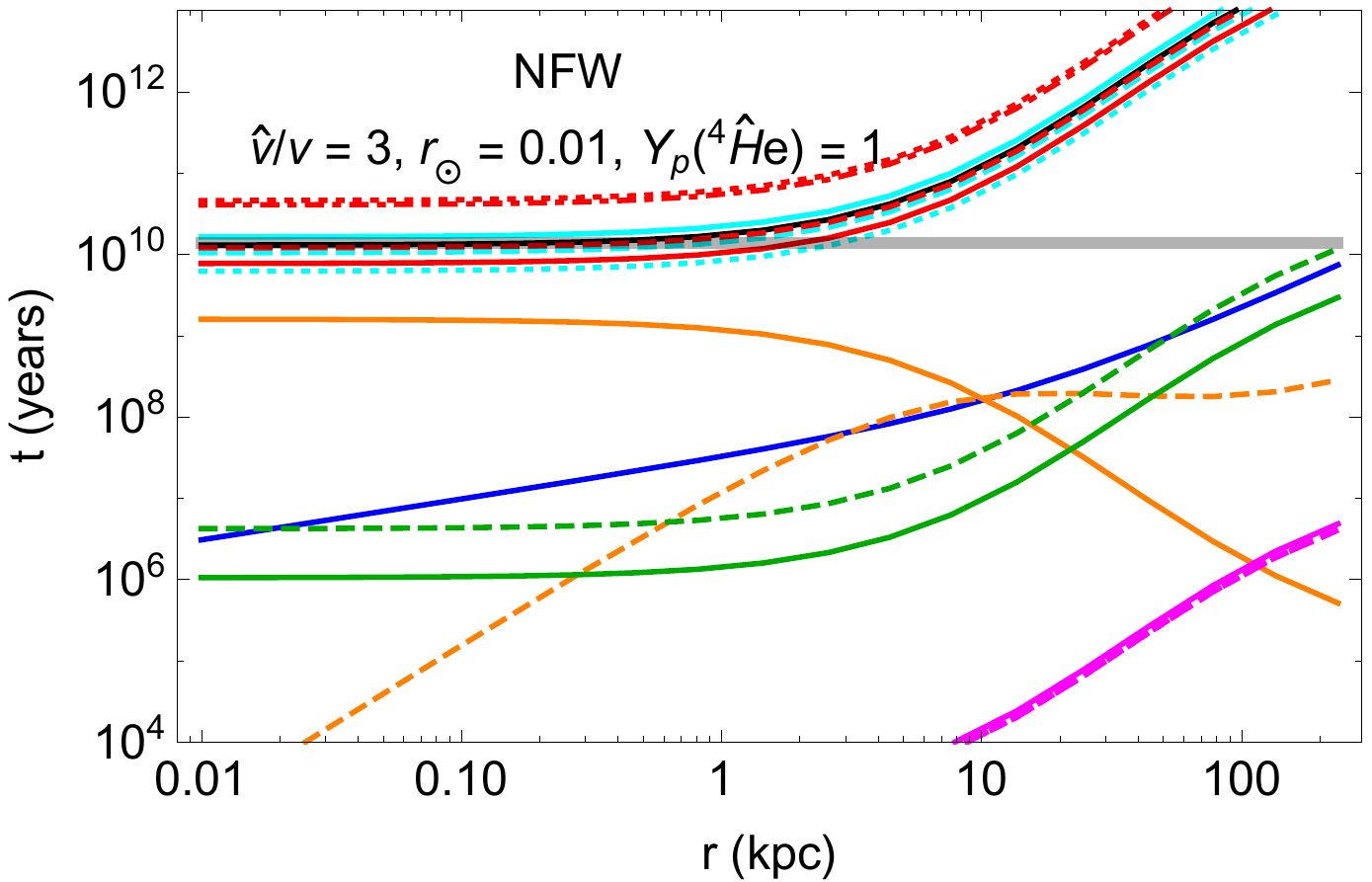} 
\end{tabular}

\vspace{2mm}
\includegraphics[width=0.9\textwidth]{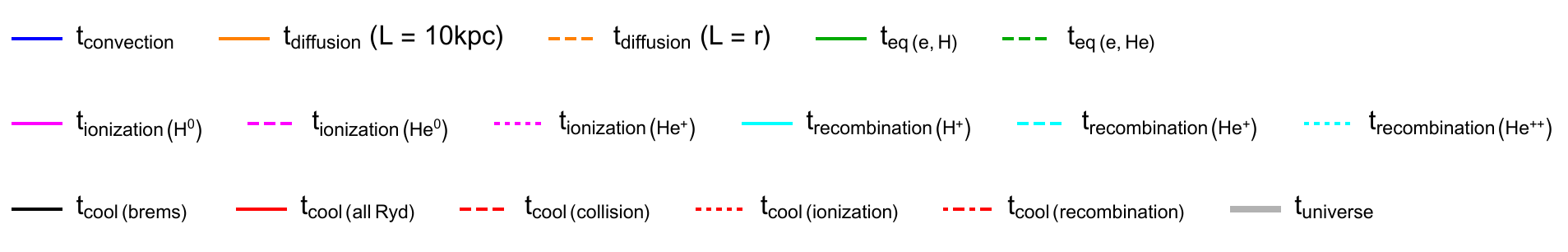}
\end{center}
\caption{
Timescales in the mirror halo for equilibration between mirror electrons and atoms (green), ionization (magenta), recombination (cyan), as well as bremsstrahlung cooling (black) and collision, ionization and recombination cooling (solid red total, various dashings individual), compared to the convection timescale dictated by the CDM halo (blue) and the age of the universe (thick grey line). 
The diffusion timescale over two length scales ($L = r$ and $L = 10$ kpc) is also shown (orange), but this scale is not required to be small for thermal equilibrium in the already isothermal halo.
The SM-like isothermal comparison halo is compared to our three MTH benchmark points from \tref{mirrorprofilesummary} assuming an NFW CDM profile. 
}
\label{f.timescalesNFW}
\end{figure}

\begin{figure}
\begin{center}
\begin{tabular}{cc}
\includegraphics[width=0.45\textwidth]{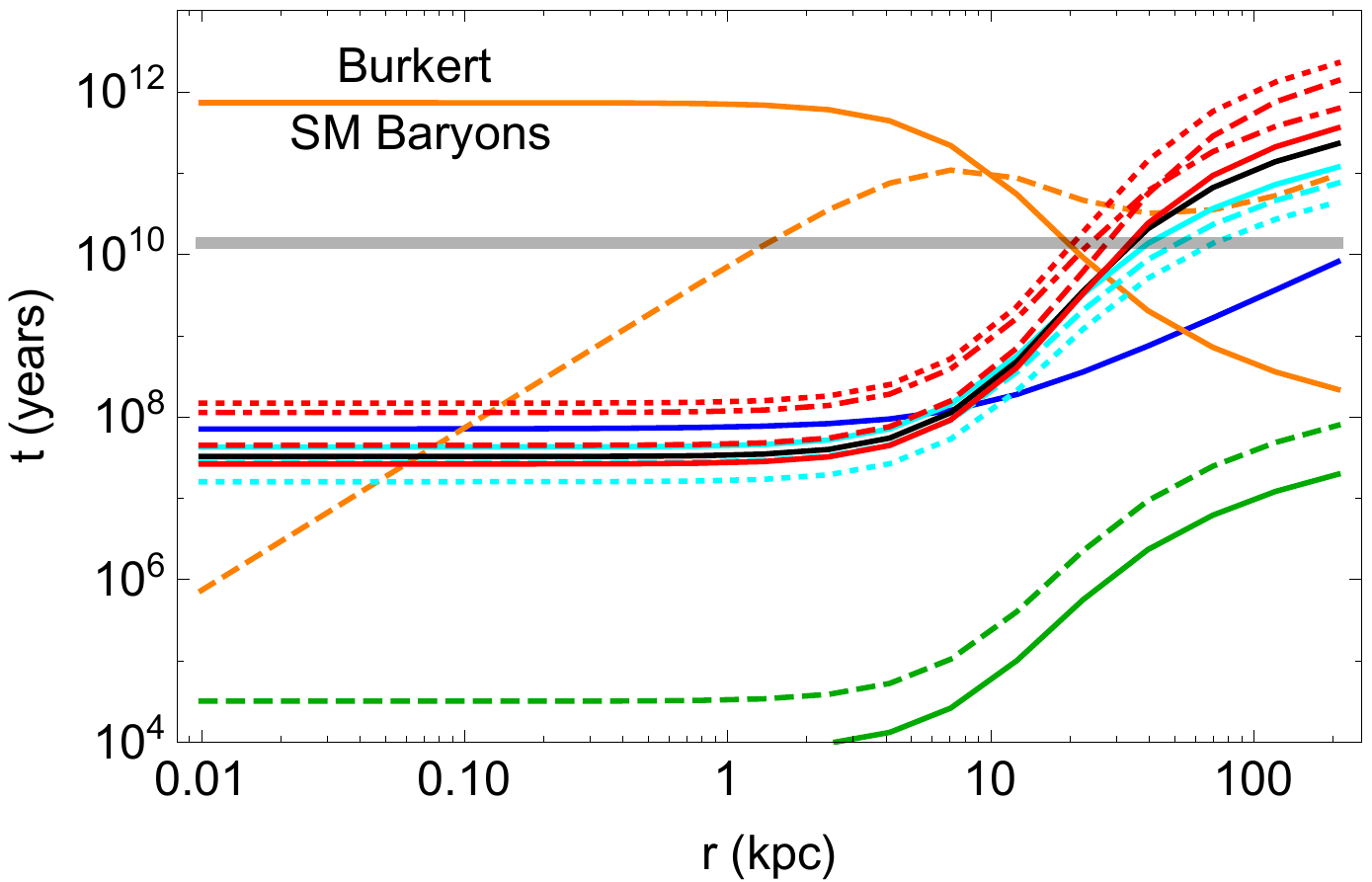} 
&
\includegraphics[width=0.45\textwidth]{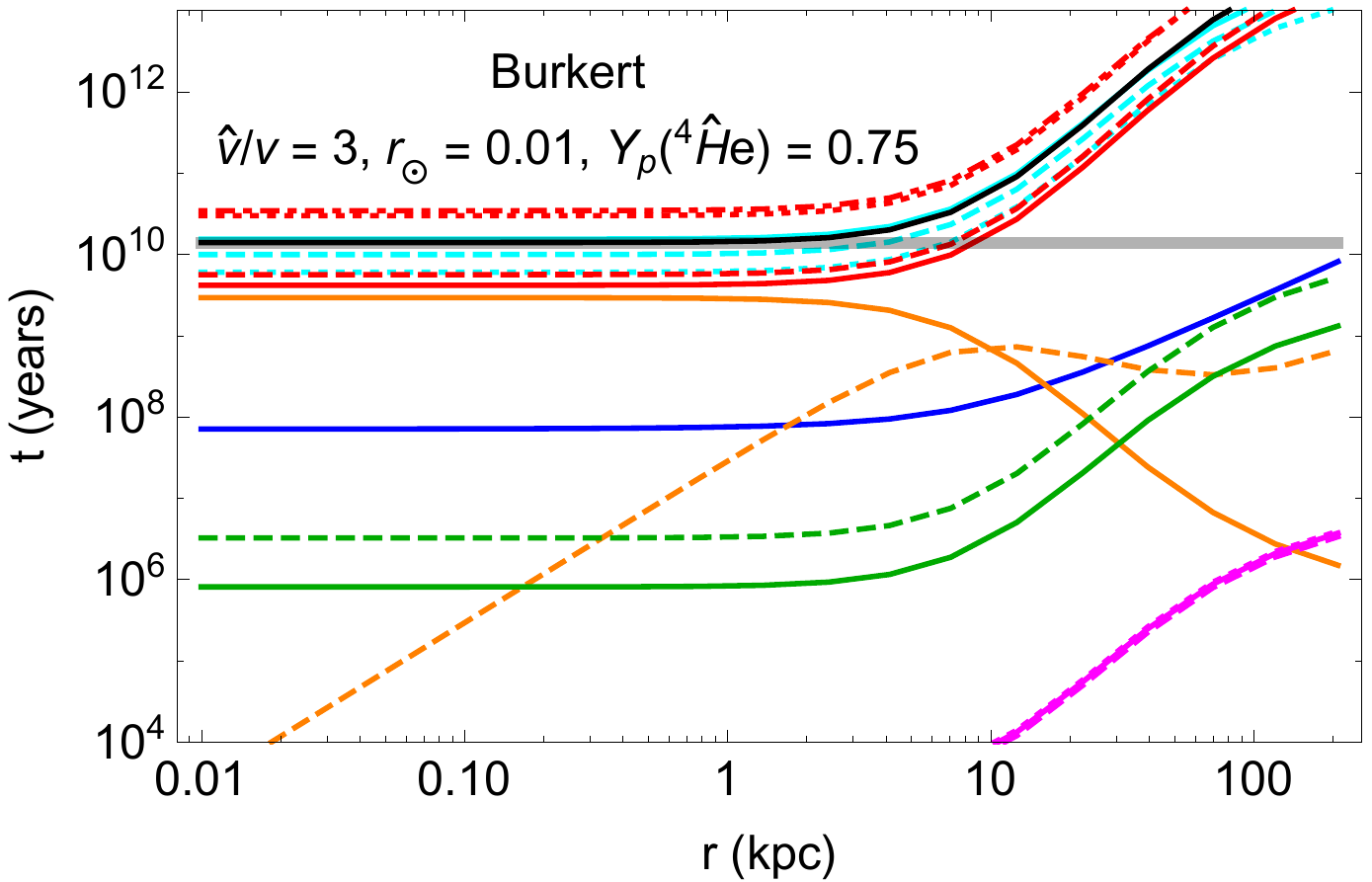} 
\\
\includegraphics[width=0.45\textwidth]{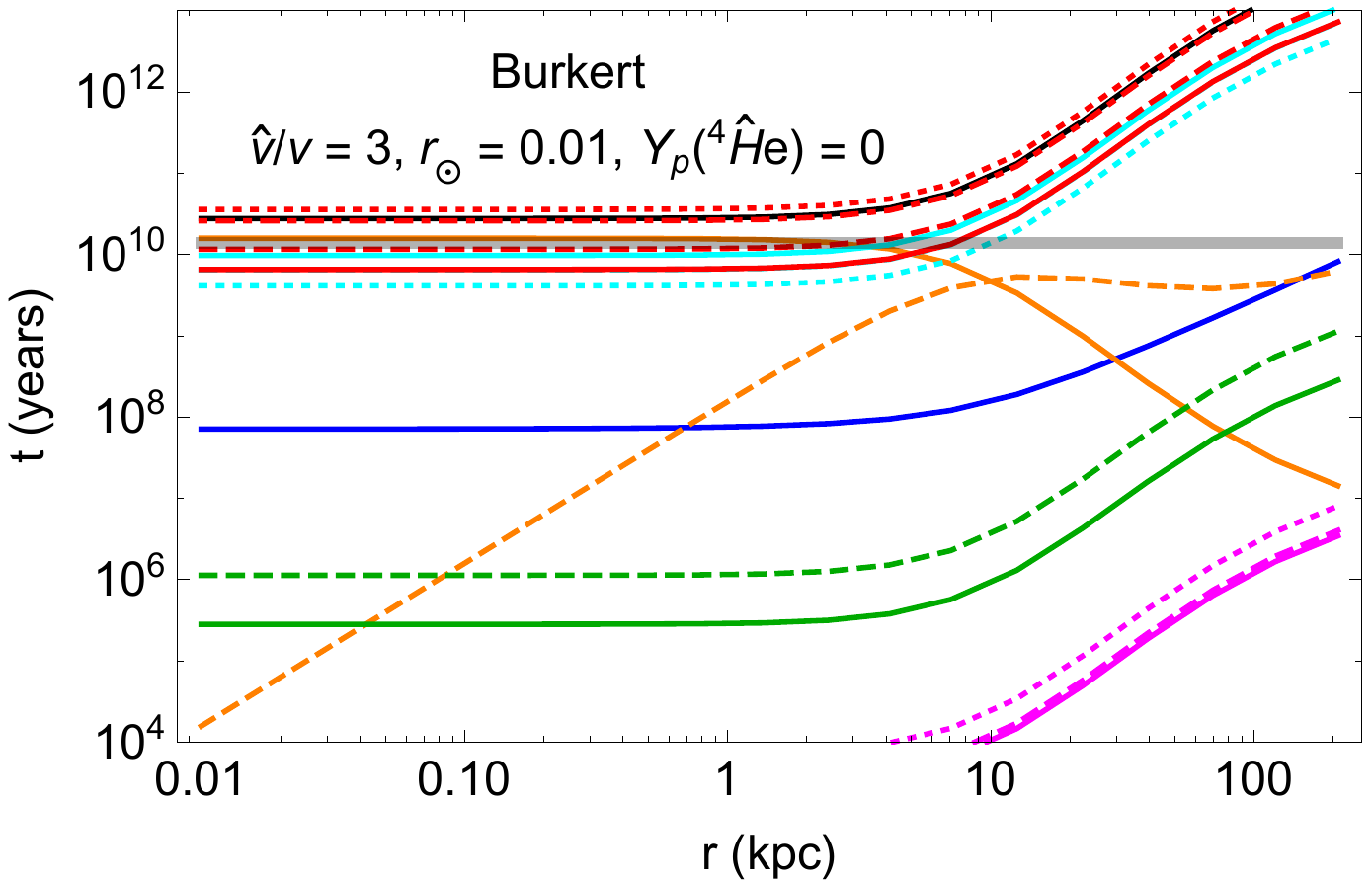} 
&
\includegraphics[width=0.45\textwidth]{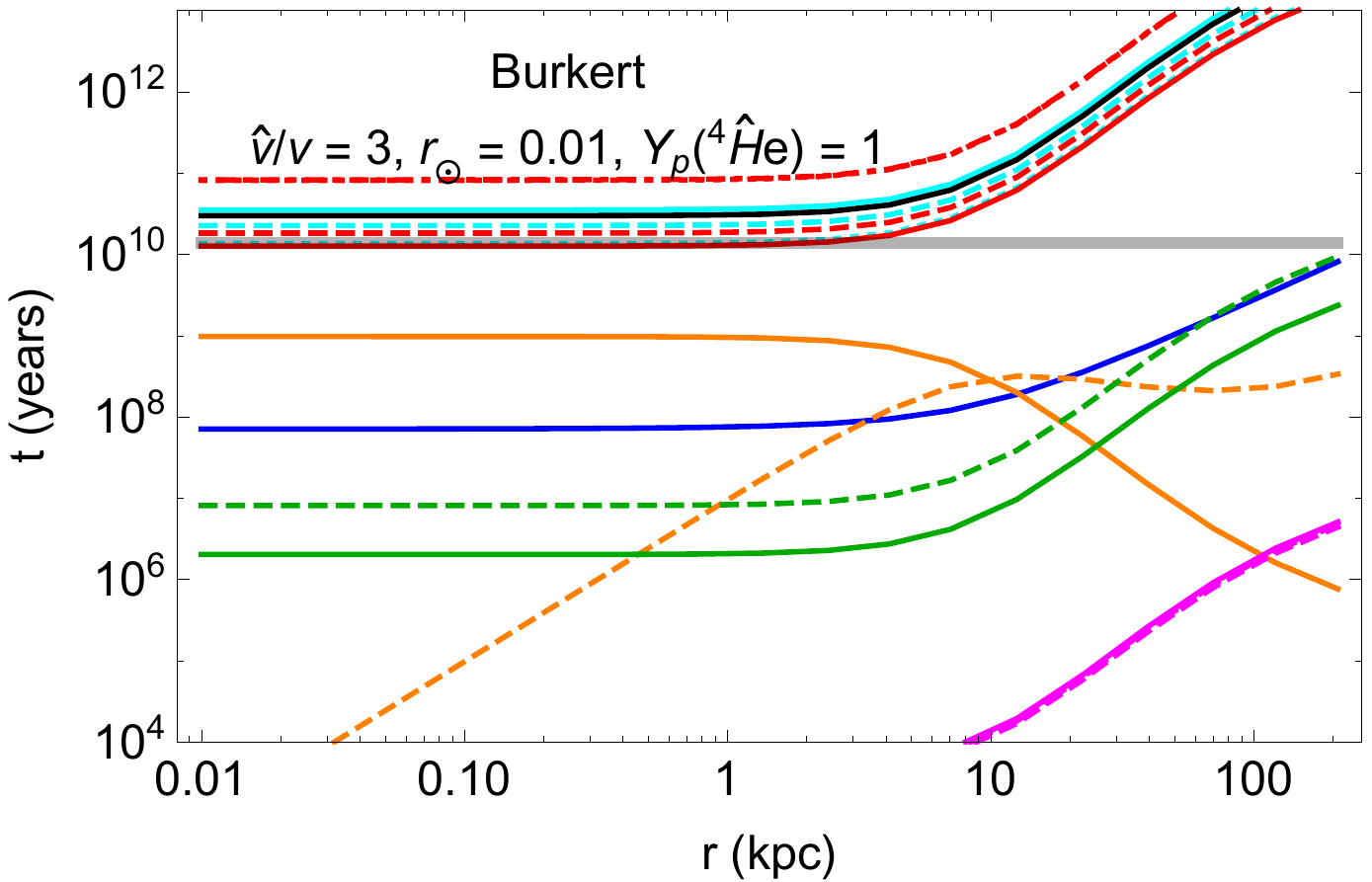} 
\end{tabular}

\vspace{2mm}
\includegraphics[width=0.9\textwidth]{PAPERPLOT_timescalelegend}
\end{center}
\caption{
Same as \fref{timescalesNFW}, but assuming the Burkert CDM profile.
}
\label{f.timescalesBUR}
\end{figure}

\subsubsection{Optical Depth}
\label{s.opticaldepth}


In the discussion of cooling and ionization above, we assumed the mirror 
halo was optically thin, so that mirror photons produced in the various 
cooling processes escape the galaxy without being recaptured. We now 
justify this assumption. The two most important processes that lead to 
absorption or scattering of mirror photons in the halo are Thomson 
scattering, $\hat e^- \hat \gamma \to \hat e^- \hat \gamma$, and 
photoionization, $\hat X_i \hat{\gamma} \to \hat X_i^+ \hat e^-$. The 
mean free path of a photon with respect to Thomson scattering is given 
by,
 \begin{equation}
\ell_T=\frac{1}{\sigma_T\,n_{\hat{e}}} = \frac{3 m_{\hat e}^2}{8 \pi \hat \alpha_{em}^2 n_{\hat e}}.
 \end{equation}
 The corresponding expression for photoionization takes the form,
 \begin{equation}
\ell_{\hat X} = \frac{1}{\sigma_{\mathrm{photo}_{\hat X}} n_{\hat X}} \; ,
 \end{equation}
 where $\hat X = \mathrm{\hat H}^0, \mathrm{\hat H e}^{0, +}$. The 
photoionization cross section is given by~\cite{Rosenberg:2017qia},
 \begin{equation} 
\sigma_{\mathrm{photo}_{\hat X}}(\omega) = \frac{2^5 \pi^2 \hat \alpha_{em}^7 m_{\hat e}^2}{3 \omega^4} \frac{e^{-4(\arctan \tau)/\tau}}{1-e^{-2\pi/\tau}},
 \end{equation}
 where $\omega$ is the mirror photon energy and $\tau \equiv 
(\omega/\omega_0-1)^{1/2}$, where $\omega_0$ is the equivalent of the 
Rydberg energy for the atom, as discussed in \sref{ionization}.

The mean free path for photoionization, $\ell_{\hat X}$, is much larger 
than the size of the galaxy. This allows us to neglect photoionization 
for the remainder of the discussion. The optical depth from Thomson 
scattering, however, can be smaller than the halo size in the dense 
inner regions. However, this does not significantly impede the cooling 
efficiency. The average fractional energy loss of a mirror photon with 
energy $\omega$ scattering once with a mirror electron at rest (a good 
approximation since $\omega$ is typically of order $T$ or of order 
$\mathrm{Ryd}$, both of which are much smaller than $m_{\hat e}$) is 
$\omega/m_{\hat e}$. After $n$ scatters, the average energy of a mirror 
photon that began with initial energy $\omega_0$ is given by,
 \begin{equation}
\omega_n \approx \frac{m_{\hat e} \omega_0}{m_{\hat e} + n \omega_0}
 \end{equation}
 The number of times the photon scatters before traveling the distance 
$d$ required to leave the halo can be estimated from a random walk, $d 
\sim \sqrt{n} \ \ell_T$, assuming the photon has an initial energy 
$\sim T \lsim \mathrm{Ryd}$.

We find that for the range of MTH parameters of interest, $\hat{v}/v \geq 
3$ and $r_{\rm all} \leq 0.1$, the mirror photons lose very little of their 
energy before leaving the halo. This is true for both the NFW and 
Burkert CDM profiles. Therefore, this has no significant effect on our 
discussion of cooling. The mirror halo is therefore optically thin, 
justifying our derivation of the ionization fractions in 
\sref{ionization}.

Even for SM-like densities, the attenuation is at most $\sim 10\%$ for photons emerging from the 
innermost regions of the halo, which does not significantly affect the 
cooling timescales. We also find that the attenuation of photons leaving 
the galaxy remains insignificant if the isothermal halo temperature is 
lower by a factor of a few compared to $T_\mathrm{iso}$, the initial 
temperature of the adiabatic halo. Therefore, the halo continues to 
remain optically thin even as cooling progresses.

\subsubsection{Results}
\label{s.mirrorhaloresultsfinal}

Figs.~\ref{f.timescalesNFW} and~\ref{f.timescalesBUR} show the cooling 
rates from bremsstrahling and $\mathcal{O}(\mathrm{Ryd})$ processes 
(collision, ionization and recombination) as a function of distance from 
the galactic center for the NFW and Burkert profiles. These cooling 
rates are to be compared to the convection timescale and to the age of 
the universe.

We begin by noting that the SM-like profile cools much faster than the 
mirror benchmark scenarios. This is primarily due to the fact that 
$\hat Y_p(^4\mathrm{\hat He}) = 0.25$, the near-optimal value for cooling, 
and also the relatively high density compared to our $r_\mathrm{all} = 
0.01$ MTH benchmarks. The fastest cooling timescale $t_\mathrm{cool} = 
t_\mathrm{cool(collision)}$ drops below the convection timescale for $r 
< r_\mathrm{collapse} \approx 10$kpc. Baryons within this radius lose 
pressure support and start collapsing. The size of $r_\mathrm{collapse}$ 
is roughly consistent with the observed size of the Milky Way visible 
disk.

The mirror baryon profiles with $r_\mathrm{all} = 0.01$ all cool 
significantly less efficiently than the SM profile, owing to their lower density and different value of $\hat Y_p(^4\mathrm{\hat He})$. Initially, the 
cooling timescale is much longer than the convection timescale, but 
still lower than (or close to) the age of the universe. This means that 
the mirror halo cools gradually without loss of pressure support. 
However, if the average temperature in the entire halo (or just the 
inner 10 kpc or so) drops by a factor of $\sim$ 2 compared to its 
initial value, the cooling time scale in the inner region will drop 
below the convection timescale for $r < r_{collapse} \sim$ few kpc. We 
have determined this via direct computation of the isothermal profile at 
various temperatures. This suggests that if the mirror halo eventually 
collapses, $r_\mathrm{collapse}$ would only be about half as large as 
the corresponding value for the SM profile. This leads us to posit that 
if the mirror baryons do form a disk, it is likely to be significantly 
smaller and younger than the visible baryonic disk of the Milky Way.

The cooling time scale scales inversely with the number density 
of mirror particles, $t_\mathrm{cool} \sim r_\mathrm{all}^{-1}$, so that 
lower mirror baryon densities are associated with slower cooling. This 
means that if the mirror baryon mass fraction $r_{\rm all}$ lies below some 
critical value, the mirror particles in the Milky Way do not undergo 
significant cooling, so that their distribution today remains in the 
form of an ionized halo. With our assumptions, we find this critical 
value to be $r_{\rm all} \sim 10^{-2}$. However, varying $T_\mathrm{iso}$ by 
a factor of 2 in either direction alters the critical value of $r_{\rm all}$ 
by 2 orders of magnitude, and so there are large uncertainties.

These observations hold for both NFW and Burkert CDM halos. 
It seems clear that although cooling is significantly less effective in 
the mirror halo than for the SM, there is still the possibility of 
forming a dark disk. However, extrapolating these results to the mirror 
baryon distribution today is not simple. We discuss this in the next 
section.

\subsection{Resulting Mirror Baryon Distribution Today}
\label{s.mirrorbaryonstoday}

The above analysis gives some idea of how a mirror matter component in 
our Milky Way might have behaved during the early stages of galaxy 
formation. This simple approach can yield useful quantitative 
information in the limit that the cooling rate is slow compared to the 
lifetime of the universe, or if complicated astrophysical processes can 
be neglected, as would be the case in a dissipative DM model 
that does not possess an analogue of nuclear 
physics~\cite{Ghalsasi:2017jna}. In the MTH model with very small mirror 
DM fraction $r_\mathrm{all} \lesssim 10^{-2}$, the cooling rate 
is so low that the mirror matter distribution would still be close to 
its original isothermal profile today. Unfortunately, for larger mirror 
DM fractions, cooling is significant and we expect that 
feedback from astrophysical processes in the mirror sector cannot be 
neglected. In this case, our only firm conclusion is that it is not 
possible to quantitatively predict the mirror matter distribution at the 
present time. Even so, we can organize the possible outcomes for the 
distribution of the mirror component in a useful manner. This will 
enable us to study the prospects for direct detection in 
\sref{directdetection}.

Naively, sufficient cooling in the halo should lead to the formation of 
an accretion disk. We have shown that cooling in the mirror sector, 
while significantly less efficient than in the visible sector, can still 
lead to a loss of pressure support. The existence of the visible disk 
implies sufficient angular momentum for disk formation, and also a relatively quiet 
merger history so that the creation of a dark disk would not be 
disrupted. However, in the visible sector, disk formation cannot be 
quantitatively reproduced without detailed magnetohydrodynamic (MHD) $N$-body 
simulations 
\cite{10.1093/mnras/stu1738,10.1093/mnras/sty674,10.1093/mnras/sty1690,10.1093/mnras/stx1160,10.1093/mnras/sty3336}. 
The size of the disk and the density profile of the SM baryons depend 
sensitively on astrophysical feedback processes.

The mirror sector in the MTH model, although similar to the SM in many 
respects, is expected to have its own version of nuclear physics. Mirror 
protons and neutrons will form elements up to mirror 
helium~\cite{Chacko:2018vss} and possibly heavier elements as well, but 
obtaining precise predictions about nuclear spectra, binding energies 
and reaction rates is extremely difficult. While it therefore seems 
quite likely that ``mirror stars''~\cite{Mohapatra:1999ih,Foot:1999hm,Foot:2000vy,Berezhiani:2005vv,Michaely:2019xpz} 
would form,  possibly giving rise to spectacular observational signatures~\cite{Curtin:2019lhm, Curtin:2019ngc, Winch:2020cju, Hippert:2021fch}, their distribution and 
characteristics, challenging to predict even if the microphysics were 
fully understood, is presently unknown. The details of mirror-baryonic feedback 
are therefore expected to be very different from the SM, making it very challenging to perform reliable MHD $N$-body simulations of the mirror sector.\footnote{This 
is in marked contrast to mirror DM models with an exact $\mathbb{Z}_2$ 
symmetry~\cite{Foot:2009mw, Ciarcelluti:2010zz, Foot:2014mia, 
Clarke:2016eac}, for which SM astrophysics can be more directly applied 
to the dark sector. Dissipative DM was recently studied in simulations~\cite{Huo:2019yhk}, but in a very different scenario featuring two nearly degenerate DM states, and without the kinds of feedback processes that arise from mirror star formation.} For appreciable DM fractions $r_\mathrm{all} 
\gtrsim 10^{-2}$, we therefore have to consider a range of possibilities 
for how the mirror matter distribution could look like today.



Our aim is to parametrize the possible distributions 
of mirror matter in our stellar neighborhood. This will be the most 
useful approach when considering direct detection signatures in the next 
section. To this end, we define the parameter,
 \begin{equation}
\label{e.localrall}
r_{\odot} = \left. \frac{\rho_\mathrm{mirror}(R_\odot)}{\rho_\mathrm{CDM}(R_\odot)}\right|_{\mathrm{today}}
 \end{equation}
 which parametrizes the fractional contribution of mirror matter to the 
total density of DM in our local neighborhood. Below, we define 
two benchmark local mirror matter distributions, one optimistic and one 
pessimistic for direct detection, which bracket the range of 
possibilities and allow us to make quantitative statements about the 
prospects for discovering mirror matter. For each benchmark, we look to 
express the direct detection limits in terms of a bound on 
$r_\mathrm{\odot}$.

\begin{enumerate}

\item \emph{Halo-Like:} This assumes that mirror matter either
 \begin{enumerate}
\item does not collapse into a disk due to inefficient cooling, or
\item does not collapse into a disk due to strong heating processes that 
keep the halo hot, or
\item does collapse into a disk, but that the Sun is outside of the 
disk radius. The outer portions of the mirror matter distribution could 
still have a diffuse halo-like form, which has an analogue in the SM.  
Star formation causes the Milky Way disk to be baryon-depleted, but 
recently those missing baryons have been discovered at large distances 
from the center of the Milky Way~\cite{2014ApJ...792....8W}, (see 
also~\cite{2018ApJ...868..108B}).
 \end{enumerate}

To represent this possibility, we choose a benchmark mirror DM 
distribution 
that is essentially that of conventional CDM, with local velocity dispersion 
$v_\odot \approx 220$ km/s, the same as that of the CDM halo at the 
Earth's location~\cite{Dehnen:1997cq}. This is not exactly the same as 
the velocity dispersion obtained from the local temperature in our 
isothermal mirror halo solutions (see \tref{mirrorprofilesummary}), but 
we choose the standard CDM value of $v_\odot = v_0$ for ease of reach comparison 
with other DM scenarios. Unlike CDM, we do not cut off the distribution 
at the galactic escape velocity $v^{gal}_{esc} \approx 544$ km/s in the 
galactic frame~\cite{Smith:2006ym}, since mirror matter is not 
collisionless. The velocity of the Earth relative to the halo is taken 
to be $v_{\rm E}  \approx 233$ km/s. The local mirror matter density is not 
fixed, but it seems reasonable that $r_\mathrm{\odot}$ is of the same 
order as the cosmological value $r_\mathrm{all}$. We consider the 
limiting cases that the local distribution of mirror baryons is either 
fully ionized or fully neutral, though the former is more likely if the 
mirror matter is in a hot halo.

\item \emph{Disk-Like:} This assumes that the mirror matter collapses 
into a disk, and that the Sun is inside that disk. Assuming that the 
dark disk is spatially aligned with our own, a pessimistic assumption is 
that the DM velocity dispersion is the same as that of stars and gas in 
our local stellar neighborhood, leading to $v_\odot = 20$ 
km/s~\cite{Dehnen:1997cq,LopezSantiago:2006xv}. Assuming no relative 
motion of the dark disk with respect to the Sun, the only motion of the 
Earth relative to the mirror matter would be due to the Earth's rotation 
around the Sun with velocity $v_{\rm E} = 30$ km/s. The local DM density 
is very difficult to estimate. On the one hand, collapse into a disk 
could concentrate the DM and could lead to $r_\mathrm{\odot} \gg 
r_\mathrm{all}$. On the other hand, mirror star formation and shedding 
of angular momentum during disk formation could deplete free mirror 
baryons from the disk. We will state direct detection constraints in 
terms of the unknown $r_\mathrm{\odot}$ in the next section. However, as 
long as the local DM fraction is within two orders of magnitude of the 
cosmic average value, the direct detection bounds on the couplings of DM 
to the SM derived under the assumption $r_\mathrm{\odot} = 
r_\mathrm{all}$ will be accurate at the level of an order of magnitude. 
Just as for the halo, we will consider two possibilities for local 
ionization, either fully ionized or fully neutral.

For simplicity we assume that even in the disk-like configuration, the 
local mirror matter distribution is approximately thermal. Of course, 
local mirror astrophysics could violate this assumption while 
introducing correlations between local ionization and temperature (which 
is certainly the case in the SM). However, the naive assumption of a 
thermal distribution, while also allowing for various possible states of 
ionization, will be sufficient to demonstrate the range of possible 
outcomes for direct detection.

 Studies of star motion in the Milky Way can be used to constrain the 
amount of dissipative DM that forms a 
disk~\cite{Kramer:2016dqu, Schutz:2017tfp,Buch:2018qdr}. Recently, 
results from the Gaia survey \cite{Schutz:2017tfp,Buch:2018qdr} have set 
an upper bound on disk DM to be at most a few percent of DM 
in the Milky Way. This limit assumes that the dark disk has 
thickness less than about $100$ pc and exhibits a radial profile similar 
to the SM disk. It also assumes that most of the DM component 
of which the disk is composed resides inside the disk. If the disk is 
thicker than about $200$ pc, the GAIA constraints lose sensitivity. In 
our mirror sector, we do not know the radial profile or the height of the
disk, if any.\footnote{One might try and estimate the disk height as was 
done in~\cite{Fan:2013yva}, but this is not appropriate given the likely
importance of feedback in our mirror sector.} Furthermore, even if 
$r_\mathrm{all} = 1\%$, it is unlikely that all of the mirror matter 
would end up in the disk, just as all the SM baryons do not end up in 
the visible disk. Therefore the Gaia constraints do not impose a robust 
constraint on $r_\mathrm{all}$, or even on $r_\mathrm{\odot}$, the 
fractional contribution of mirror matter to the local DM 
density, in the MTH framework.

\item \emph{Nothing in our Neighborhood:} 

One could in principle imagine that the entire mirror halo collapses 
into the center of the galaxy, and that no appreciable abundance is left 
in our stellar neighborhood. However, the fact that a significant 
fraction of SM baryons reside outside the disk makes this a somewhat 
implausible scenario. Nevertheless, it is worth keeping in mind that 
even a local mirror matter abundance that is significantly reduced 
compared to the galactic average could still lead to a direct detection 
signal.

Even so, it is amusing to consider that the central black hole of our 
Milky Way has a mass of $\sim 10^6 M_\odot$, which is of order $10^{-6}$ 
of the mass of the full DM halo. Given the approximate nature 
of our cooling estimates, one might envisage the possibility that the 
central black hole was formed from a mirror matter halo that collapsed 
relatively slowly and adiabatically within about a billion years or so, 
without significant mirror star formation or loss of pressure support. 
(A subdominant dissipative DM component that could seed central 
black hole production via gravithermal collapse has previously been 
considered, for example in~\cite{Pollack:2014rja, Choquette:2018lvq, 
Essig:2018pzq, DAmico:2017lqj}.) However, it is far beyond the scope of this paper to 
carefully examine this possibility.

\end{enumerate}

With the maximally pessimistic option (3) being somewhat unlikely, the 
information at our disposal does not allow us to say whether the 
halo-like option (1) or the disk-like option (2) is more favored. We 
therefore examine direct detection in both of these scenarios, noting 
that they represent an optimistic and pessimistic scenario respectively 
from the point of view of direct detection rates. We expect that the 
true direct detection prospects are likely to lie somewhere between 
these two extremes.

\section{Direct Detection of sub-nano-charged Mirror Matter}\label{s.directdetection}

Current cosmological constraints allow up to about $10\%$ of the DM 
density to consist of mirror matter in the MTH 
framework~\cite{Chacko:2018vss}. In the near future, improved 
measurements of large scale structure are expected to be able to 
constrain this fraction to the sub-percent level. In this section we 
discuss how an even more subdominant component can still give rise to 
distinctive signals in direct detection experiments, which may allow 
this class of theories to be distinguished from other models.

Given the large uncertainties involved in the cooling of the halo 
discussed in the previous section, we focus on the limiting cases of a 
mirror DM distribution that is either halo-like or disk-like, 
and either fully ionized or fully atomic. Our analysis finds that by 
combining the results of different direct detection experiments, it may 
be possible to differentiate between these different possibilities for 
the mirror matter distribution and ionization. Furthermore, it may be 
possible to discern the multi-component nature of mirror DM, 
and thereby distinguish this class of models from other theories.

In the MTH, the SM and mirror sectors interact through the Higgs portal. 
Mirror DM can therefore scatter off SM particles through Higgs 
exchange. Unfortunately, this interaction is far too small to allow for 
direct detection of mirror hydrogen or mirror helium nuclei. This can be 
easily seen by comparison to the FTH scenario \cite{Craig:2015pha}, in 
which the mirror tau is a potential candidate for weakly interacting 
massive particle (WIMP) DM \cite{Craig:2015xla, 
Garcia:2015loa}. For masses below 10 GeV, the mirror tau-nucleon direct 
detection cross section is $\lesssim 10^{-45} \mathrm{cm}^2$, which is 
below the neutrino floor and already very challenging to detect. The 
coupling of mirror protons to the Higgs is an order of magnitude smaller 
than that of the mirror tau. This, together with the lower mirror baryon 
relic density, makes it clear that Higgs exchange is not expected to 
generate an observable direct detection signal in the foreseeable 
future. Similarly, possible contributions to the scattering via scalar 
twin-bottomonium exchange~\cite{Garcia:2015toa} do not increase the 
cross section above the neutrino floor.

The only other renormalizable interaction between the two sectors 
allowed by the gauge symmetries is a kinetic mixing term between the 
hypercharge gauge boson of the SM and its mirror counterpart,
 \begin{equation}
\frac{\epsilon}{2 {\rm cos} \; \theta_W} B_{\mu \nu} B^{\prime \mu \nu} \;.
 \end{equation}
 At low energies this leads to kinetic mixing between the SM photon and 
its mirror counterpart (for a review, see e.g. \cite{Essig:2013lka}). 
Since both U(1) gauge groups are unbroken, mirror baryons acquire an 
electric charge proportional to $\epsilon$, and can be detected in 
electron recoil (ER) and nuclear recoil (NR) direct detection 
experiments through photon exchange. 

 It is crucial for the viability of the MTH framework that the hidden
and visible sectors remain out of equilibrium with each other after
asymmetric reheating has taken place. Avoiding recoupling of the mirror
sector at temperatures of order a few MeV via $e\hat{e}$ scattering
leads to an upper bound on  the kinetic mixing parameter~\cite{Vogel:2013raa},
 \begin{equation}
\epsilon \lesssim 10^{-9} \;.
\label{epsilonbound}
 \end{equation}
 In the MTH model, no kinetic mixing is generated through 3-loop
order~\cite{Chacko:2005pe}, and therefore even such small values of
$\epsilon$ are radiatively stable. 
Therefore, this bound can naturally
be satisfied provided that the contributions to $\epsilon$ from UV
physics are also small{\footnote{There is also a constraint on
$\epsilon$ arising from the distortions in the CMB that result from
energy transfer between the two sectors at temperatures below
$\mathcal{O}(100)$ eV, through scattering of the residual $e$ and
$\hat{e}$. However, this effect is suppressed by the temperature
asymmetry between the SM and mirror sectors, resulting in a weaker bound on
$\epsilon$ than the one in Eq~(\ref{epsilonbound}).}}.

In general, the size of $\epsilon$ depends on details of the UV 
completion of the MTH model, but in the asymmetrically reheated 
scenario, it has to satisfy Eq~(\ref{epsilonbound}). Encouragingly, 
gravity-mediated interactions between the two sectors at the Planck scale
are expected to 
generate $\epsilon \sim 10^{-13}$ \cite{Gherghetta:2019coi}. DM with a 
tiny electric charge, of order $10^{-9}$ or less, may therefore 
constitute a key feature of MTH models. Such {\emph {sub-nano-charged DM}} cannot be probed at colliders or fixed-target 
experiments~\cite{Essig:2013lka}, but provides a natural sensitivity 
goal for direct detection experiments, in particular if the expected 
size of the gravity-mediated contributions to $\epsilon$ 
is realized.

It is worth emphasizing that several constraints that have been applied 
to millicharged DM in the past are not applicable to the MTH 
scenario. It has been argued that any subcomponent of DM with a 
detectable electric charge will be expelled from the disk by galactic 
magnetic fields \cite{Chuzhoy:2008zy,McDermott:2010pa}, and so cannot 
give rise to a direct detection signal. However, this argument does not 
apply in the case of the MTH because, just like SM ions, the twin ions 
radiate and interact with each other with large cross sections and 
quickly dissipate the energy they obtain from the magnetic 
field.\footnote{From Eq.~(3.2) of \cite{Chuzhoy:2008zy}, the relaxation 
time of twin baryon scattering is only of order $100$ years, which is 
much shorter than the time scale $\tau_{acc}$ for the momentum increase 
due to the galactic magnetic field. Also see the discussion in 
Ref.~\cite{Foot:2010yz}} DM carrying a sizable unbroken dark 
charge might also be expected to be severely constrained by Bullet 
Cluster measurements, since the long-range Coulomb interaction induces 
instabilities in the DM plasma~\cite{Damico:2009tep}. However, 
the small DM mass fractions we consider, $r_{\rm all} \lesssim 10\%$, means 
the MTH framework is not affected by these constraints.

\subsection{Local Mirror Baryon Ionization and Velocity Distribution}
\label{ss.ionizationfv}

We will follow the road map laid out in \sref{mirrorbaryonstoday} to 
study the prospects for direct detection of mirror DM in the 
MTH framework. We focus on the limiting cases when the mirror matter is 
either in a halo-like distribution or has collapsed into a disk. The 
\emph{local} mirror DM fraction $r_{\odot}$, defined as per 
\eref{localrall}, is the parameter that direct detection searches 
constrain or measure. For the local DM density we take 
$\rho_\mathrm{CDM}(R_\odot) \approx 0.3 \gev/\mathrm{cm}^3$. Note that 
this is slightly different from the value of $\rho_\mathrm{CDM}(R_\odot) 
\approx 0.5\gev/\mathrm{cm}^3$ assumed in the simulated CDM profiles 
used in the previous section. The value of $0.3 \gev/\mathrm{cm}$ is 
used for ease of comparison with various existing direct detection 
limits,  
but its precise value does not meaningfully affect our discussion.

The signal also depends on the local mirror helium mass fraction 
$\hat{Y}_\odot$. Note that, in general, this can 
be very different from the cosmic value of the helium fraction 
$\hat Y_p(^4\mathrm{\hat He})$ (for $\hat Y_p(^4\mathrm{\hat He}) \neq 0, 1$), as 
can be seen in \tref{mirrorprofilesummary}. It is hence necessary to 
consider different possibilities for $\hat{Y}_\odot$ even if the cosmic 
value of $\hat Y_p(^4\mathrm{\hat He})$ is close to the asymmetrically 
reheated MTH expectation of $\sim0.75$. Therefore, we consider three 
benchmark values for our sensitivity estimates; $\hat{Y}_\odot = 0, 1$, 
and $0.75$. For each case, we derive sensitivities for the local 
$r_{\odot}$ in the natural range of MTH parameters, $\hat{v}/v \sim 3 - 
5$.

We focus on the limiting cases that the mirror matter in our local 
neighborhood is either fully ionized or fully atomic. Intermediate 
ionizations, although possible, do not significantly change our 
conclusions regarding the range of sensitivity. 
The mass 
fractions of mirror hydrogen and helium in our local neighborhood are
 \begin{equation}
\nonumber 
r_\mathrm{\hat H, \odot} = (1- \hat{Y}_{\odot}) r_{\odot}\ , 
\ \ \ 
r_\mathrm{\hat He, \odot} = \hat{Y}_{\odot} r_{\odot} \ , \\
\label{e.ri}  \\
 \end{equation}
 In the limit of complete ionization the mass fraction contributed by 
free mirror electrons is given by,
 \begin{equation}
r_{\hat e, \odot} =
r_{\odot} \left( \frac{m_{\hat e}}{m_{\mathrm{\hat H}}}  \right) 
\left[
1 - \frac{\hat{Y}_{\odot}}{2}
\right] 
 \ .
 \end{equation}
The direct detection signal from mirror matter can then be determined 
once the local mirror baryon velocity dispersion $v_\odot$ and the velocity 
of the Earth relative to the mirror baryons $v_{\rm E}$ are specified. 
Since, in general, mirror particles scatter many times with each other 
before they have a chance to escape the galaxy, this distribution is not 
cut off at the galactic escape velocity $v_\mathrm{esc}$.

An important feature of multi-component DM whose subcomponents are in 
thermal equilibrium with each other is that the individual components 
$\mathrm{\hat H}, \mathrm{\hat He}$ and $\hat e$ have very different 
velocity dispersions that depend on their masses,
 \begin{equation}
\label{e.v0i}
v_{\odot i} =
 v_\odot \sqrt{\frac{\bar{m}_\odot}{m_i}}  =  
 v_\odot \sqrt{\frac{\bar{m}_\odot}{m_\mathrm{\hat H}}}   
 \sqrt{\frac{m_\mathrm{\hat H}}{m_i}} 
  \ \ \ \ , \ \ \ \ i = \mathrm{\hat H}, \mathrm{\hat He}, \hat e \ ,
 \end{equation}
 Here $\bar{m}_\odot \equiv \bar{m}(R_{\odot}) \sim \mathcal{O}(m_{\mathrm{\hat 
H}})$ is the average mass per mirror particle near 
the location of the Earth, and is fully determined in terms of $\hat{Y}_\odot$ and the local 
ionization. For the $\hat Y_\odot = 0.75$ benchmark the mirror hydrogen velocity dispersion is close to 
the standard $v_\odot$, but only half that big for mirror helium and 
enhanced by a factor of 20-30 for free mirror electrons (if present). As 
we shall see, these fast mirror electrons are a promising target for 
direct detection.

With these ideas in mind we now specify the four scenarios for which we
will determine the direct detection signal. 

\begin{enumerate}
\item \emph{Ionized Halo:} 
$ \chi_{\mathrm{{\hat H}^{+}}} = \chi_{\mathrm{{\hat He}^{++}}} = 1$, 
$v_\odot = 220$ km/s, $v_{\rm E} = 233$ km/s. \\
In this scenario, mirror electrons will be very fast, $v_{0{\hat e}} 
\sim 6000$ km/s, and can lead to spectacular signals at ER detectors. 
Mirror hydrogen and helium can both show up in ER and NR detectors, and 
can potentially be distinguished from each other.

\item \emph{Ionized Disk:}
$ \chi_{\mathrm{{\hat H}^{+}}} =  \chi_{\mathrm{{\hat He}^{++}}} = 1$, $v_\odot = 20$ km/s, $v_{\rm E} = 30$ km/s. \\
 This case is similar to the halo, but with much lower $v_\odot$ and $v_{\rm E}$. 
As a result, all recoil energies are reduced, and mirror baryons become 
invisible to NR detectors. Mirror electrons now have a velocity 
distribution similar to that of standard CDM in a halo. While mirror 
electrons can still be detected in ER experiments, ER detection of the 
slow mirror nuclei would require detectors with much lower threshold 
than what is likely to be available in the near future.

\item \emph{Atomic Halo:} 
$\chi_i = 0$, $v_\odot = 220$ km/s, $v_{\rm E} = 233$ km/s. \\
 As compared to the ionized halo, the velocity dispersions of the mirror 
baryons are only slightly different (due to the absence of free 
electrons), and so NR detection of $\mathrm{\hat H}, \mathrm{\hat He}$ 
proceeds almost identically. ER detection of mirror baryons is 
suppressed by a mirror atomic form factor, and there is no separate 
signal from mirror electrons. Note that a mirror matter distribution 
that survives in the halo-like state until today is expected to be very 
hot and therefore fully ionized. Although we believe that this 
distribution is rather unlikely, we include it for completeness.

\item \emph{Atomic Disk:} 
$\chi_i = 0$, $v_\odot = 20$ km/s, $v_{\rm E} = 30$ km/s. \\
 A very challenging scenario without NR signals (just as in the case of 
the ionized disk), and also without an ER signal from free electrons. 
Mirror baryons only show up in ER detectors with very low recoil, with a 
rate that is further suppressed by the mirror atomic form factor. Direct 
detection of this scenario may require ultra-low-threshold ER 
experiments, and may not be possible in the near future.

\end{enumerate}
 We see  that the ionized halo offers the most promise for direct 
detection, while the atomic disk is by far the most pessimistic. The 
true sensitivity is expected to lie somewhere between these different 
limiting cases.
 Fortunately, the sensitivity of direct detection is highly complementary to astrophysical probes of dark mirror baryons. In particular, dark disk scenarios are most likely to lead to the formation of mirror stars, which may provide an alternative discovery 
channel~\cite{Curtin:2019lhm, Curtin:2019ngc, Winch:2020cju, Hippert:2021fch}. White dwarf cooling bounds are also most sensitive for disk-like mirror baryon distributions, probing $\epsilon$ as low as $10^{-12} - 10^{-11}$ for a DM fraction of 10\%~\cite{Curtin:2020tkm}.

 The plasma-like nature of this DM component and its coupling 
to ordinary matter via the photon portal, which causes cross section 
enhancements at low momentum transfer, give rise to several 
complications. The most important of these is capture of mirror 
particles in the Earth. This can potentially affect the direct detection 
signal in two distinct ways. Firstly, a captured population of mirror 
nuclei could collisionally shield direct detection experiments from 
incoming mirror particles~\cite{Foot:2018jpo}. Secondly, in the case of 
the accumulation of a net mirror charge in the Earth, the resulting 
electrostatic repulsion could result in a suppression of the flux of 
incoming mirror particles that carry same charge, along with a reduction 
in their velocities. The sizes of these effects depends on the number of 
captured mirror particles of various species. This in turn depends, not 
just on the various capture processes, but also on the evaporation of 
captured mirror particles from the Earth and the Debye screening of the 
accumulated mirror charge by the ambient mirror plasma.

We perform a detailed study of the effects of mirror matter capture on 
direct detection in \aref{capture}, for kinetic mixings in the range of 
interest, $\epsilon \lsim 10^{-9}$. We find that the captured mirror 
nuclei are primarily composed of $\mathrm{\hat He}$, since the 
evaporation of $\mathrm{\hat H}$ is much more efficient. Mirror 
electrons are ejected extremely efficiently from the Earth, and so the 
Earth eventually acquires a net positive mirror electric charge from the 
excess $\mathrm{\hat He^{++}}$. This net charge is efficiently screened 
by the ambient mirror plasma within a few 100 km of the Earth's 
surface. Our investigation is, to the best of our knowledge, the first 
such study to take the effects of this Debye screening into account. Our 
analysis shows that collisional shielding never plays a significant role 
in suppressing direct detection of mirror matter in our framework. The 
effects of electrostatic shielding are also negligible for $\epsilon 
\lesssim 10^{-11}$. For larger values of the kinetic mixing parameter, 
$\epsilon \gtrsim 10^{-10}$, electrostatic shielding can only modestly weaken the 
projected bounds on the kinetic mixing parameter $\epsilon$ by less than 50\%~(25\%) for mirror helium (hydrogen), while the mirror electron signal is unaffected or slightly enhanced.
 For $r_{\odot} \sim 0.01$, we find that future experiments probe values 
of $\epsilon$ much smaller than $10^{-10}$ and are therefore unaffected. 
Our sensitivity projections for $\epsilon \sqrt{r_{\odot}}$ from mirror 
nuclear recoils will therefore have at most a factor of 2 uncertainty, 
and only if $r_{\odot}$ is so small that the sensitivity boundary lies 
near or above $\epsilon \sim 10^{-10}$. The mirror-plasma effects of 
capture warrant further study, particularly in the context of more 
general dissipative DM models. However, since they do not affect our 
conclusions for MTH models, we neglect them in our analysis of direct 
detection below.

It is also possible for the self-interaction between two mirror 
particles in the solar system to result in one of the two particles 
becoming gravitationally bound in the Sun's gravitational well. However, 
for $r_{\odot} \lesssim 0.1$, the 
scattering length for two mirror baryons is so much larger than the size 
of the solar system that this effect is unlikely to be important. 
Finally, focusing of mirror DM in the gravitational wells of the Sun and 
Earth has the effect of increasing its local velocity at the Earth's 
surface. The low-velocity tails of distributions dominate capture, and 
so we take the speed gain when falling into the Earth's gravitational 
well into account in \aref{capture}, but neglect all such effects in our 
direct detection calculations. Therefore, our sensitivity estimates 
below are somewhat conservative with regard to this effect.

\subsection{Direct Detection via Nuclear Recoils}

We first consider the direct detection of mirror DM via nuclear 
recoils. We begin by reviewing the basic kinematics involved in the 
scattering of sub-nano-charged DM off nuclei. We then determine the 
reach of nuclear recoil experiments in the MTH framework.

\subsubsection{Review}

We begin by computing the cross section for a DM particle $X$ 
of mass $m_X$, mirror electric charge $Q_X$ and kinetic mixing parameter 
$\epsilon$ scattering off a single SM proton of mass $m_p$. Since the 
particles being scattered are nonrelativistic, the velocity of the 
incoming DM particle satisfies $v_X \ll 1$ in the lab frame. 
This then implies $E_r \ll m_X$, where $E_r$ is the recoil kinetic 
energy of the target particle after the collision. The matrix element 
for the $X p \to X p$ process, averaged over initial and summed over 
final particle spins is given by
 \begin{equation}
\label{e.Mbarsq}
|\mathcal{\overline M}|^2 \ \ \ =  \ \ \ 4 e^4 \epsilon^2 Q_X^2 \frac{m_X^2}{E_r^2}
 \end{equation}
(This formula is applicable to both electron and nuclear scattering.) The 
corresponding differential scattering cross section\footnote{Note that 
the total integrated cross section $\int_0^{E_r^{max}}d \sigma_p/d E_r$ 
is divergent, reflecting the infrared singularity expected in Rutherford 
scattering. In direct detection experiments this divergence is regulated 
by the minimum detectable recoil energy.} is given by,
 \begin{eqnarray}
\label{e.dsigman}
\frac{d \sigma_p}{d E_r}  &=& 
\frac{2 \pi \alpha_{em}^2 \epsilon^2 Q_X^2 }{m_p v_X^2 E_{r}^2} \ .
 \end{eqnarray}
 This is to be contrasted with the case of WIMP DM, where nuclear 
scattering via a contact operator leads to a differential cross section 
that is independent of the recoil energy $E_{r}$. A typical collision 
with sub-nano-charged DM therefore produces much less recoil than a typical 
WIMP collision. The maximum nuclear recoil energy for a given DM 
velocity in the target rest frame is given by,
 \begin{equation}
\label{e.Ermax}
E_{r}^{max} = \frac{2 m_p m_X^2 v_X^2}{(m_p + m_X)^2}\ .
 \end{equation}
 Eqns.~(\ref{e.dsigman}) and (\ref{e.Ermax}) can be applied to 
scattering off nuclei $N$ in the usual way, by replacing $m_p$ with $m_N$ and 
multiplying the LHS of \eref{dsigman} by $Z^2 F^2(E_r)$, with $F$ 
being the Helm form factor~\cite{Lewin:1995rx}. The mass range of 
interest for mirror hydrogen and mirror helium DM in the MTH 
model is from about $1 - 5 \gev$. Mirror electrons are too light to be 
detected via NR but can show up in ER experiments.

\begin{figure}
\begin{center}
\includegraphics[height=7cm]{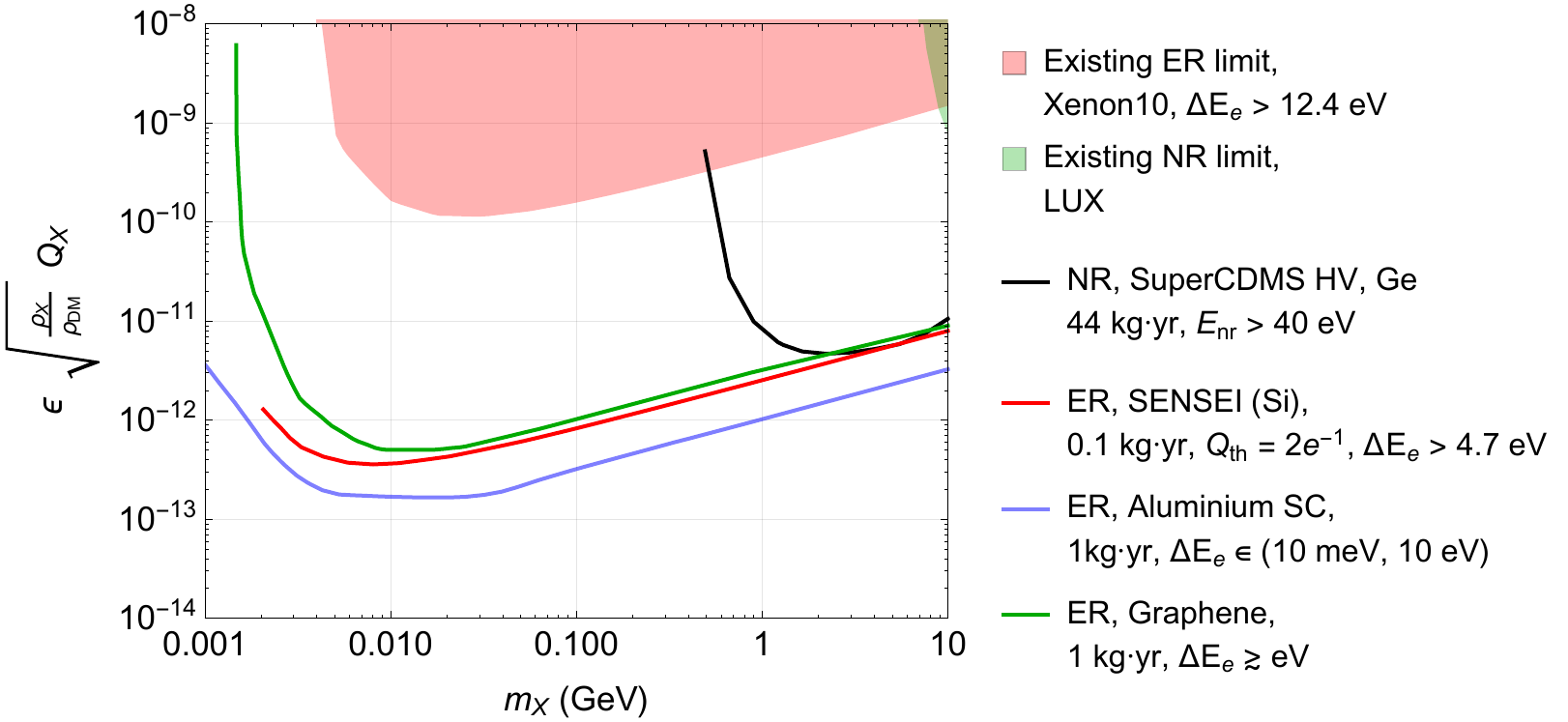}
\end{center}
\caption{
Overview of current direct detection exclusion limits (shaded regions for Xenon10 and LUX) and the most relevant limit projections (lines) on sub-nano-charged DM, assuming the DM is distributed in a fully ionized, single-component standard halo. 
These limits do not directly apply to the MTH model due to the 
different velocity distribution and ionization of mirror baryons.
$\rho_X/\rho_\mathrm{DM}$ refers to the fraction of DM that is made up by the particular constituent $X$.
Current Xenon10 limits from \cite{Essig:2017kqs}. LUX~\cite{Akerib:2013tjd} limits from~\cite{Essig:2015cda}.
We derived SuperCDMS NR limit projections using information from \cite{Agnese:2016cpb}.
ER limit projections: 
SENSEI~\cite{Tiffenberg:2017aac} limits from \cite{Essig:2015cda}.
Superconducting aluminum target with 1 kg$\cdot$year exposure from \cite{Hochberg:2015fth},
Graphene target with 1 kg$\cdot$year exposure from \cite{Hochberg:2016ntt}, 
}
\label{f.millichargedlimits}
\end{figure}	

For completeness and ease of comparison to previously computed limits, 
we first consider the case of the standard collisionless 
single-component DM halo. For DM masses in the GeV range, the largest 
possible recoil energies are $\mathcal{O}(100 \ev)$ on silicon, 
germanium or xenon targets. Most direct detection experiments sensitive 
to nuclear recoil search for WIMPs with masses above $\sim 10 \gev$ and 
have energy thresholds in the keV range, severely limiting their 
sensitivity to sub-nano-charged DM. The next-generation SuperCDMS 
SNOLAB detectors~\cite{Agnese:2016cpb} are the exception, with nuclear 
recoil energy thresholds as low as 40 eV.\footnote{Both the HV and iZIP 
detectors are also sensitive to electron recoil, but present thresholds 
are too high to be useful for mirror DM detection. A possible 
exception are the fast mirror electrons in the ionized disk case, but 
even in that scenario the detectors have considerable background near 
the lower limit of their recoil sensitivity. We discuss dedicated 
semiconductor-based ER detectors in the next subsection.} Using the 
information provided in \cite{Agnese:2016cpb} on minimum thresholds, 
signal efficiency and nuclear recoil spectra of backgrounds after 
imposing selection criteria, it is straightforward to compute exclusion 
limit projections for sub-nano-charged DM from the SuperCDMS SNOLAB 
Si/Ge HV/iZIP detectors.\footnote{We derive these limits by assuming the 
given background distributions are accurate and optimizing, for each 
DM mass, the choice of a single $E_{r}$ interval which 
maximizes signal significance over the background. This gives a cross 
section limit that is a factor of a few better than the conservative 
``optimum interval method'' used by \cite{Agnese:2016cpb}, which does 
not make use of background subtraction. Our method is appropriate for a 
projection of the best possible reach.} The Ge HV detector has the best 
sensitivity due to its low 40 eV threshold, and we show that limit 
projection as the black curve in \fref{millichargedlimits}. We stress, 
however, that the limits shown in \fref{millichargedlimits} cannot be 
directly applied to the MTH model even in the case of an ionized halo, 
because they do not take into account the fact that the velocity 
distributions of the mirror particles are different from that of a 
standard collisionless CDM species.

\subsubsection{Nuclear Recoils in the Mirror Twin Higgs}

It is straightforward to determine the limits on the MTH in the cases of 
ionized or atomic halos or disks, after taking into account their 
different velocity distributions as explained in \ssref{ionizationfv}. 
The projected sensitivity of SuperCDMS HV Ge to mirror hydrogen and 
helium is shown in \fref{MTHepsilonlimits}, for the ionized halo and 
ionized disk scenarios. For local $r_{\odot} \approx 0.01$, mixings as 
small as $\epsilon \sim 10^{-11}$ can be probed in the ionized halo 
case. Note that these experiments have greater sensitivity to mirror 
helium than to mirror hydrogen. For the ionized disk, mirror baryon 
recoil energies are two orders of magnitude lower than for the halo, 
since velocities go down by roughly a factor of ten. As a result, there 
is no NR signal. The momentum transfer for a NR collision with 
$\mathrm{\hat H}$ or $\mathrm{\hat He}$ is $q \approx \sqrt{2 m_N E_r} 
\sim \mathcal{O}(1-10 \mev)$, which is much larger than the energy scale 
corresponding to the size of a mirror atom $\hat a_0 = 
(\alpha_\mathrm{em} m_{\hat{e}})^{-1} \sim (\hat{v}/v \times 4 
\kev)^{-1}$. Therefore, the NR sensitivities for atomic disk and halo 
scenarios are very similar to the corresponding ionized cases.

\begin{figure}
\begin{center}
\hspace*{-10mm}
\begin{tabular}{ccc}
\includegraphics[height=6.5cm]{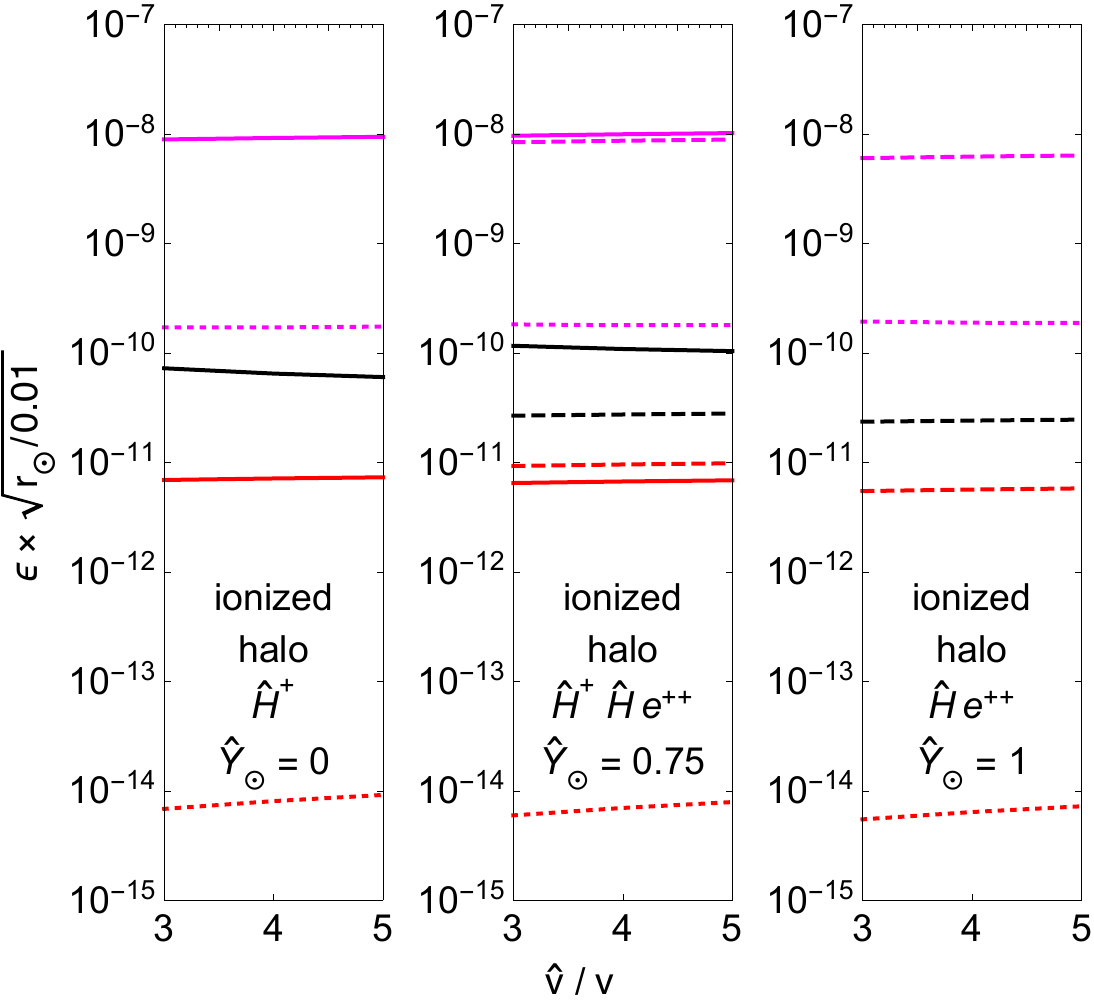}
&
\includegraphics[height=6.5cm]{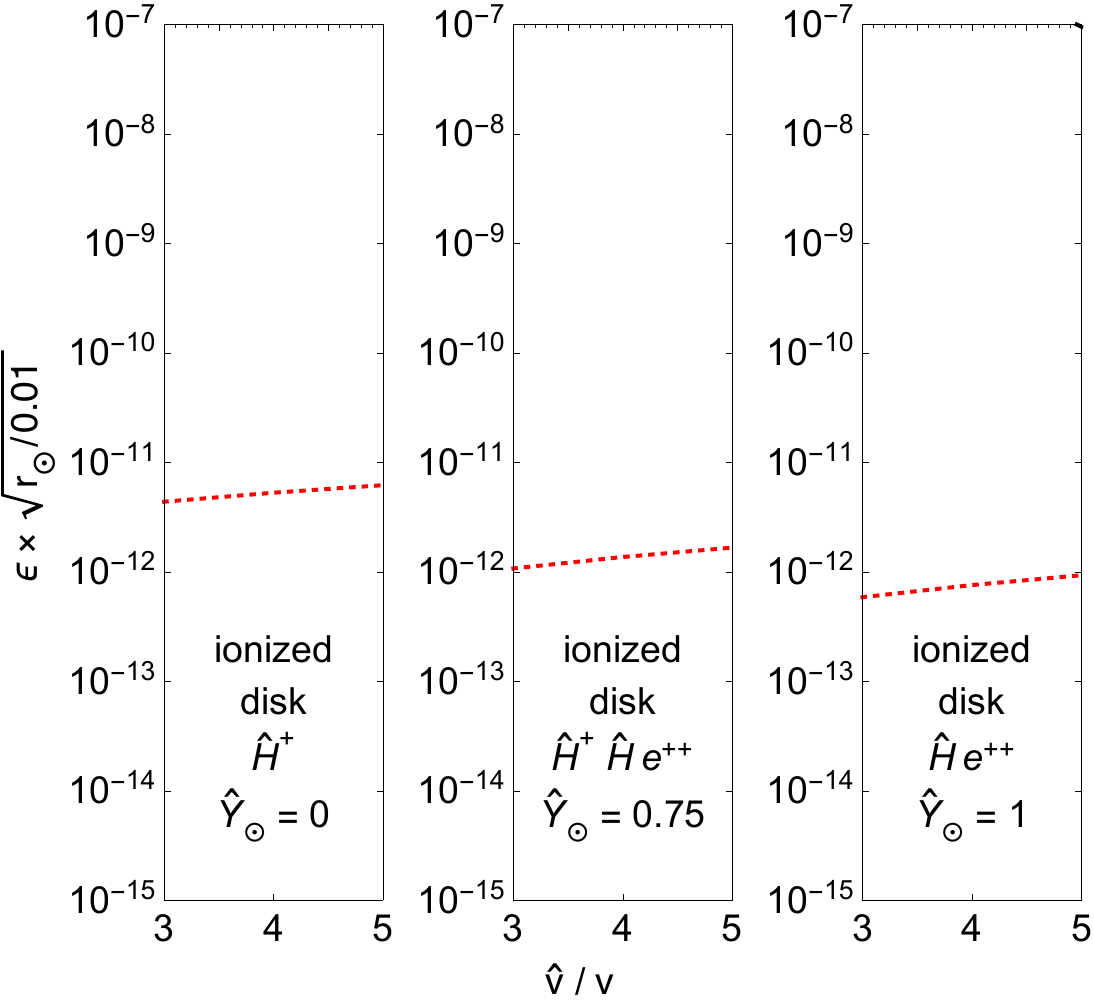}
&
\hspace*{-3mm}
\includegraphics[height=6.5cm]{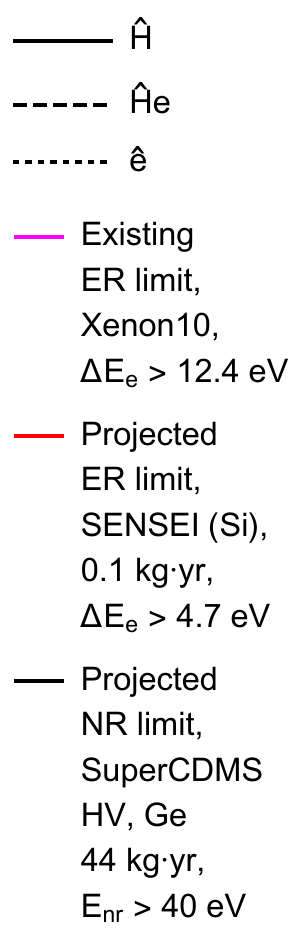}
\\
(a) & (b) & 
\end{tabular}
\end{center}
\caption{
Constraints on photon kinetic mixing $\epsilon$ from direct detection of mirror H, He and electrons at Xenon10 (ER, existing constraint) or SuperCDMS and SENSEI (NR and ER, projected constraints)
for the ionized halo (a) and the ionized disk (b) distributions. 
The mirror particle masses and velocity distributions depend on $\hat{v}/v$ and the mirror helium fraction $\hat Y$, see \ssref{ionizationfv} for details. The limits for partial ionization are very similar to the shown cases of full ionization.
Note that the disk limits do not take the increased local DM density into account.}
\label{f.MTHepsilonlimits}
\end{figure}

\subsection{Direct Detection via Electron Recoils}

We now turn our attention to the direct detection of mirror DM 
via electron recoils. We first review the basic kinematics and existing 
detector technologies before computing the reach of these experiments in 
the MTH framework.

\subsubsection{{Review}}

DM direct detection via electron recoil has been studied in a 
variety of detector materials, including noble 
gases~\cite{Essig:2017kqs, Essig:2011nj, Essig:2012yx}, 
semiconductors~\cite{Tiffenberg:2017aac, Essig:2015cda, Lee:2015qva, 
Essig:2011nj, Graham:2012su}, scintillators~\cite{Derenzo:2016fse}, 
graphene~\cite{Hochberg:2016ntt}, and 
superconductors~\cite{Hochberg:2015pha, Hochberg:2015fth}. ER 
experiments have much lower energy thresholds than NR detectors and 
therefore have the potential to set very stringent limits on 
sub-nano-charged DM. Sensitivity projections for light vector 
mediators are typically expressed in terms of an effective interaction 
cross section $\bar \sigma_e$, which is related to the photon mixing 
parameter via the relation (see e.g. \cite{Essig:2015cda}),
 \begin{equation}
\bar \sigma_e \equiv \frac{\mu_{Xe}^2}{16 \pi m_X^2 m_e^2}
|\mathcal{\overline M}|^2_{q^2 = \alpha^2 m_e^2}
=
\frac{16 \pi m_X^2 Q_X^2 \epsilon^2}{m_e^2 (m_e + m_X)^2 \alpha_\mathrm{em}^2} 
\;,
 \end{equation} 
 where we have made use of \eref{Mbarsq}. Here $m_e$ is the elctron mass 
while $\mu_{Xe}$ represents the reduced mass of the DM-electron 
system. Existing ER limits from Xenon10, as well as projected 
sensitivities of future experiments that have been proposed, are shown 
in \fref{millichargedlimits}. Note that these limits assume a standard 
single-component DM halo. To understand these limits, and how they are 
changed by the different mirror matter distributions in the MTH 
framework, some discussion of ER kinematics and detector technology is 
required.

The kinematics of DM-electron scattering is very different from 
DM-nuclear scattering, since for a standard halo, the electron 
is both the fastest and the lightest particle in the problem. 
Consequently, the typical momentum transfer in a collision is much 
smaller than in the case of DM-nucleus scattering, and so the 
fact that the electron forms a bound state in an atom or a bulk material 
must be taken into account. In particular, the electron does not have a 
definite momentum and very large momentum transfers $q$ are possible. 
Following the discussion in~\cite{Essig:2015cda}, the energy imparted to 
a bound electron can be obtained from energy conservation as
 \begin{equation}
\label{e.DeltaEe}
\Delta E_e = \vec q \cdot \vec v_X - \frac{q^2}{2 \mu_{XN}} \ ,
 \end{equation}
 where $\mu_{XN}$ is the reduced mass of the DM and the nucleus.
 The likelihood of a given $q$ depends, in general, on a (possibly very 
complicated and material-dependent) form factor, but is typically of the 
order
 \begin{equation}
\label{e.qtyp}
q_\mathrm{typ} \sim \mu_{Xe} v_\mathrm{rel}  \sim m_e v_e \sim \mathcal{O}(\mathrm{few} - 10\ \mathrm{keV})
 \end{equation}
 where we have assumed $m_X \gg m_e$. The relative velocity between the 
DM particle and the electron, $v_\mathrm{rel}$, is dominated by 
the velocity of the bound state electron $v_e \sim Z_\mathrm{eff} 
\alpha_\mathrm{em} \sim 10^{-2} Z_\mathrm{eff}$ in atoms (relevant for 
semiconductors, graphene and noble gases) or $v_e \sim v_F \sim 10^{-2}$ 
in Fermi-degenerate materials. (The effective charge $Z_\mathrm{eff}$ is 
1 for outer shell electrons and larger for inner shells.) This is much 
larger than the DM velocity $v_X \sim 10^{-3}$.

For GeV-scale sub-nano-charged DM the second term in \eref{DeltaEe} 
can be neglected, and the typical energy of a scattered electron depends 
\emph{linearly} on the DM velocity. Assuming a standard halo 
profile with $v_X \sim 10^{-3}$, the typical energy imparted to the SM 
electron from a collision with mirror H or He is in the few eV range. 
For sub-nano-charged DM in a standard halo with mass closer to that 
of the electron, the second term in \eref{DeltaEe} can no longer be 
neglected, and \eref{qtyp} is no longer valid. Instead the typical 
electron recoil energy now scales quadratically with DM 
velocity, $\Delta E_e^\mathrm{max} = \frac{1}{2} \mu_{XN} v_X^2$. Then 
the typical energy imparted to a SM electron from a collision with a 
DM particle with the mass of a mirror electron in a standard 
halo is only of order 0.1 eV.

With this parametric understanding of ER kinematics we now consider the 
various experiments in turn and discuss their differences.
 \begin{itemize} 
 \item \emph{Ionization in noble gases}~\cite{Essig:2017kqs, 
Essig:2011nj, Essig:2012yx} \emph{(shaded red region in 
\fref{millichargedlimits})}: In detectors based on noble gases such as 
xenon, NR from DM collisions is detected via prompt scintillation as the 
excited atom returns to its ground state (``S1'' signal) as well as 
ionization, where the liberated electrons are accelerated by a strong 
external electric field, escape the liquid phase of the detector and 
release scintillation light as they traverse the gaseous phase (``S2'' 
signal). These detectors can also be sensitive to electron recoil if the 
S1 signal requirement is dropped. The best current limits on 
sub-nano-charged DM for masses below 10 GeV were derived 
in~\cite{Essig:2012yx} (refined in \cite{Essig:2017kqs}) using an 
S2-only Xenon10 dataset with a single-electron ionization 
threshold~\cite{Angle:2011th}. This allows ER events to be detected as 
long as the collision imparts at least 12.4 eV of energy to the electron 
and liberates it from the outer shell. Higher energy recoils can be 
distinguished by higher levels of ionization. The relatively high levels 
of detector specific background limit sensitivity. Current limits were 
obtained without a background model, assuming all the observed events to 
arise from DM scattering.\footnote{ While this paper was in preparation, 
LUX~\cite{Akerib:2019diq} and Xenon1T~\cite{Aprile:2019xxb} published 
analyses based on searches for electron recoils in mirror models. The 
LUX analysis relied on assumptions about collisional shielding that do 
not apply in our case (see \aref{capture}). Nevertheless, it may have 
better sensitivity than Xenon10 to the halo case. The Xenon1T analysis 
improves the reach in $\epsilon$ by one order of magnitude for GeV DM 
masses compared to Xenon10, but the projected reach of SuperCDMS via nuclear 
recoil that we compute is still more sensitive. For mirror electrons in 
the MeV range, Xenon10 is more sensitive than Xenon1T.
}

\item \emph{Ionization in Semiconductors}~\cite{Essig:2015cda, 
Lee:2015qva, Essig:2011nj, Graham:2012su} \emph{(red curve in \fref{millichargedlimits})}:
 In silicon (germanium), the minimum electron energy required to eject 
an electron is 1.11 eV (0.67 eV). As in noble gas detectors, the 
resulting ionization(s) are picked up by accelerating the liberated 
electrons in an external electric field. Detecting a single ionization 
is very challenging, but the SENSEI 
collaboration~\cite{Tiffenberg:2017aac, Abramoff:2019dfb} was recently funded 
to build a 
100-gram silicon detector capable of detecting ER, with an ionization 
threshold that could be as low as $Q_\mathrm{th} = 2$, corresponding to 
$\Delta E_e > 4.7 \ev$. The resulting sensitivity~\cite{Essig:2015cda} 
is shown as the red curve in \fref{millichargedlimits} for the standard single-component halo. Note that the sensitivity extends down to DM masses in the mirror electron 
range, relying on the tail of the mirror DM velocity distribution to 
achieve an electron recoil above threshold.

\item \emph{Scintillators}~\cite{Derenzo:2016fse}: 
 An alternative path to low-threshold ER detection makes use of 
scintillators. Here the experimental observable is the scintillation 
light emitted as the excited atoms relax to their ground state. While 
this has slightly worse sensitivity than ionization in semiconductors 
due to the $\sim 6 \ev$ thresholds of the readily available 
scintillation materials such as NaI, it may allow for lower backgrounds 
since no electric field is required to manipulate the liberated 
electron.

\item \emph{Graphene}~\cite{Hochberg:2016ntt} \emph{(green curve in \fref{millichargedlimits})}:
 Graphene is a very attractive target for ER DM direct detection. It has 
an energy threshold $\Delta E_e \gtrsim$ eV comparable to 
semiconductors, and the momentum of the ejected electron can be directly 
determined without relying on secondary excitations, allowing for 
\emph{directional} DM detection as well as a very precise 
measurement of the ER spectrum. Furthermore, this proposal could be 
realized in the near future by running the PTOLEMY 
experiment~\cite{Betts:2013uya} with bare rather than tritium-holding 
graphene surfaces. Achievable sensitivity, calculated 
by~\cite{Hochberg:2016ntt} assuming backgrounds can be rejected, is 
shown as the green curve in \fref{millichargedlimits}.

\item \emph{Superconductors}~\cite{Hochberg:2015pha, Hochberg:2015fth} \emph{(blue curve in \fref{millichargedlimits})}:
 A DM collision with electrons in a superconductor could 
disrupt a Cooper pair and create two propagating quasiparticle 
excitations above the Fermi sea. The band gap for this transition is 
tiny, of order $10^{-3}$ eV, allowing in principle for DM 
detection with extremely low thresholds. Once these excitations are 
produced in a large volume superconducting substrate, they must be 
concentrated and collected in a small volume absorber, and read out with 
sensors like Transition Edge Sensors (TES) or Microwave Kinetic 
Inductance Devices (MKID). One promising approach discussed 
in~\cite{Hochberg:2015fth}, (see also~\cite{Hochberg:2019cyy}), is the 
use of a single aluminum crystal in the superconducting state, allowing 
for efficient propagation and collection of the produced excitations. 
Estimated reach for sub-nano-charged DM with 1 kg$\cdot$year of 
exposure is shown as the blue curve in \fref{millichargedlimits} for a 
readout sensor dynamic range of 10 meV - 10 eV. The solar neutrino 
background has been included in this sensitivity estimate, which
scales linearly with ER energy in almost all of the relevant energy range. 
The time scale for implementing superconductors as ER DM 
detectors is probably longer than for the other technologies discussed 
here. The required $\mathcal{O}(1 \; \mathrm{meV})$ sensitivities have not 
yet been achieved, though they are theoretically possible in TES and 
MKID sensors, and various engineering approaches for improving 
sensitivity have been proposed.

Recent ideas like polar target 
materials~\cite{Knapen:2017ekk, Griffin:2018bjn} could play an 
important role similar to superconductors due to their very low thresholds.
 \end{itemize}
 \fref{millichargedlimits} makes it clear that superconductors, graphene 
and semiconductors have comparable sensitivities to sub-nano-charged DM in 
a standard single-component halo. We now discuss the role each of these 
detection technologies could play in the detection of mirror baryons and 
electrons.

\subsubsection{Electron Recoils in the Mirror Twin Higgs}
\label{s.ERMTH}

We now compute the sensitivities of ER experiments to mirror baryons and 
electrons for the different benchmark distributions defined in 
\ssref{ionizationfv}. For our quantitative analysis we focus on Xenon10, 
on which the best current constraints are based, and SENSEI (ionization 
in silicon), which serves as an example of the sensitivity achievable in 
the near future. We also give a qualitative discussion of the role that 
other technologies, such as graphene and superconductors, can play.

For the Xenon10 ER constraint, we compute the limits in the same manner 
as in Ref.~\cite{Essig:2017kqs}, but taking into account the different 
velocity distributions of the individual mirror electron and baryon 
components.\footnote{We are very grateful to Tien-Tien Yu for supplying 
us with the necessary code, which includes the atomic form factors for 
xenon.} As is clear from the magenta curves in \fref{MTHepsilonlimits}, 
Xenon10 already constrains the nano-charged regime in the ionized halo 
scenario. The signals from mirror hydrogen and helium in this case are 
not very different from those of a standard halo, leading to a bound of 
$\epsilon \lesssim 10^{-8}$ for $r_{\odot} \sim 0.01$, as expected from 
\fref{millichargedlimits}. For mirror electrons, the situation is quite 
different. They would not be detectable at Xenon10 if they exhibited a 
standard halo velocity distribution, due to the high single ionization 
threshold of 12.4 eV compared to the typical recoil energy. However, in 
the ionized halo, their increased speed allows them to easily liberate 
electrons from the outer shells of xenon. It is worth noting that the 
kinematics of this collision are quite different from the discussion 
following \eref{qtyp}. Since the mirror electron is now the fastest 
particle in the problem, we might naively expect $q_\mathrm{typ} \sim 
m_e v_{\hat e}$. However, the atomic form factor of xenon, as well as 
the $1/q^4$ suppression in the scattering cross section, still favor 
momentum transfers at or below $\alpha m_e$. Instead, the increased 
mirror electron speed allows sizable $\Delta E_e$ to be generated from 
collisions with very low momentum transfer, at or much below $\alpha 
m_e$. The same $1/q^4$ cross section dependence then leads to a huge 
rate enhancement for fast mirror electrons, allowing Xenon10 to set 
limits on $\epsilon$ of order $10^{-10}$ for the ionized halo with 
$r_{\odot} \sim 0.01$.

Xenon10 does not set limits on the ionized disk scenario; mirror 
electrons now have a speed comparable to that expected from a standard 
halo, and as shown in \fref{millichargedlimits}, this is not sufficient 
to ionize xenon. Mirror baryons are slower by an order of magnitude, 
leading to a corresponding decrease in recoil energy, which is also 
insufficient to ionize xenon.

We repeat this calculation for the ionization of silicon in the SENSEI 
experiment~\cite{Tiffenberg:2017aac}, using the public \texttt{QEDark} 
code made available by the authors of~\cite{ Essig:2015cda} to compute 
the signal rate, which includes the fully pre-computed crystal form factor. 
The corresponding background-free ER limit projections (4 expected 
events) are shown as red lines in \fref{MTHepsilonlimits}. The 
discussion of kinematics and rate enhancement for fast mirror electrons is very 
similar to the case of xenon. An important difference is that, due to the 
lower ionization threshold in silicon, the signal from fast mirror 
electrons in the ionized halo is even more enhanced than in xenon, 
making it possible to see them even in the ionized disk scenario.

Mirror baryons can be discovered in SENSEI for $\epsilon \gtrsim 
10^{-11}$ in the ionized halo with $r_{\odot} \sim 0.01$, while mirror 
electrons give rise to a detectable signal even for tiny $\epsilon \sim 
10^{-14}$, a truly remarkable sensitivity that can probe mixings even 
smaller than the expected gravity-mediated 
contributions~\cite{Gherghetta:2019coi}. In the ionized disk scenario, 
mirror electrons are detectable for $\epsilon \gtrsim 10^{-12}$. This 
will allow a very effective probe of the sub-nano-charged regime. Mirror 
baryons in the ionized disk are very challenging to detect due to their 
low recoil, both in NR and ER experiments. This represents a great 
opportunity for a future superconductor or polar material based ER 
detector, which would be able to probe this scenario very effectively 
due to its tiny meV thresholds.

Future graphene-based detectors~\cite{Hochberg:2016ntt} are likely to 
have a sensitivity comparable to SENSEI for the same exposure due to 
their similar ionization energies. However, graphene has the unique 
ability to detect the direction of the DM impact, which could provide 
another useful handle for diagnosing the mirror baryon distribution. For 
example, in the ionized disk scenario as described in 
\ssref{ionizationfv}, we do not expect significant annual modulation in 
the strength of the signal, but to the small extent that directional 
bias exists in the impact of mirror electrons, events would be sensitive 
to the direction the Earth is currently heading around its orbit. This 
would constitute a striking signal of a mirror baryonic disk.
 
What if the mirror particles form atoms either in a disk or a halo? 
Since the typical momentum transfer in the collision of a mirror atom 
with a nucleus is much larger than the binding energy of a mirror atom, 
the NR signal is expected to be very similar to that in the ionized 
case. However, the situation with regard to ER signals is very 
different. As discussed above, the typical momentum transfer in 
collisions between mirror baryons and a visible bound electron is 
$q_\mathrm{typ} \sim$ few - 10 keV, see \eref{qtyp}. The characteristic 
size of mirror atoms is given by the mirror Bohr radius, $\hat a_0 = 
(\alpha_\mathrm{em} m_{\hat{e}})^{-1} \sim (\hat{v}/v \times 4 \kev)^{-1} $. 
Since the momentum transfer corresponds to length scales similar to or 
larger than a mirror atom, the total scattering rate will be suppressed 
by mirror atomic form factors. (The velocity distributions would also 
change slightly due to the absence of free mirror electrons but, just as 
for NR, this is a less important effect.) For $q_\mathrm{typ}\ll \hat 
a_0^{-1}$, this cross section suppression is $\sim \hat 
a_0^4\,q_\mathrm{typ}^4$~\cite{Cline:2012is}. Compared to the ionized 
case, the corresponding reduction in sensitivity to $\epsilon$ is 
roughly
 \begin{equation}
\label{e.atomicDMsuppression}
\hat a_0^2\,q_\mathrm{typ}^2 \sim \left(\frac{v}{\hat{v}}\right)^2 \left( \frac{v_e}{\alpha_\mathrm{em}}\right)^2 \ .
 \end{equation}
 The first term is $\sim \mathcal{O}(0.1)$ for our parameters of 
interest, while $v_e/\alpha_\mathrm{em}$ is $\sim \mathcal{O}(1)$ for 
outer ionization electrons and in superconductors. Therefore, the 
$\epsilon$ sensitivities for the atomic disk/halo are at most an order 
of magnitude or so weaker than the corresponding sensitivity for the 
ionized disk/halo. Compared to the ionized case, the ER spectrum will 
also be modified. Since the scattering proceeds via the dipole moment of 
the mirror atom, there is no separate mirror electron signal.

The atomic halo would therefore be clearly discoverable via NR and ER 
detection of mirror baryons, though with reduced ER sensitivity. Both 
the different signal rate and recoil spectrum shape can be used to 
distinguish this scenario from the ionized halo or disk. The atomic disk 
is an extremely challenging case. The absence of free and fast mirror 
electrons means that only a future superconductor or polar material based detector with 
very low threshold has a chance to detect mirror atoms. Even then the 
rate would be suppressed by DM atomic form factors. That being 
said, the absence of events in all other detectors would make a 
discovery at such a low-threshold detector a striking signal of an 
atomic disk.

\subsection{Characterization of the Dark Sector}

\begin{table}
\begin{center}
\begin{tabular}{|l||l|m{23mm}|m{26mm}|m{39mm}|}
\hline
& ionized halo & ionized disk & atomic halo & atomic disk \\
\hline \hline
ER $\mathrm{\hat H}$, $\mathrm{\hat He}$
& $\epsilon \sim 10^{-8}$
& no signal
& AFF: $\epsilon \sim 10^{-7}$
& no signal
\\
\hline
ER $\hat e$
& $\epsilon \sim 10^{-10}$
& no signal
& no signal
& no signal
\\
\hline
\end{tabular}
\end{center}
\caption{
Summary of \emph{existing constraints} on dark photon mixing $\epsilon$ from direct detection of mirror baryons and electrons via electron recoil at Xenon10. 
Here we assume $r_{\odot} \sim 1 \%$ for all cases. Limits scale with $r_{\odot}^{-1/2}$. Note that for a given cosmic mirror baryon abundance $r_{all}$,  the local density $r_{\odot}$ is likely higher in the disk cases compared to the halo.
In the left three columns, the reason for the sensitivity reduction compared to the ionized halo is given. AFF = Atomic Form Factor, RR = Reduced Recoil, see text for details. 
}
\label{t.ddsummaryxenon}
\end{table}

\begin{table}
\begin{center}
\begin{tabular}{|l||l|m{23mm}|m{26mm}|m{39mm}|}
\hline
& ionized halo & ionized disk & atomic halo & atomic disk \\
\hline \hline
NR $\mathrm{\hat H}$, $\mathrm{\hat He}$
& $\epsilon \sim 10^{-11}$ 
& RR:  no signal 
&  $\epsilon \sim 10^{-11}$
& RR: no signal
\\
\hline
ER $\mathrm{\hat H}$, $\mathrm{\hat He}$
& $\epsilon \sim 10^{-11}$
& RR: SC only?
& AFF: $\epsilon \sim 10^{-10}$
& RR and AFF: SC only?
\\
\hline
ER $\hat e$
& $\epsilon \sim 10^{-14}$
& $\epsilon \sim 10^{-12}$
& no signal
& no signal
\\
\hline
\end{tabular}
\end{center}
\caption{
Summary of \emph{projected sensitivities} to dark photon mixing $\epsilon$ from direct detection of mirror baryons and electrons via nuclear recoil at SuperCDMS HV Ge, and via electron recoil at SENSEI (or a hypothetical superconductor detector).
Here we assume $r_{\odot} \sim 1 \%$ for all cases. Limits scale with $r_{\odot}^{-1/2}$. Note that for a given cosmic mirror baryon abundance $r_{all}$,  the local density $r_{\odot}$ is likely higher in the disk cases compared to the halo.
In the left three columns, the reason for the sensitivity reduction compared to the ionized halo is given. AFF = Atomic Form Factor, RR = Reduced Recoil (meaning detection may require a superconductor or polar material based detector) , see text for details.
}
\label{t.ddsummary}
\end{table}

In Tables~\ref{t.ddsummaryxenon} and~\ref{t.ddsummary} we summarize the 
present limits and projected future sensitivities of NR and ER direct 
detection experiments to mirror dark matter. In the case of an 
ionized halo, the existing Xenon10 constraints on ER already probe some 
of the nano-charged regime, but the other scenarios are presently 
unconstrained. With future experiments, the ionized halo, ionized disk 
and atomic halo distributions can all be effectively probed, with 
clearly distinct patterns of detection. This may allow these different 
distributions to be distinguished. The atomic disk scenario is however 
very challenging, and will rely on the future development of 
superconductor-based ER detectors with extremely low 
thresholds.\footnote{Direct searches of mirror 
stars~\cite{Curtin:2019lhm, Curtin:2019ngc, Winch:2020cju, Hippert:2021fch}, which are more likely to 
form if the mirror baryons have collapsed into a cold disk, may provide 
a more immediate probe of this scenario.} Once that capability exists, 
it can also be discovered, and distinguished from the other mirror 
matter distributions. Directional detection in graphene-based ER 
detectors, as well as the characteristic annual modulation of any 
detected signal at different detectors, would provide additional 
information that could help resolve any remaining degeneracy between 
different distributions of the sub-nano-charged DM component.

We close this section with the argument that a detailed study of the 
distribution of recoil energies in signal events can be used to further 
characterize the dark sector. In particular, once the nature of the 
mirror baryon distribution has been determined by correlating data from 
different detectors as shown in \tref{ddsummary}, the detailed recoil 
energy spectra can potentially be used to establish that DM is 
multi-component, and also to distinguish the signal from that of a 
primary WIMP DM component. As a demonstration we consider 
mirror baryons distributed in an ionized halo giving rise to a signal in 
a NR detector, such as SuperCDMS SNOLAB Ge HV. \fref{recoilspectrum} 
shows the nuclear recoil spectra for $\mathrm{\hat H}$ and $\mathrm{\hat 
He}$ for $\epsilon = 3 \times 10^{-10}$, as well as WIMP signals of 
comparable statistical significance and identical masses. This 
corresponds to of order a thousand signal events from mirror baryons. 
The recoil spectra from $\mathrm{\hat H}$ and $\mathrm{\hat He}$ are 
clearly very different. Even though the mirror hydrogen signal is much 
smaller than the mirror helium signal, their combined recoil spectrum 
can be distinguished from either individual component.  This can be used 
to establish that the mirror DM consists of more than one type 
of nucleus, and allows for direct measurement of the local mirror helium 
fraction $\hat{Y}_{\odot}$. Furthermore, by establishing that the masses 
and charges of mirror hydrogen and helium are integer multiples of each 
other, these experiments may be able to distinguish the mirror nature of 
the theory.

For comparison, we show the distribution of recoil energies that would 
be expected from WIMPs of the same masses (1.3 or 5.1 GeV in this case). 
It is clear that the WIMP and mirror baryon signals can also be 
distinguished, since the distribution of WIMP events goes out to much 
larger $E_{r}$ than the sub-nano-charged DM component. We have 
further verified that given a few hundred signal events, the nuclear 
recoil spectra of mirror hydrogen and mirror helium can be reliably 
distinguished from that of a WIMP without any prior assumptions about 
the WIMP mass.

\begin{figure}
\begin{center}
\includegraphics[width=\textwidth]{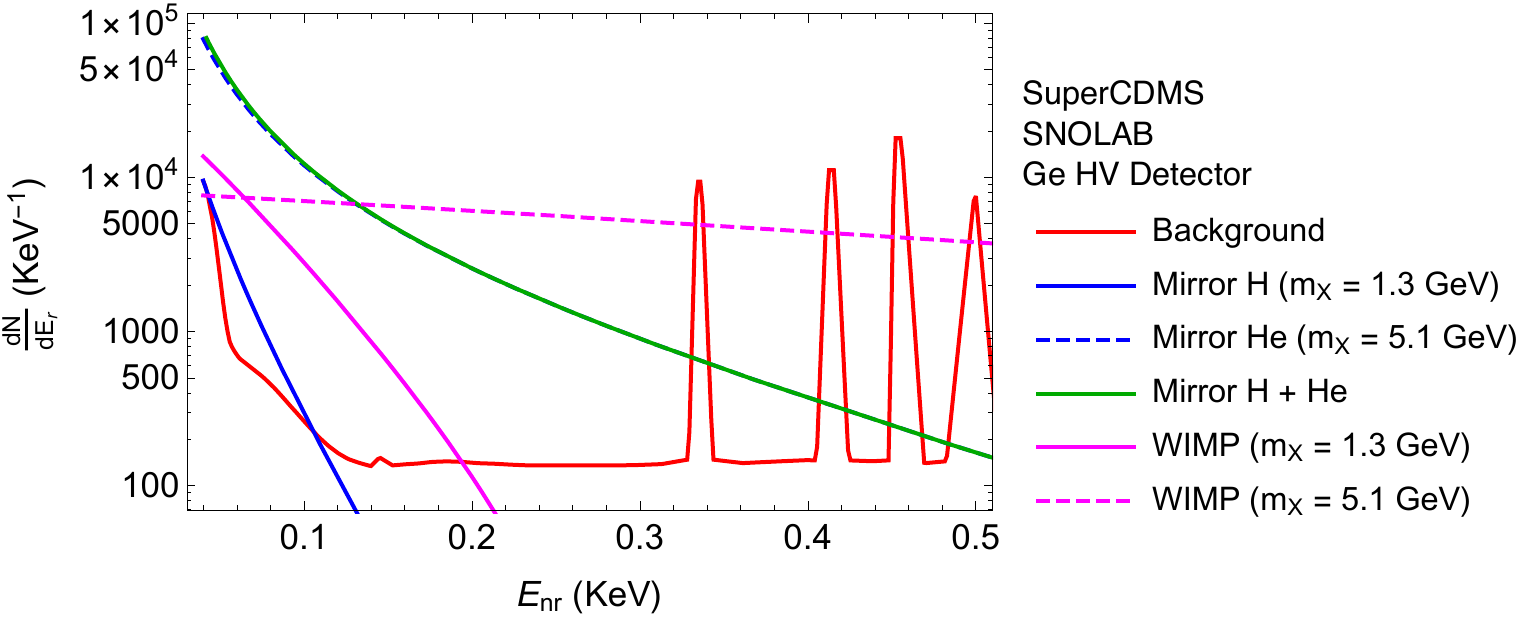}
\end{center}
\caption{
Recoil spectrum in the SuperCDMS SNOLAB Ge HV detector assuming ionized halo DM distribution. 
For mirror H and He, assume local values 
$r_\odot = 0.01, \hat Y_\odot = 0.75, \hat{v}/v = 4$, 
and $\epsilon = 3 \times 10^{-10}$. 
We compare to a WIMP with $m_X = 1.3~(5.1) \gev$ and $\sigma_{nX} = 0.6~(1.1) \times 10^{-42} \mathrm{cm}^{-2}$. 
}
\label{f.recoilspectrum}
\end{figure}

Combining and correlating data from different detectors can reveal 
additional information that could help distinguish mirror DM 
from WIMPs. For example, in an ionized halo the fast mirror electron 
signal in ER detectors would stand out because of its high recoil 
energies. By contrast, a dominant WIMP-like DM component may 
not even produce a signal at ER experiments. Shape analysis of the 
spectrum of signal events could also reveal, for example, whether the 
sub-nano-charged DM is atomic or ionized. 

This study demonstrates 
the extraordinary power of direct detection experiments in discovering 
and probing a rich dark sector. In the future, a determination of the 
distribution, ionization, and multi-component nature of sub-nano-charged 
matter at direct detection experiments could provide a multi-pronged 
verification of the mirror nature of the MTH model.

\section{Conclusions}\label{s.conclusion}

The MTH framework connects the solution of the Higgs hierarchy problem 
to striking cosmological signatures from the early universe. A crucial 
aspect of this scenario is the likely existence of an asymmetric mirror 
matter component that constitutes a subdominant but dynamically rich 
fraction of DM. In this work we have studied the behavior of these relic 
mirror particles during galaxy formation, and their resulting unique 
multi-component signatures in DM direct detection searches.

In contrast to conventional mirror matter models, the requirement of 
solving the Higgs hierarchy problem and satisfying cosmological 
constraints places limits on the mass and temperature of the mirror 
particles. Although detailed $N$-body simulations incorporating 
magnetohydrodynamic effects would be necessary to obtain a precise time 
evolution of the mirror plasma distribution, we can still use the known 
properties of twin particles to estimate their cooling rates and obtain 
a qualitative understanding the current distribution of this DM 
component. We find that the mirror matter distribution today is right near the 
threshold of being either halo- or disk-like, and we consider the possibility that local mirror baryons are either fully ionized or fully atomic due to unknown mirror astrophysics. 
These distinct possibilities for twin profiles today generate very 
distinct signals in different types of DM detectors, assuming the twin 
photon mixes with the SM photon at the levels expected from 
gravitational effects~\cite{Gherghetta:2019coi}. Measurements at various 
experiments can then establish a unique fingerprint of the twin sector, 
allowing us to probe its multi-component nature, ionization, local distribution, and 
also MTH model parameters such as $\hat v/v$. Our analysis shows that the 
relic MTH mirror particles in the universe can give rise to distinctive 
signatures that are sensitive to the detailed properties of the twin 
sector.

We find that direct detection is especially sensitive in scenarios in 
which the mirror matter remains hot and distributed in the form of a 
halo down to the present day.
This makes direct detection 
complementary 
to signals from white dwarf cooling~\cite{Curtin:2020tkm}
and mirror stars~\cite{Curtin:2019lhm, Curtin:2019ngc, Winch:2020cju, Hippert:2021fch}. 
 These astrophysical searches for mirror baryons are particularly 
sensitive in dark disk scenarios, since this leads to more accumulation 
of dark matter in SM stars and favors the formation of mirror stars. In 
this case, direct detection searches are still important but 
significantly more difficult.
Clearly, the combination of direct detection experiments and astrophysical searches greatly enhances our chances 
of discovering or excluding the asymmetrically reheated MTH and other 
mirror matter scenarios.
If any or several of these signals were observed
and correlated with the expected Higgs decay 
signal $Br(h \to \mathrm{invisible}) \sim (v/\hat v)^2$ at the LHC or a 
future collider,
 the night sky would illuminate a picture of naturalness 
that establishes the existence of the twin universe.

 \section*{Note added}
 While this work was being completed, the Xenon1T experiment reported an 
excess of a few-keV electronic recoil events~\cite{Aprile:2020tmw}, 
which admits a variety of DM interpretations. It is interesting to note 
that mirror electrons within the MTH model might be able to account for 
such an excess~\cite{Zu:2020idx}, due to their higher velocity in the 
mirror plasma relative to mirror nuclei. However, we defer a careful 
study of this possibility for future work.

 \begin{acknowledgments}

 We thank 
Asimina Arvanitaki,
Masha Baryakhtar,
John Dubinski,
Daniel Egana-Ugrinovic,
 Rouven Essig, 
 Akshay Ghalsasi,
 Junwu Huang,
 Rabindra Mohapatra,
 Norman Murray,
 Shmuel Nussinov,
 Jessie Shelton,
 Thomas Quinn,
 Tien-Tien Yu,
 Yiming Zhong,
 and Yue Zhao
for helpful discussions. 
 ZC is supported in part by the US National Science Foundation under 
Grant Number PHY-1914731. The research of DC is supported by a Discovery 
Grant from the Natural Sciences and Engineering Research Council of 
Canada, and by the Canada Research Chair program. In the early stages of this study, the work of DC and MG was also supported by the by the Maryland Center for Fundamental Physics. MG is supported in 
part by the Israel Science Foundation (Grant No. 1302/19). The work of 
YT was supported in part by the National Science Foundation under grant 
PHY-1914731 and by the Maryland Center for Fundamental Physics. YT was also supported in part by the National Science Foundation under grant PHY-2014165. ZC and MG are also supported in part by the US-Israeli BSF grant 2018236. YT thanks the Aspen Center for Physics, which is supported by National 
Science Foundation grant PHY-1607611. 
 \end{acknowledgments}
\appendix 
\section{Capture of Mirror Matter in the Earth}
\label{a.capture}

In this appendix we consider the accumulation of mirror matter in the 
Earth and its effects on direct detection in the scenario discussed in 
this paper. Our analysis shows that for $\epsilon \lesssim 10^{-9}$, the 
accumulation of mirror matter does not have a significant impact on the 
direct detection prospects of either mirror nuclei or mirror electrons. 
The assumptions we make in this analysis are chosen such as to 
overestimate the rate of capture of mirror matter, and consequently its 
effects on direct detection. Our results are therefore somewhat 
conservative.

In what follows we study the accumulation of mirror matter for $\epsilon 
\lesssim 10^{-9}$, $\hat v/v \in (3,5)$ and local mirror helium 
fractions $\hat Y_\odot \in (0,1)$, considering both the disk and halo 
distributions as outlined in \ssref{ionizationfv}. The equilibrium 
accumulated mirror particle number densities are determined by the rates 
of capture and evaporation, which in turn depend on the accumulated 
mirror electric charge of the Earth and the resulting screening by the 
ambient mirror plasma. In our analysis we assume that all mirror matter 
is fully ionized. This assumption is conservative, since capture is 
suppressed for mirror atoms.

Mirror nuclei are captured by scattering off SM nuclei in the Earth. 
Mirror hydrogen evaporates fairly efficiently but mirror helium 
accumulates, resulting in the Earth acquiring a net positive mirror 
electric charge. The resulting repulsive force arrests further capture 
of mirror baryons and gives rise to an equilibrium population of 
captured mirror helium nuclei. Mirror electrons are mainly captured by 
inelastic scattering with bound atomic SM electrons in the Earth. After 
being captured, mirror electrons are very efficiently evaporated by 
scattering off conduction band electrons in the earth's metallic core. 
This results in an equilibrium number of captured mirror electrons that 
is fairly small and quite insensitive to the number density of captured 
mirror helium unless the net positive charge of the Earth from the 
captured mirror nuclei is very large. The total number of accumulated 
mirror particles is always small enough that the capture process is 
dominated by interactions with SM matter in the Earth, rather than 
interactions with mirror particles that have already been captured.

The positive mirror electric charge arising from the captured mirror 
nuclei is screened by the ambient mirror plasma. The characteristic 
length scale for this screening is at most of order $\sim 10^{-1} 
R_{Earth} \sim 500$ km, and mirror particles that are further 
away from the Earth's surface than this do not experience a large 
electric field. Nevertheless, for $\epsilon \gtrsim 10^{-10}$ it is very 
important to take this mirror electric screening effect into account, 
since it greatly modifies the mirror electric potential in the 
neighborhood of the Earth and thereby affects the equilibrium population 
of captured mirror nuclei.

A captured population of mirror nuclei could impact direct detection in two 
distinct ways:
 \begin{enumerate}
 \item
 \emph{Collisional shielding:} collisions of incoming mirror particles with the population of 
accumulated mirror baryons could act as a shield, preventing the 
incoming mirror particles from reaching direct detection experiments. 
This collisional shielding effect was taken into account by the recent 
LUX analysis for $\mathbb{Z}_2$-symmetric mirror DM~\cite{Akerib:2019diq}, based on the analysis 
in~\cite{Foot:2018jpo}. However, in the framework we are considering, we 
find that this effect is negligible for $\epsilon \lesssim 
10^{-9}$.
 \item 
 \emph{Electrostatic shielding:}
 one might expect that the accumulated positive mirror charge due to 
captured $\mathrm{\hat He}$ would electrostatically repulse incoming 
mirror nuclei, suppressing their direct detection signal to negligible 
levels. However, we show that electrostatic effects only matter for 
$\epsilon \gtrsim 10^{-10}$ and also act to suppress the population of 
captured mirror particles, therefore resulting in only a modest 
$\mathcal{O}(1)$ reduction in the signal rate at direct detection 
experiments. Since the proposed experiments we consider probe far 
smaller kinetic mixings than $10^{-10}$ for $r_\odot \sim 0.01$, this 
effect only introduces a $\lsim 50\%$ uncertainty in the projected reach 
for $\epsilon \sqrt{r_\odot}$ if $r_\odot \ll 0.01$.
 \end{enumerate}
 The projections for mirror baryon direct detection that we present in 
this paper are therefore at most modestly affected by capture inside the 
Earth. While these effects are interesting and deserve future study, 
particularly in the context of more general dissipative DM models, we 
are justified in neglecting them in our analysis of direct detection in 
the MTH framework in Section~\ref{s.directdetection}.
 
We now proceed to discuss the capture of mirror matter, mirror plasma 
screening, and mirror matter evaporation in detail. Throughout, we 
denote the free, or ambient, mirror particle densities in the local 
mirror plasma far away from the Earth by $n_i^F$ (completely determined 
in terms of $r_\odot$ and $\hat Y_\odot$ for $i = \mathrm{\hat e}, 
\mathrm{\hat H}, \mathrm{\hat He}$). Number densities of mirror 
particles captured in the Earth are denoted $n_i^C$. We focus our 
discussion on the limit where the mirror baryons are entirely composed 
of twin helium, $\hat Y_\odot = 1$, since their evaporation is less 
effective than for twin hydrogen, and consequently their effect on 
direct detection is larger. Other values of $\hat Y_\odot$ do not 
qualitatively affect our conclusions. The Maxwell-Boltzmann 
distributions of mirror helium and mirror electrons are determined by 
their velocity dispersions, as discussed in \ssref{ionizationfv}. For 
the disk (halo) case, $v_{\odot \hat{H}e} \sim 11$ km/s (120 km/s) and 
$v_{\odot \hat{e}} \sim 500 - 700$ km/s (5000 - 7000 km/s) for $\hat v/v 
\sim 3-5$. We neglect the speed of the Earth relative to the mirror 
plasma in this discussion, since it is not expected to qualitatively 
alter our conclusions. In all cases, our baseline assumption is that 
mirror baryons make up 5\% of the local DM density, i.e. $r_\odot = 
0.05$, but we discuss how our results can be extended to more general 
$r_\odot$ values as well.

\subsection{${\epsilon \lesssim 10^{-11}}$}

We first estimate the effects of capture on direct detection for 
relatively low values of the kinetic mixing, $\epsilon \lesssim 
10^{-11}$, for which the accumulated mirror charge in the Earth is small 
compared to the ambient density of the mirror plasma.

\subsubsection{Capture of Mirror Helium}

We begin by considering the capture of mirror helium. Depending on the 
velocity of the incoming mirror particles, capture may primarily arise 
either through multiple soft scatterings with the material in the Earth, 
or through a single hard scattering process. 

For an incoming nano-charged particle $X$, the rate of kinetic energy 
loss per unit distance traveled inside the Earth due to \emph{multiple 
soft scatterings} is given by,
 \begin{eqnarray}
\nonumber
\frac{dE_k}{dr} &=& -n_{T} \int \frac{d\sigma}{dE_R}E_R dE_R =- n_{T} \int \frac{m_X\pi \alpha^2 \epsilon^2 Q^2_X Z^2 }{m_T E_k E_R}dE_R =-n_{T} \frac{m_X\pi \alpha^2 \epsilon^2 Q^2_XZ^2 }{m_T E_k}\log\frac{E_R^{max}}{E_R^{min}}\,.
\\
\label{e.dEkdr}
 \end{eqnarray}
 Here we assume that the energy loss mainly arises from Rutherford 
scattering with nuclei. The parameters $n_T$, $m_T$, and $Z$ correspond 
to the number density, mass, and charge of the atoms in the Earth 
representing the scattering target. The generalization to a more 
realistic material composition is straightforward. The parameter $E_R$ 
represents the recoil energy and $E^{max}_R$ its maximum value, given by
 \begin{equation}
E^{max}_R = \frac{4 m_T m_X E_k}{(m_T + m_X)^2} \;.
 \end{equation}
  The parameter $E^{min}_{R}$ denotes the infrared scale at which the 
finite size of the atom cuts off the interaction, given roughly by 
$E_R^{min} \sim {(m_e\alpha)^2}/{m_T}$.  Assuming a constant density and 
composition of the Earth and neglecting the energy dependence of the 
logarithm, the energy loss with distance travelled is given by
 \begin{equation}
 \label{e.eloss}
E_k(r) =E_k(0) \sqrt{\frac{r^{Rut}_E-r}{r^{Rut}_E}}\,,
 \end{equation}
 where $r^{Rut}_E$ corresponds to the penetration depth needed to lose 
all of the initial energy,
 \begin{eqnarray}
r^{Rut}_E &\sim& \frac{1}{4} m_X v^2_X  \left(\frac{dE_k}{dr} \right)^{-1} \sim m_X m_T v^4_X\left(8\pi \alpha^2 \epsilon^2 Q^2_X Z^2n_{T}  \log\frac{E_R^{max}}{E_R^{min}}\right)^{-1}, \label{e.pendepth}
\\ &\sim& R_{Earth}\left(\frac{v_X}{20\,{\rm km/s}}\right)^4 
\left(\frac{2 \cdot 10^{23}{\rm cm}^{-3}}{n_T} \right)
\frac{m_T}{16\text{ GeV}} \frac{m_X}{5\text{ GeV}} \left(\frac{8}{Q_T}\right)^2   \left(\frac{2}{Q_X}\right)^2    \left(\frac{10^{-10}}{\epsilon}\right)^2\,  \nonumber
 \end{eqnarray}
 Here $v_X$ is the mirror particle velocity when it enters the Earth, 
and we use the representative values of the charge $Q_X=2$ and mass $m_X 
\sim 5 \gev$ of mirror helium as an example. The most effective target 
inside the Earth is oxygen, since it is fairly abundant and light. We have checked that a more 
careful integration of \eref{dEkdr} yields a comparable result.

 \begin{figure}
\center{\includegraphics[width=7.5cm]{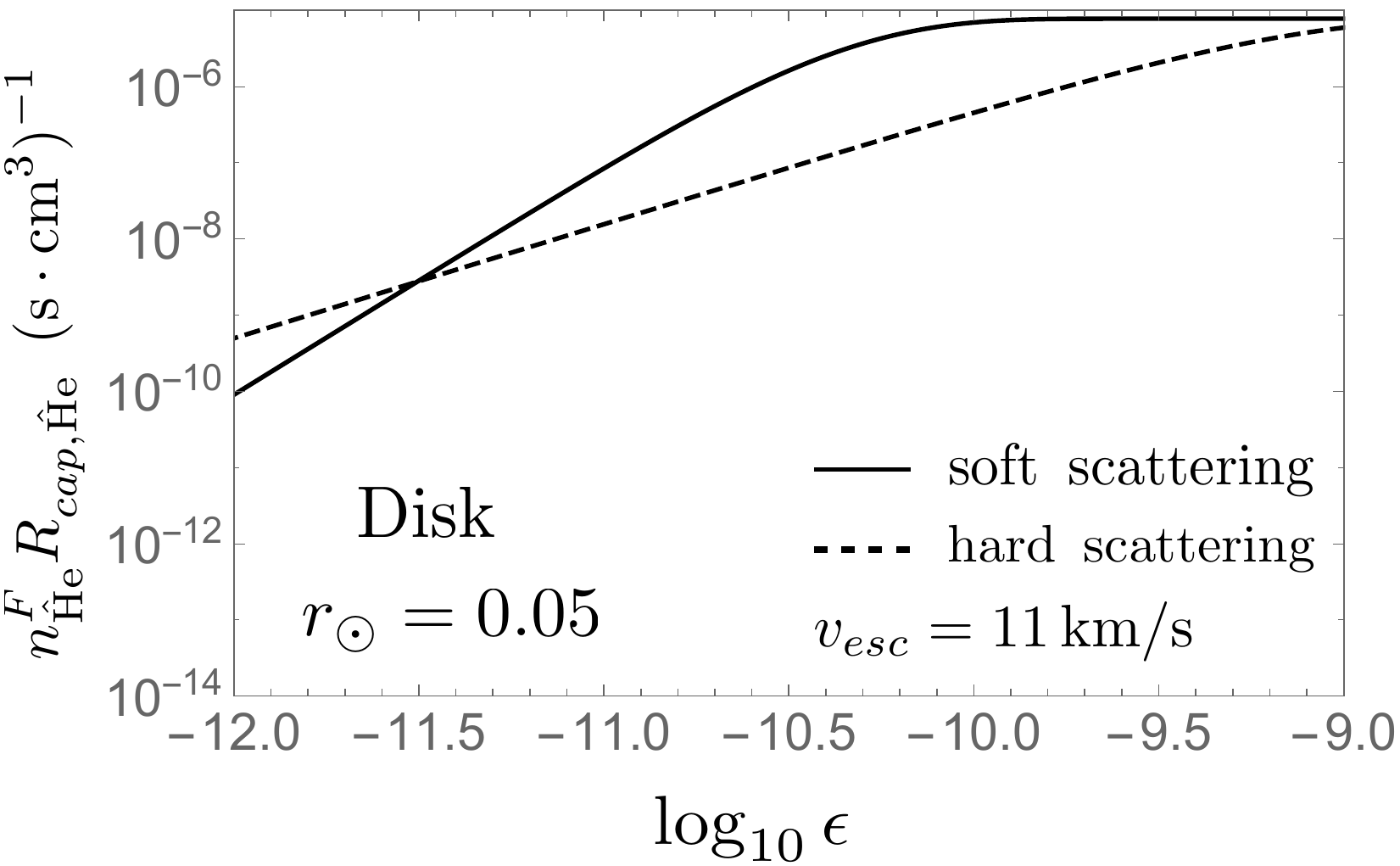}\quad\includegraphics[width=7.5cm]{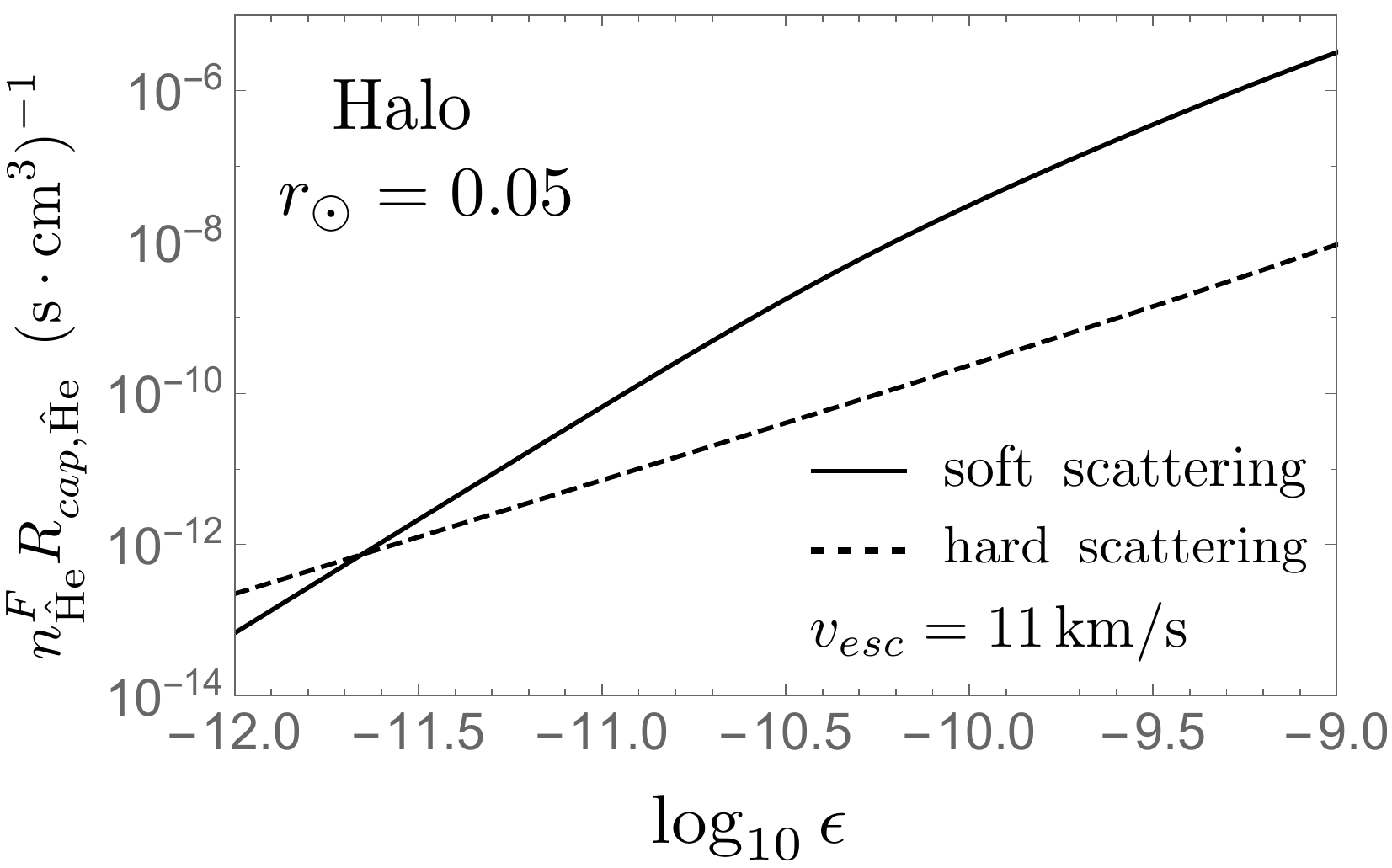}
}
 \caption{Mirror helium capture rates from multiple soft scattering 
(solid) and single hard scattering (dashed) processes as a function of the 
kinetic mixing parameter $\epsilon$ for $r_{\odot} = 0.05$. This plot 
assumes that the net mirror charge of the Earth is small, so that its 
effect on the escape velocities of mirror particles is negligible.
}
\label{fig.capturerate}
 \end{figure}

We regard a mirror nucleus as captured if it enters the Earth and loses 
enough energy such that it either gets stuck inside the Earth, or 
emerges from the Earth with velocity less than than the escape velocity, 
$v_{esc,i}$. As we shall see, the Earth eventually acquires a net 
positive mirror electric charge from the capture of ${\rm\hat{H}e}$, and 
the escape velocities of mirror particles depend on the net charge. In 
the absence of any net captured charge the escape velocity is the same 
for all species and is given by $v_{esc} \approx$ 11 km/s. If enough 
positive mirror charge is accumulated to overcome the Earth's 
gravitational attraction, then $v_{esc,{\rm\hat{H}e}}^2 < 0$. In this 
case, getting stuck inside the Earth is necessary for capture. To 
quantify this, we define $v_{\rm cap}$ as the maximum velocity an 
infalling mirror helium nucleus can have \emph{at the time it enters the 
Earth} if it is to be captured by soft scatterings.
Substituting $E_k(0) = \frac{1}{2} m_X v_{cap}^2, E_k(R_{Earth}) = \mathrm{max}[0, \frac{1}{2} m_X v_{esc, \mathrm{\hat He}}^2]$
into \eref{eloss} yields
 \begin{equation}
 \label{e.rRutEandvcap}
r^{Rut}_E(v_{\rm cap}) =  \left\{
\begin{array}{lll}
\left(1- \frac{v_{esc,{\rm\hat{H}e}}^4}{v_{\rm cap}^4}\right)^{-1}R_{Earth}
& \mathrm{for} & v_{esc,{\rm\hat{H}e}}^2 > 0\\
R_{Earth} & \mathrm{for} & v_{esc,{\rm\hat{H}e}}^2 < 0
\end{array}
\right.
 \end{equation}
 For $\epsilon \approx 10^{-10}$, in the absence of any accumulated 
mirror charge on the Earth, $v_{\rm cap}^2 \approx (14\,{\rm km/s})^2$, 
while for $\epsilon \approx 10^{-11}$, $v_{cap}$ is very close to 
$v_{esc}$, meaning that only nuclei much slower than 11 km/s far away 
from the Earth get captured.

As an $\hat{{\rm H}}{\rm e}^{++}$ nucleus falls towards the Earth, it gains 
kinetic energy $\frac{1}{2} m_{\hat{{\rm H}}{\rm e}} 
v_{esc,{\rm\hat{H}e}}^2$, where $v_{esc,{\rm\hat{H}e}} \approx 11$ km/s 
in the absence of any accumulated mirror charge. Therefore, if far away 
from the Earth the $\hat{{\rm H}}{\rm e}^{++}$ initially has velocity $v 
< \sqrt{v_{\rm cap}^2 - v_{esc}^2}$, it will become bound to the Earth. 
The \emph{capture rate of mirror helium} from soft scatterings can 
therefore be estimated as $\sim n_{{\rm\hat{H}e}}^F \pi 
R_{Earth}^{2}\langle v_{\rm \hat{H}e}\rangle\,$, leading to a capture 
rate per unit volume in the Earth of
 \beq
n_{{\rm\hat{H}e}}^FR^{<v_{\rm  cap}}_{cap,{\rm \hat{H}e}}\sim n_{{\rm\hat{H}e}}^F R_{Earth}^{-1}\langle v_{\rm \hat{H}e}\rangle\,.
 \eeq
 Here $\langle v_{\rm \hat{H}e}\rangle$ represents the average speed, 
far away from the Earth, of incoming mirror particles that can be 
captured
 \begin{equation}
 \label{eq:vHe}
\langle v_{{\rm \hat{H}e}}\rangle =  \int_{v_{min}}^{\sqrt{v^2_{\rm cap}-v_{esc,{\rm\hat{H}e}}^2}}dv \,v f_{\rm \hat{H}e}(v) \,. 
 \end{equation}   
 Here $f_{\rm \hat{H}e}(v)$ is the local velocity distribution of mirror 
helium in the Earth frame. The lower limit of integration $v_{min} = 0$ 
as long as $v_{esc,{\rm\hat{H}e}}^2 > 0$. For $v_{esc,{\rm\hat{H}e}}^2 < 0$, only particles above a 
certain initial speed can even reach the Earth surface, and so $v_{min} 
= \sqrt{-v_{esc,{\rm\hat{H}e}}^2}$.

While capture of slow mirror helium nuclei proceeds via multiple soft 
scatterings, capture of $\hat{{\rm H}}{\rm e}^{++}$ that enter the Earth 
with speed $v>v_{\rm cap}$ can still proceed through hard scatterings. 
In this case we estimate the capture rate by determining the probability 
of having a single scattering process that takes away a significant 
fraction of the energy of the mirror helium nucleus.

In our estimate of the capture rate from hard scattering, we assume that 
the energy transfer between the incoming $\mathrm{\hat He}$ and the SM 
nucleus in the Earth is always maximal. This is a conservative 
assumption, since it overestimates the true capture rate and the 
resulting suppression in the direct detection signal. For a fixed mass 
$m_T$ of the target nucleus, kinematics places an upper limit on the 
velocities of mirror helium nuclei that can be captured through a single 
scattering. In particular, only mirror helium particles with velocity $v 
< v_{max} = v_{esc,{\rm\hat{H}e}} (m_\mathrm{\hat He} + m_T) / 
|m_\mathrm{\hat He} - m_T|$ when they enter the Earth have a chance of 
getting captured. Here $v_{esc,{\rm\hat{H}e}}$ is the escape velocity at 
the surface of the Earth. It takes value 11 km/s in the absence of any 
accumulated mirror charge but is reduced if the Earth has a net charge.

We find that $\hat{{\rm H}}{\rm e}^{++}$ capture via single hard 
scatterings is also dominated by oxygen.  We compare the scattering 
length $\ell\sim (\sigma_{XT}n_T)^{-1}$ to the Earth's radius, since we 
expect to have only one chance to scatter and capture the particle. The 
mirror capture rate per unit volume arising from single hard scatterings 
can then be estimated as
 \beq
n_{{\rm\hat{H}e}}^F R^{>v_{\rm cap}}_{cap,{\rm \hat He}}\sim n_{{\rm\hat{H}e}}^F \langle\sigma_{{\rm \hat{H}e\,O}}\,v_{{\rm \hat{H}e}}\rangle\,n_{\rm O}\,,
 \eeq
 where
 \beq\label{eq:sigmav}
\langle\sigma_{{\rm \hat{H}e\,O}}\,v_{{\rm \hat{H}e}}\rangle = 
\int_{\sqrt{v^2_{\rm cap}-v^2_{esc,{\rm\hat{H}e}}}}^{\sqrt{v_{max}^2 - v_{esc,{\rm\hat{H}e}}^2}}dv\,\sigma_{{\rm \hat{H}e\,O}}v f_{\rm \hat{H}e}(v) 
\ \ \ \ , 
 \ \ \ \ \  \sigma_{{\rm \hat{H}e\,O}} \approx \frac{4\pi\alpha^2\epsilon^2}{m_{{\rm \hat{H}e}}^2(v^2+v_{esc,{\rm\hat{H}e}}^2)^2}\,.
 \eeq
 The capture rates of mirror helium arising from soft and hard 
scattering as a function of $\epsilon$ are shown in 
Fig.~\ref{fig.capturerate} for both halo and disk distributions. We see 
that capture by soft scattering dominates for $\epsilon \gtrsim 
10^{-12}$.

In most of our estimates, we neglect \emph{self-capture} of incoming 
mirror helium nuclei by the captured $\mathrm{\hat He}$ population 
already inside the Earth. It is important to understand when this is a 
valid approximation. Self-capture is dominated by multiple soft 
scatterings. We can obtain an estimate for the penetration depth of an 
incoming mirror helium nucleus due to interactions with captured 
$\mathrm{\hat He}$ by substituting $n_T \to N_{\mathrm{\hat He}} / 
(\frac{4}{3} \pi R_{Earth}^3), m_{T,X} \to m_{\mathrm{\hat He}}, Q_{T,X} 
\to 2, Z\to 2, \epsilon \to 1$ into \eref{pendepth},
 \begin{equation}
L_{\mathrm{\hat He} - \mathrm{\hat He}} \sim R_{Earth}  \left(\frac{2\cdot10^{31}}{N_{\mathrm{\hat He}}}\right) \left( \frac{v_X}{20 \mathrm{km/s}} \right)^4 \ .
 \end{equation}
 In order for our estimates neglecting self capture to be trustworthy, 
capture must be dominated by scattering off SM particles in the Earth,
 \begin{equation}
L_{\mathrm{\hat He} - \mathrm{\hat He}}  \gg r_E^{Rut} \ .
\end{equation}
 This corresponds to the condition,
 \begin{equation}
\label{e.ignoreselfcapture}
N_{\mathrm{\hat He}} \ll 10^{31} \left( \frac{\epsilon}{10^{-10}} \right)^2 \;,
 \end{equation}
 which specifies the regime of validity of our estimates.
 It is also interesting to consider the regime of runaway self-capture, 
where $L_{\mathrm{\hat He} - \mathrm{\hat He}} \sim R_{Earth}$ and most 
incoming mirror helium nuclei get captured. Taking $v_X \sim v_{\odot, 
\mathrm{\hat He}}$ we see that we enter this regime if $N_{\mathrm{\hat 
He}} \gtrsim 10^{30} (10^{34})$ for the disk (halo) distributions.

\subsubsection{Capture and Evaporation of Mirror Electrons}\label{sec.electron}

In this subsection we discuss the capture and evaporation of mirror 
electrons in the halo and disk scenarios. Our analysis shows that the 
capture of mirror electrons has a negligible effect on direct detection 
for small values of the kinetic mixing, $\epsilon \lesssim 10^{-11}$.

In the halo distribution, mirror electrons have an average velocity 
$v_{\odot,\hat e}\sim 7\times 10^3$ km/s, corresponding to a kinetic energy 
of $\sim 400$ eV. This is comparable to the binding energy of inner 
shell electrons in atoms. Therefore capture primarily arises from the 
scattering of mirror electrons with inner shell electrons. This process 
is inelastic since the atom is either left in an excited state or the 
electron is simply ejected from the atom leaving behind an ion, and may 
be accompanied by the emission of additional (mirror) photons. Although 
this process can result in the mirror electron losing enough energy to 
be captured, the phase space for capture is very limited since the 
velocity of the incoming mirror electron is orders of magnitude larger 
than the escape velocity from the Earth.

Since iron is the most abundant element in the Earth, for concreteness 
we will focus on scattering from iron atoms in the Earth's core. A 
process that can result in the loss of the required amount of energy 
involves $\hat e$ kicking out electrons from the $2p$ state of an iron 
atom ($\Delta E\approx 700$ eV). We can therefore estimate the 
scattering length by assuming each iron atom has $6$ useful electrons 
for the capture process. A detailed calculation of the capture that 
takes into account the details of the atomic structure is beyond the 
scope of this work. We will instead place an upper bound on the number 
of captured mirror electrons, where the limit is obtained by assuming 
that $\hat e$ is \emph{always} captured in scatterings with $2p$ 
electrons regardless of the actual momentum transfer. We will later show that 
even if the bound is saturated the number of mirror electrons in the 
Earth is still too small to affect the direct detection signals.
From the density of iron inside the core $\approx 13\,\,{\rm 
g\,cm^{-3}}$, we can place a lower bound on the capture length as,
 \begin{equation}
 \label{e.lcaphalo}
\ell^{cap}_{\hat e,{\rm Halo}} \gtrsim 
(\sigma_{\hat e e}6n_{\rm Fe})^{-1} \sim 10^8{\rm km}\left(\frac{\epsilon}{10^{-9}}\right)^{-2}\,.
 \end{equation}

We now turn our attention to the disk distribution. Here mirror 
electrons have velocities of order $v_{0,\hat e}\sim 6\times 10^2$ km/s, 
corresponding to kinetic energies of order $3$ eV. This energy is close 
to the thickness of the valence band of iron below the Fermi surface. 
Therefore, an efficient scattering process involves $\hat e$ kicking out 
an electron from this band. Nevertheless the phase space for capture is 
limited, since the velocity of the incoming mirror electrons is still 
much more than the escape velocity from the Earth. Again, a calculation 
of the capture that takes into account the precise dispersion relation 
of the electrons inside the metal is beyond the scope of this work. We 
instead estimate the scattering length by rescaling the mean free path 
of the electrons in iron at room temperature~\cite{Ashcroft}, for which 
the associated energy transfer between electrons is not far from the eV 
scale. We again limit ourselves to placing an upper bound on the number 
density of captured mirror electrons, where the bound is reached if 
$\hat e$ is \emph{always} captured in scatterings with conduction 
electrons regardless of the actual momentum transfer. We will later show 
that even if the bound is saturated, the number of captured mirror 
electrons is too small to affect the signal. The lower bound on the 
capture length of disk mirror electrons is given by\footnote{
Since the iron in the earth's core is at a higher temperature and 
pressure, the mean free path is expected to be somewhat shorter in the core, 
leading to a smaller capture length. However, as we show below, a 
shorter mean free path also corresponds to more efficient evaporation of 
mirror electrons. Therefore, as long as the mean free paths for capture 
and evaporation are much larger than the Earth's radius, the shorter 
$\ell^{cap}_{\hat e,{\rm Disk}}$ will not significantly affect the 
equilibrium $\hat e$ abundance in the disk case and will only decrease 
the abundance in the halo case.}
 \begin{equation}\label{e.elle}
 \ell^{cap}_{\hat e,{\rm Disk}}
 \gtrsim 
 10^{7}\,{\rm km}\left(\frac{\epsilon}{10^{-9}}\right)^{-2}\,.
 \end{equation}

From the lower bound on the capture length, we can obtain an upper bound 
on the the capture rate of mirror electrons per unit volume in the 
Earth in the halo and disk cases,
 \begin{equation}
n_{\hat e}^F R_{cap,\hat e}\lsim n_{\hat e}^F R_{core}^{-1} v_{\odot,\hat e}\left(\frac{R_{core}}{\ell^{cap}_{\hat e}}\right)
=
n_{\hat e}^F \left(\frac{ v_{\odot,\hat{e}}}{\ell^{cap}_{\hat e}}\right) \ \ .
 \end{equation} 
  After capture, the $\hat e$ has a velocity $v \lsim v_{esc,\hat{e}}$, 
where $v_{esc,\hat{e}}$ is the escape velocity at the Earth's core. We 
expect that the mirror electrons will tend to thermalize with the matter 
in the Earth's core. If the net accumulated mirror charge of the Earth 
is small, so that the thermal velocity of the mirror electrons is larger 
than their escape velocity, any captured $\hat{e}$ are efficiently 
evaporated from the Earth. It is only if the escape velocity of the 
mirror electrons is greater than or comparable to their thermal velocity 
that there is any significant accumulation of $\hat{e}$.

 An efficient evaporation process for the $\hat e$ involves scattering 
off conduction band electrons in metals. We will once again focus on 
iron. When scattering off mirror electrons, electrons in the conduction 
band can impart energy that is comparable to the width of the valence 
band ($\sim 0.6$ eV) that corresponds to the iron temperature inside the 
Earth's core ($T_{\rm Earth}\sim 4000-6000\,$K). This means that a 
captured mirror electron will be promptly expelled by collision with a 
conduction electron unless the net positive mirror charge of the Earth 
is so large that the escape velocity of $\hat{e}$ at the Earth's core is 
greater than the thermal velocity $v_{th,\hat e}$, which is of order 
$260$ km/s. We can estimate the mean free path for $\hat e$ evaporation 
by rescaling the mean free path in Eq.~(\ref{e.elle}) after accounting 
for the change of the $\hat{e}$-$e$ scattering cross section with the 
energies of the incoming and outgoing particles. The Fermi velocity of 
electrons in iron, $v_{F} \approx 2\times 10^3~$ km/s in iron, is much 
greater than $v_{th,\hat e}$. Then the scattering cross section for a 
fast moving electron with velocity $v_F$ to evaporate a captured mirror 
electron of velocity $v_{th,\hat e}$ by injecting a recoil energy $E_R 
\gsim \frac{1}{2} m_{\hat e} v_{esc,\hat e}^2$ can be estimated as
 \begin{equation}
 \sigma_{evp} \approx 
 \frac{4\pi\alpha^2\epsilon^2}{m_{\hat e}^2 v_{F}^2 v_{esc,\hat e}^2} 
 \propto 
 (m_{\hat e}^2 v_{F}^2 v_{esc,\hat e}^2)^{-1}\,.
 \end{equation}
 From this cross section, for $v_{esc,\hat e} \lsim v_{th,\hat e}$, we 
obtain an upper bound for the mean free path for evaporation,
 \begin{equation}
\ell^{eva}_{\hat e,{\rm Disk\,\&\, Halo}}\lsim 10^{-1}\ell^{cap}_{\hat e,{\rm Disk}}\,.
 \end{equation}
 In the absence of significant charge accumulation, the evaporation rate 
of mirror electrons per unit volume is given by
 \begin{equation}
 \label{e.NecRevp}
n_{\hat e}^C R_{evp,\hat e}\sim n_{\hat e}^C
\left(\frac{v_{F}}{\ell^{eva}_{\hat e}}\right) \; .
 \end{equation}
 When the system is in equilibrium $n_{\hat e}^F R_{cap,\hat e} = 
n_{\hat e}^C R_{evp,\hat e}~$. From the upper bound on the capture rate 
we can obtain an upper bound on the total number of captured mirror 
electrons,
 \begin{eqnarray}\label{eq.Ne_th}
N_{\hat e}\lsim n^F_{\hat e}\left(\frac{v_{\odot, \hat e}}{v_{F}}\right)\left(\frac{\ell^{eva}_{\hat e}}{\ell^{cap}_{\hat e}}\right)R_{core}^3\sim 10^{23}\,\,({\rm Disk\,,\,\,Halo})\,.
 \end{eqnarray}
This bound is valid for $v_{esc,\hat{e}} \lsim v_{th, \hat{e}}$. However, since in this regime the velocity of incoming mirror electrons is still much greater than the escape velocity, $v_{\odot, \hat{e}} \gg v_{esc,\hat{e}}$, the phase space for capture is very limited and so the actual number of captured electrons is expected to be several orders of magnitude less than this upper bound.

If the Earth acquires a large net positive charge the evaporation rate 
is exponentially suppressed because of the much larger velocity needed 
to escape the Earth. Under the assumption that the mirror electrons 
thermalize with the matter in the Earth's core, Eqn.~(\ref{e.NecRevp}) 
generalizes to
 \begin{equation}
n_{\hat e}^C R_{evp,\hat e}\sim n_{\hat e}^C 
\left(\frac{v_{F}}{\ell^{eva}_{\hat e}}\right) \exp\left(-\frac{m_{\hat{e}} v^2_{esc,\hat e}}{2\,T_{\rm Earth}}\right)\,.
 \end{equation}
In equilibrium we get the upper bound,
 \begin{eqnarray}\label{eq.Ne}
N_{\hat e}&\lsim& n_{\hat e}^F\left(\frac{v_{\odot, \hat e}}{v_{F}}\right)\left(\frac{\ell^{eva}_{\hat e}}{\ell^{cap}_{\hat e}}\right)\exp\left(\frac{m_{\hat{e}} v^2_{esc,\hat e}}{2\,T_{\rm Earth}}\right)R_{core}^3\,, \nonumber
\\
&\sim& \exp\left(\frac{m_{\hat{e}} v^2_{esc,\hat e}}{2\,T_{\rm Earth}}\right)\times10^{23}\left({\rm Disk},\,\,{\rm Halo} \right),
 \end{eqnarray}
 for the disk and halo cases. This bound is only saturated when the captured charge is large enough that the escape velocity is greater than or of order the velocity of incoming mirror electrons,  $v_{esc,\hat{e}} \gtrsim v_{\odot, \hat{e}}$. 
 However, for $\epsilon \lesssim 10^{-11}$, we shall see that the number of captured mirror nuclei is too small to affect the evaporation of mirror electrons. 

Just as for mirror helium capture, we now determine the condition for 
mirror electron self-capture to play a significant role. Assuming that 
the captured mirror electrons in the Earth are roughly uniformly 
distributed, the mean free path for scattering of incoming mirror 
electrons with the captured population can be estimated as,
 \begin{equation}
L_{\hat e - \hat e}
\sim \left(\frac{4 \pi \alpha^2}{m_{\hat e}^2v^4_{\odot, \hat e}}n^C_{\hat e}\right)^{-1}
\sim
\left( 
\frac{10^{23}}{N_{\hat e}}
\right)
\times
\left\{ \begin{array}{ll}
10^{13}\  \mathrm{km} & \mathrm{(disk)}\\
10^{17}\ \mathrm{km}& \mathrm{(halo)}
\end{array} \right.
 \end{equation}
 Requiring self capture to be negligible compared to scattering off SM 
electrons in the Earth corresponds to the conditions $L_{\hat e - \hat e} \gg 
\ell_{\hat e, \mathrm{Halo}}^{cap}, \ell_{\hat e, \mathrm{Disk}}^{cap}$. 
Using Eqns.~(\ref{e.lcaphalo}) and (\ref{e.elle}), this translates to an upper
bound on the number of captured mirror electrons,
 \begin{equation}
N_{\hat e} \ll 
\left( \frac{\epsilon}{10^{-9}} \right)^2
\left\{ \begin{array}{ll}
\label{e.ignoreelectronselfcapture}
10^{29} & \mathrm{(disk)}\\
10^{32} & \mathrm{(halo)}
\end{array} \right. \ .
 \end{equation}
 We will later see that self capture is never important for kinetic mixings $10^{-12} \lesssim \epsilon \lesssim 10^{-9}$ where capture is non-negligible. (Capture is negligible for much smaller kinetic mixings.) We are even further 
away from the runaway self-capture regime, corresponding to $L_{\hat e - 
\hat e} \sim R_{Earth}$, which is only reached for $N_{\hat e} \gtrsim 
10^{32} (10^{36})$ assuming a disk (halo) distribution.

\subsubsection{Screening and the Escape Velocity}\label{s.debye}

The Earth will eventually accumulate more mirror helium nuclei than 
mirror electrons and acquire a net positive charge. This net charge is 
screened by the ambient mirror plasma within a short distance of the 
Earth's surface. We discuss this phenomenon here, since it can have a 
significant impact on the capture and evaporation processes. To simplify 
the discussion, we neglect the effects of gravity on the mirror electric 
potential. We have verified that including these effects does not alter 
our conclusions.

The mirror electric potential $\phi(r)$ close to the Earth follows the 
Poisson equation,
 \begin{equation}
\nabla^2 \phi(r) =   - \sum_i q_i n^C_{i}(r) - \sum_i q_i n^{S}_i(r) \,, \label{e.electric_potential}
 \end{equation}
 where $r$ is the distance from the Earth center, $n^C_{i}$ is number density of the captured mirror particle species (dominated by $i = \mathrm{\hat He}$), and $n^{S}_i$ is 
the free number density of the mirror particle species $i$ with mirror 
charge $q_i$. This ambient but uncaptured density near the Earth reacts 
to the accumulated charge to screen it, and asymptotes to the 
unperturbed free mirror number densities $n_i^F$ far away from the 
Earth. Solving \eref{electric_potential} in generality is quite 
complicated, since the free charge and captured charge distribution (which adjusts itself through diffusion in the Earth, 
Section~\ref{sec.helium.evaporation}) are themselves functions of the 
potential. Here we will derive a simplified solution that is 
conservative in the sense that it underestimates mirror helium 
evaporation and therefore overestimates the equilibrium accumulated 
$\mathrm{\hat He}$-abundance and its effects on direct detection. We 
start by expressing $n_i^S$ as a function of $\phi$, since the free 
mirror-charge distributions near the Earth adjust quickly to the 
existence of any mirror-electric potential. This allows us to solve for 
$\phi$ within our pessimistic but simplified assumptions.

 It is convenient to parametrize the potential $\phi$ in terms of its 
contribution to the escape velocity of a particle species $i$ at a 
radius $r$ from the center of the Earth,
 \begin{equation}
v_{esc,i}^2(r) = - \frac{2 (e q_i \phi(r)+\phi_{gr}(r))}{m_i} \; .
 \end{equation}
 As noted above we will ignore gravity in this discussion, i.e.  
$\phi_{gr} = 0$. Recall that $v^2_{esc,i}<0$ signifies a repulsive 
rather than attractive force on the particle. To determine $n_i^S(r)$, 
we consider particle trajectories that approach the Earth from far away. 
By calculating the fraction of this incoming flux that reaches a 
distance $r$ from the center of the Earth, we can obtain an expression 
for $n_i^S(r)$.

 \begin{figure}
\center{\includegraphics[width=6.5cm]{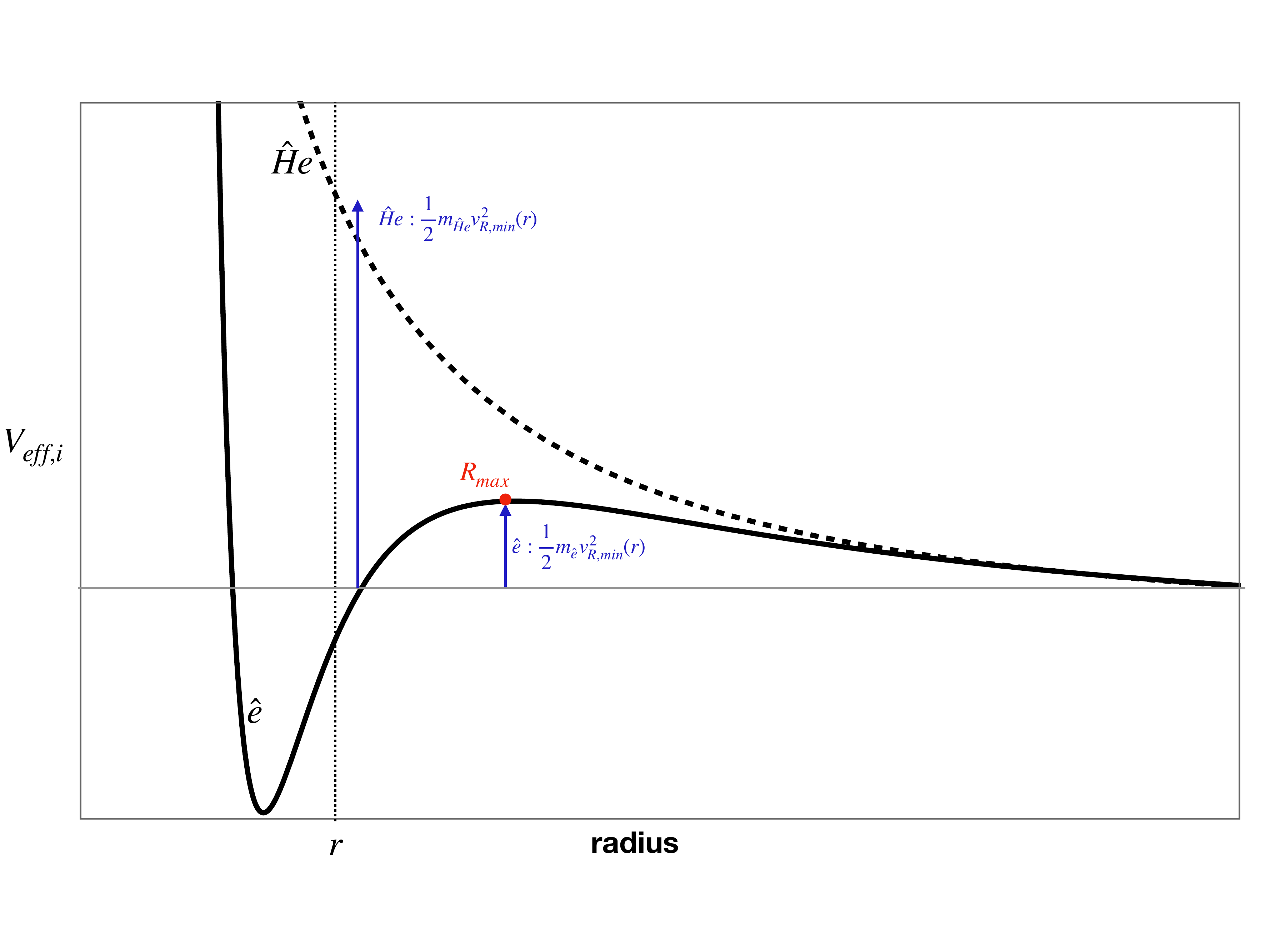}
\includegraphics[width=6.5cm]{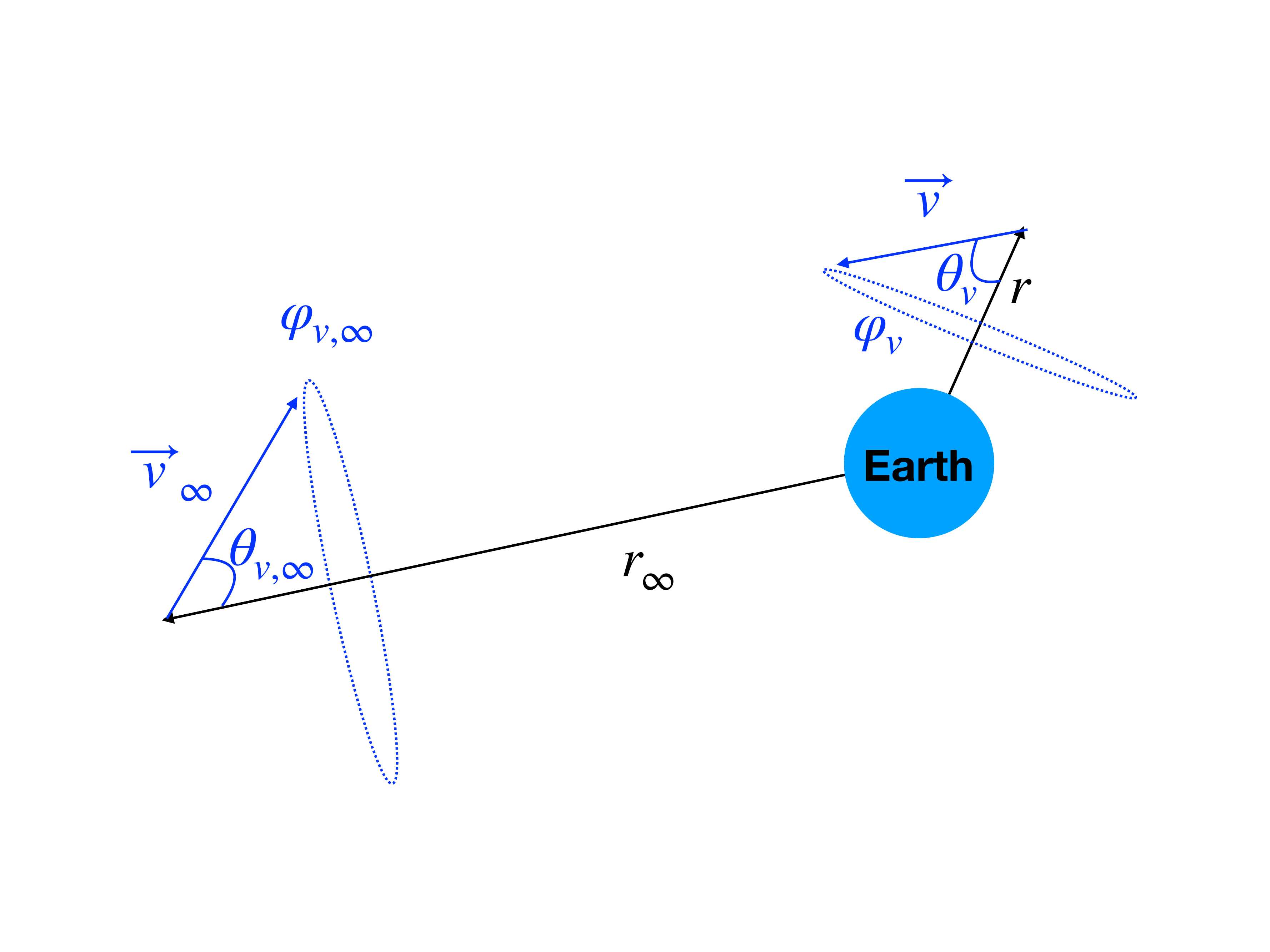}
}
 \caption{On the left is an illustration of the one dimensional 
effective potentials for positively and negatively charged mirror 
particles induced by a net captured positive charge inside the Earth 
(shown for ${\rm \hat{H}e}^{++}$ as a dashed line and for $\hat{\rm 
e}^-$ as a solid line). The one dimensional effective potential for 
negatively charged particles exhibits a local maximum where the 
repulsive force from angular momentum exactly balances the mirror 
electrostatic attraction. In this case for an incoming particle to reach 
any point deeper than $R_{max}$, the location of the maximum of the 
barrier, the radial kinetic energy at infinity must be greater than the 
height of the barrier at $R_{max}$. For the case of positively charged 
particles no such barrier exists. Then, to reach any point $r$, it 
suffices for the radial kinetic energy of the incoming particle at 
infinity to be greater than the effective potential at $r$. On the right 
we illustrate our choice of coordinates for the velocity distribution.}
 \label{fig.VeffRmax}
\end{figure}

 We first consider all the trajectories that stretch from some large but 
finite $r_{\infty}$ to $r$, where $r_{\infty}$ will be taken to infinity 
at the end. We denote the speeds of the particles at $r_{\infty}$ and 
$r$ by $v_{\infty}$ and $v$ respectively. We employ the coordinate 
system shown in Fig.~\ref{fig.VeffRmax}, in which the expressions for 
the radial and tangential velocities at $r_{\infty}$ and $r$ take the 
form,
 \begin{eqnarray}
v_{r,\infty} &=& v_{\infty} \cos \theta_{v,\infty}~,~v_r = v \cos \theta_v\nonumber \\
v_{\rm tan,\infty} &=&\sqrt{v^2_{\infty}- v^2_{r,\infty} }~,~v_{\rm tan} = \sqrt{v^2- v^2_r } 
  \end{eqnarray}
 For each species $i\in (\mathrm{\hat He},\mathrm{\hat H},\hat e)$, 
these quantities are related by energy and angular momentum 
conservation,
 \begin{equation}
\label{eq:energy_angular_momentum}
v^2 - v_{esc,i}^2 = v_\infty^2  ~,~ r^2\left(v^2-v^2_r\right) =r_\infty^2 \left(v^2_{\infty}- v^2_{r,\infty}\right)
 \end{equation}
 where we have assumed that the escape velocity at $r_\infty$ is 
negligible. 

The total radial flux of particles that enter a sphere of radius $r$ 
centered at the Earth with speeds between $v$ and $v+dv$ and radial 
velocities between $v_r$ and $v_r+dv_r$ is equal to the radial flux of 
particles which enter a sphere of radius $r_\infty$ with speeds between 
$v_\infty$ and $v_{\infty}+dv_{\infty}$ and radial velocities between 
$v_{r,\infty}$ and $v_{r,\infty}+dv_{r,\infty}$, provided that the 
velocities satisfy the relations in 
Eq.~(\ref{eq:energy_angular_momentum}). (This assumes that there is no 
potential barrier in between $r_\infty$ and $r$ that affects the flow of 
particles between these two points.) We take $f_i(\vec{r},\vec{v})$ to 
be the number of mirror particles of species $i$ that have positions 
lying within a volume element $d^3 \vec{r}$ about $\vec{r}$ and 
velocities lying within a velocity space element $d^3 \vec{v}$ about 
$\vec{v}$. The spherical symmetry of the problem implies that 
$f_i(\vec{r},\vec{v})$ depends only on the radial coordinate $r$, the 
projection of the velocity in the radial direction $v_r$ and the speed 
$v$, $f_i(\vec{r},\vec{v}) = f_i(r,v,v_r)$. This allows us to write
 \begin{equation}
\label{eq:flux_equality}
v_r \frac{dv_r}{v} v^2 dv  \int f_i(r,v,v_r) r^2 d\Omega d\phi_v = v_{r,\infty} \frac{dv_{r,\infty}}{v_{\infty}} v_{\infty}^2 dv_{\infty}  \int f_i(r_\infty,v_\infty,v_{r,\infty} ) r_{\infty}^2 d\Omega_{\infty} d\phi_{v,\infty}  \; .
 \end{equation}
 If we neglect capture inside the Earth, an identical relation holds for 
particles flowing outward from $r$ ro $r_{\infty}$. Integrating over all 
$v_r$ from $v_r =-v$ to $v_r = v$, and $\phi_v$ from 0 to $2 \pi$ we can 
find the speed distribution of incoming particles $f_i(\vec{r},v)$ about 
the point $\vec{r}$,
 \begin{equation}
f_i(\vec{r},v) =  
\int f_i(\vec{r},\vec{v}) v_r \frac{dv_r}{v} v^2  d\phi_v \; .
 \end{equation} 
 The speed distribution is related to the number density as $\int dv 
f_i(\vec{r},v) = n_i(\vec{r})$. Because of the radial symmetry of the 
problem, $f_i(\vec{r},v) = f_i(r,v)$. Performing the integration we 
obtain for the speed distribution at radius $r$,
 \begin{eqnarray}
 \label{eq:speed_distribution}
f_i(r,v) &=&  \frac{1}{4\pi}  \int f_i(r,v,v_r) v^2 d\Omega d\phi_v \frac{dv_r}{v} =\nonumber \\
&=& \frac{1}{4\pi}  \int  f_i(r_\infty,v_{\infty},v_{r,\infty}) 
\frac{r_{\infty}^2}{r^2} \frac{v_{r,\infty}}{v_r} v_{\infty}^2 
\gamma\left(\vec{r}_\infty,r,\vec{v}_{\infty}\right)  d\Omega_{\infty} d\phi_{v,\infty}\frac{dv_{\infty}}{dv} \frac{dv_{r,\infty}}{v_{\infty}} \; .
 \end{eqnarray}
 where we have used Eq.~(\ref{eq:flux_equality}) in the last line. Here 
$\gamma\left(\vec{r}_\infty,r,\vec{v}_{\infty}\right) $ is defined to be 
equal to one if a trajectory starting at $\vec r_{\infty}$ with velocity 
$\vec{v}_{\infty}$ reaches radius $r$ and zero if that is not the case. 
This factor is included to account for the possibility that there is a 
potential barrier somewhere between $r$ and $r_\infty$ that prevents the 
flow of particles between them if their energy is too low.  Spherical 
symmetry ensures that
 \begin{equation}
\gamma\left(\vec{r}_\infty,r,\vec{v}_{\infty}\right) = \gamma\left(v_\infty, v_{r,\infty},r_\infty ,r \right).  
 \end{equation}
 This allows us to rewrite Eq.~(\ref{eq:speed_distribution}) as,
 \begin{equation}
\label{eq:speed_distribution_derived}
f_i(r,v) = \frac{r_{\infty}^2}{r^2} \frac{dv_{\infty}}{dv} f_i(r_\infty,v_\infty) \int \frac{v_{r,\infty}}{v_r}   \gamma\left(v_\infty, v_{r,\infty},r_\infty ,r \right) \frac{dv_{r,\infty}}{v_{\infty}} \;, 
 \end{equation}
 where we have assumed that that $r_\infty$ is large enough that the 
velocity distribution there is approximately isotropic.

 To determine $ \gamma\left(v_\infty, v_{r,\infty},r_\infty ,r \right)$ 
we consider the equivalent one dimensional problem with the effective 
potential
 \begin{equation}
V_{eff,i}(r) = -\frac{1}{2} m_i v_{esc,i}^2(r) + \frac{1}{2} \frac{L^2}{m_i r^2} \; ,
 \end{equation}
 where $L$ is the angular momentum. For mirror nuclei the force is 
always repulsive and so the condition that a particle with radial 
velocity $v_{r,\infty}$ reaches a point a distance $r$ from the center 
of the Earth is given by,
 \begin{equation}
 \frac{1}{2} m_i v^2_{r,\infty} \geq V_{eff,i}(r) \; .
 \end{equation}
 This can be rewritten as
 \begin{equation}
\label{eq:angular_barriernuc}
v^2_{r,\infty} \geq v^2_{r,min} \equiv  v_\infty^2 - \frac{r^2}{r^2_{\infty}}\left(v_{r,\infty}^2 + v^2_{esc,i}(r) \right) \to  v_\infty^2 - \frac{r^2}{r^2_{\infty}}\left(v_{\infty}^2 + v^2_{esc,i}(r) \right) \; ,
 \end{equation}
 where $i\in \mathrm{\hat He},\mathrm{\hat H}$ and in the last line we 
have taken the $r_\infty\to \infty$ limit. We integrate 
Eq.~(\ref{eq:speed_distribution_derived}) with respect to $v_{r,\infty}$ 
with the lower limit of integration set to $v_{r,min}$ and find that for 
the nuclei,
 \begin{equation}
\label{eq:speed_dist_nuc}
f_i(r,v) =
\frac{v^2}{v^2_{\infty}}
f_i(r_\infty,v_{\infty}) = \frac{v^2}{v^2-v_{esc,i}^2} f_i\left(r_\infty,\sqrt{v^2-v_{esc,i}^2}\right) \ .
 \end{equation}

For the mirror electrons the situation is more complicated since the 
force is attractive, so that the effective potential can acquire a local 
maximum as shown in Fig.~\ref{fig.VeffRmax}. If the the trajectory is to 
reach a distance $r$ from the center of the Earth the radial component 
of the kinetic energy must be greater than or equal to the effective 
potential at all points for which the radial coordinate $R \geq r$,
 \begin{equation}
\frac{1}{2} m_i v^2_{r,\infty}  \geq  \max_{R>r} V_{eff,i}(R)
 \end{equation}
 We can rewrite this condition as,
 \begin{equation}
\label{eq:angular_barrier}
v^2_{\infty} - v^2_{r,\infty}  \leq \min_{R>r} \frac{R^2}{r_{\infty}^2} \left(v_{r,\infty}^2 + v^2_{esc,\hat{e}}(R) \right) \to \min_{R>r}   \frac{R^2}{r_{\infty}^2} \left(v_{\infty}^2 + v^2_{esc,\hat{e}}(R) \right)  \, ,
 \end{equation}
 where the last step is valid in the $r_{\infty} \to \infty$ limit. For 
any given $v_{\infty}$ we can find the radius $R_{max} = 
R_{max}(v_{\infty})$ that maximizes the right hand side in 
Eq.~\eqref{eq:angular_barrier} (without the restriction $R >r$),
 \begin{equation}
\label{eq:Rmax_cond}
\partial_r v^2_{esc,\hat{e}}(R_{max}) R_{max}+2 \left(v_{\infty}^2 + v^2_{esc,\hat{e}}(R_{max}) \right) =0 \; .
 \end{equation}
 We now can see that $ \gamma\left(v_\infty, v_{r,\infty},r_\infty ,r 
\right) = 1$ for the electrons when
 \begin{equation}
\label{eq:angular_barrier2}
v^2_{r,\infty} \geq  v^2_{r,min} \equiv
\left\{
\begin{array}{lll}
 v_\infty^2-\frac{R_{max}^2}{r_{\infty}^2} \left(v_{\infty}^2 + v^2_{esc,\hat{e}}(R_{max}) \right)
 & \mathrm{for} & r < R_{max}(v_{\infty})
 \\
 v_\infty^2 - \frac{r^2}{r^2_{\infty}}\left(v_{\infty}^2 + v^2_{esc,\hat{e}}(r) \right)
 & \mathrm{for} & r \geq R_{max}(v_{\infty})
 \end{array}
 \right.
 \end{equation}
 Here $R_{max}(v_\infty)$ corresponds to the position of the maximum of 
the angular momentum barrier for an initial speed $v_{\infty}$ at 
infinity. Integrating Eq.~(\ref{eq:speed_distribution_derived}) with 
respect to $v_{r,\infty}$, with $v_{r,min}$ as the lower limit of 
integration, we find for the mirror electron distribution,
 \begin{eqnarray}
\nonumber \frac{f_{\hat{e}}(r,v)}{f_{\hat{e}}(r_\infty,v_{\infty})} &= &\left\{\begin{array}{ll}
\frac{v}{v^2_{\infty}}
\left(v - {\rm Re\,}\sqrt{v^2-\frac{R_{max}^2}{r^2}\left(v^2-v^2_{esc,\hat{e}}(r)+v^2_{esc,\hat{e}}(R_{max}(v_{\infty}))\right)}\right)&r<R_{max}(v_\infty)\\ \frac{v^2}{v^2_{\infty}} &r \geq R_{max}(v_\infty)\end{array}\right.
\\
\label{eq.speed_distribution}
 \end{eqnarray}
 In order to simplify the discussion we first consider a toy model in 
which the distribution $f_i(r_\infty,v_{\infty})$ takes the form of a 
delta function with speed $\frac{3}{2}v_{\odot,i}$ for each species. As 
we shall see, this simpler case captures the main features of the 
more complicated Maxwell-Boltzmann distribution. With this we find that 
the number density of electrons is given by
 \begin{equation}
\frac{n^{S}_{\hat{e}}(r)}{n^F_{\hat{e}} } = 1 + \frac{3v_{esc,\hat{e}}^2}{2v^2_{\odot,\hat{e}}}  
- \left\{ \begin{array}{l l} {\rm Re\,}\sqrt{1+\frac{3v_{esc,\hat{e}}^2}{2v^2_{\odot,\hat{e}}} }\sqrt{1+\frac{3v_{esc,\hat{e}}^2}{2v^2_{\odot,\hat{e}}}-\frac{R^2_{max}}{r^2}\left( 1+\frac{3v_{esc,\hat{e}}^2(R_{max})}{2v^2_{\odot,\hat{e}}}\right) }&  r<R_{max} 
\\ 0 & r \geq R_{max} 
\end{array} \right.
 \label{eq.focusing}
 \end{equation} 
 while for the nuclei $i\in \mathrm{\hat He},\mathrm{\hat H}$ we obtain,
 \begin{equation}
\frac{n^{S}_{i}(r)}{n^F_{i} } =1+ \frac{3v_{esc,i}^2}{2v^2_{\odot,i}} \; .
\label{eq.focusing_nu}
 \end{equation}
 Inserting this into the Poisson equation for the potential $\phi(r)$, 
we can rewrite it in terms of just the captured number density and the free number density far away from the 
Earth,
 \begin{eqnarray}
-\lambda^2_D\nabla^2 \varphi_{esc}(r)  &=& -\beta -\alpha_{He} (1-2 \varphi_{esc}) -\alpha_{H} (1-\varphi_{esc}) \nonumber\\ &-& \alpha_e \left(1+\varphi_{esc} - \left\{ \begin{array}{l l} {\rm Re\,}\sqrt{1+\varphi_{esc}  }\sqrt{1+\varphi_{esc} -\frac{R^2_{max}}{r^2}\left(1+\varphi_{esc}(R_{max})\right) }& r<R_{max} \\ 0 & r>R_{max} \end{array}\right. \right)
\nonumber\\
\label{e.poissoneqnrewritten}
 \end{eqnarray}
 We have defined a dimensionless potential 
 \begin{eqnarray}
\label{e.fesc}
\varphi_{esc}(r) &=& \frac{e \phi}{T_{\rm mirror}} = -
\frac{3 v^2_{esc,i}(r)}{2q_i  v^2_{\odot,i}}  
\ \ \ \forall \ i \ ,
 \end{eqnarray}
 and introduced the dimensionless parameters, 
 \begin{eqnarray}
\alpha_i &=& \frac{q_i n_i^F}{\sum{q^2_j n_j^F}}  \ \ \ \mbox{for} \ \ \ i = \mathrm{\hat He}, \mathrm{\hat H}, \hat e,
\\
\beta(r) &=& \frac{\sum q_i n^C_{i}(r)}{\sum{q^2_j n_j^F}} \ .
 \end{eqnarray}
 Note that $\beta$ parametrizes the size of the 
captured mirror charge density relative to that of the ambient 
plasma far away from the Earth.
Charge neutrality of the external plasma requires $\sum \alpha _i =0$.  
The Debye length $\lambda_D$ is a constant of the problem, defined as 
 \begin{eqnarray}
\label{e.lambdaD}
\lambda_D  
&=& \left(4\pi \alpha \frac{\sum q_i^2 n^{F}_{i}}{T_{\rm mirror}}\right)^{-1/2} 
\ \ \ 
\sim  \ \ \ 
\left(\frac{r_\odot}{0.05}\right)^{-1/2} \times 
\left\{
\begin{array}{ll}
100 \mathrm{m}& \mbox{(disk)}
\\
1000 \mathrm{m} & \mbox{(halo)}
\end{array}
\right.
 \end{eqnarray}
 The potential must vanish as $r \to \infty$, and in fact vanishes 
exponentially within a Debye length of $R_{max}$. For $r>R_{max}$ the 
equation simplifies to the standard Debye-Huckel equation which has the 
solution,
 \begin{equation}
\label{eq.r_above_Rmax}
\varphi_{esc}(r) = 
\varphi_{esc} (R_{max}) {\rm exp}\left\{-\frac{r-R_{max}}{\lambda_D}\right\}
\frac{R_{max}}{r} \ \ \ \mathrm{for} \ \ \ r > R_{max}.
 \end{equation}
 Since $\lambda_D$ is at most of order a few kms and much smaller than 
the size of the Earth, we can work in the approximation where $\lambda_D 
= 0$ and the right hand side of \eref{poissoneqnrewritten} vanishes.

We can now insert the full solution at $r>R_{max}$ of 
Eq.~(\ref{eq.r_above_Rmax}) into
Eq.~(\ref{eq:Rmax_cond}) to find the boundary condition at $r=R_{max}$ 
for the $r<R_{max}$ solution. This is justified because the full 
solution is continuous and has a continuous derivative at $r=R_{max}$.  
In Eq.~(\ref{eq:Rmax_cond}) the derivative term dominates in the limit that 
$\lambda_D\to 0$, because the derivative of $\varphi_{esc}(r)$ in 
Eq.~(\ref{eq.r_above_Rmax}) is proportional to ${\varphi_{esc} 
(R_{max})}/{ \lambda_D}$. Therefore, in this limit the boundary condition 
at $R_{max}$ is simply $\varphi_{esc}(R_{max}) =0$.
 
\begin{figure}
\center{\includegraphics[width=8cm]{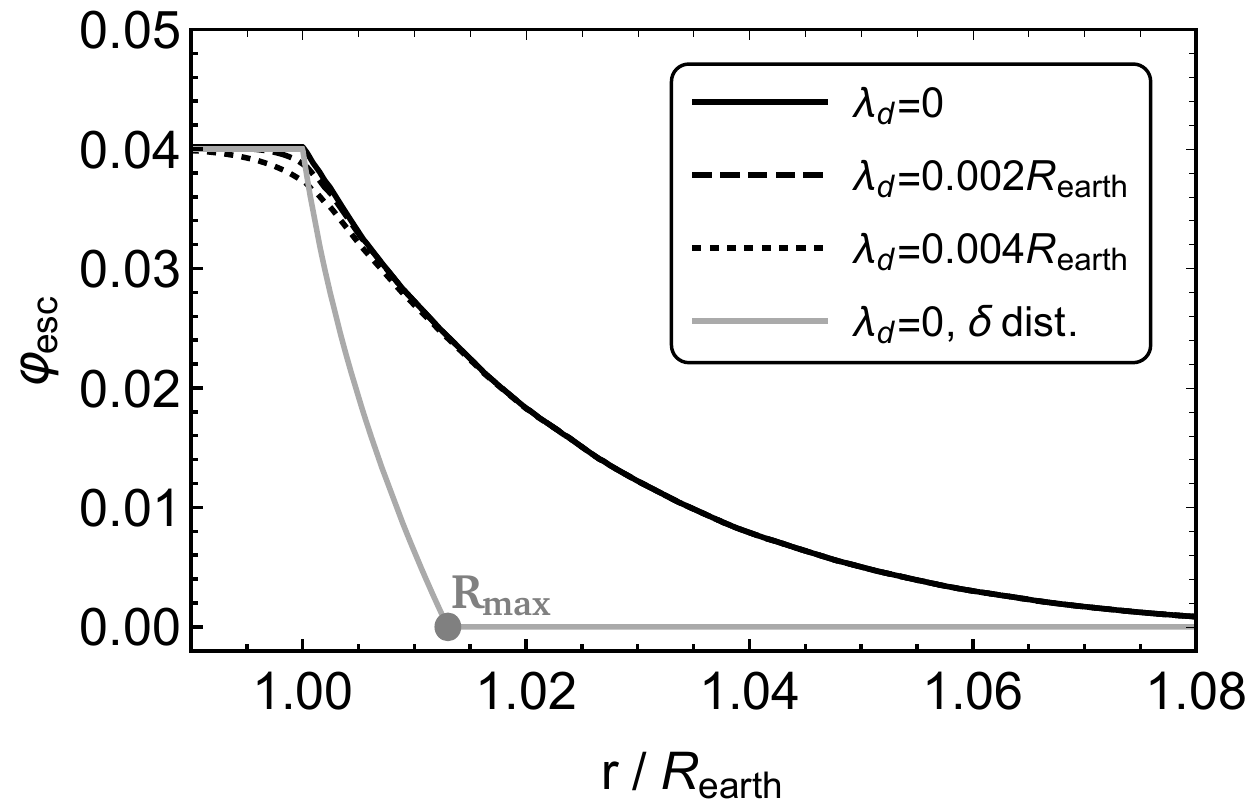}
}
 \caption{Mirror electric potential $\varphi_{esc}(r)$ in the disk case 
for $\epsilon=10^{-11}$ (the small charge accumulation regime), showing 
the effects of mirror helium capture inside the Earth and screening by 
the ambient mirror plasma. Shown are the solutions for $\varphi(r)$ for 
a delta function distribution and for a Maxwell-Boltzmann distribution 
of speeds at infinity, in the limit that the Debye length $\lambda_D = 
0$. Also shown is $\varphi_{esc}(r)$ in the case of a Maxwell-Boltzmann 
distribution for two nonvanishing values of $\lambda_D$, showing how the 
potential is smoothed out. (The actual transition region is extremely 
narrow as $\lambda_D<10^{-3} R_{Earth}$.)
 }\label{fig.phi}
\end{figure}

To find $\varphi_{esc}(r)$ for $r < R_{max}$, we first study the regime 
of small charge accumulation, where $\beta \ll 1$. 
This is a good approximation for $\epsilon \leq 10^{-11}$. We will make 
the simplifying assumption that the potential is constant inside the 
Earth, and the captured mirror helium abundance has a profile that will 
ensure this is maintained. In reality, $\phi(r)$ must decrease with 
increasing $r$, since it would push the accumulated charges away from 
each other and towards the Earth surface. By assuming the potential to 
be constant, we are therefore underestimating mirror helium evaporation 
and overestimating the effects of accumulation. As $r \to 0$, the 
$\mathrm{Re}$ term in \eref{poissoneqnrewritten} vanishes and we can 
algebraically solve with the right hand side set to zero to obtain the 
value of the potential near the center of the Earth,
 \begin{equation}
\label{e.frHesmall}
\varphi_{esc}(r \to 0) \equiv \varphi_{esc}^{\rm Earth}  = \frac{\beta_{0}}{2 \alpha_\mathrm{\hat H} + 3 \alpha_{\mathrm{\hat He}}}  \ll 1 \ \ \ \mathrm{for} \ \ \ \beta_{0}\ll1 \ ,
 \end{equation}
 where $\beta_0$ is the value of $\beta$ at the center of the Earth. Note that $\beta(r)$ 
adjusts to ensure that the right hand side vanishes in 
\eref{poissoneqnrewritten} while keeping
 \begin{equation}
\varphi_{esc}(r) =  \varphi_{esc}^{\rm Earth}  \ \ \ \ \mathrm{for} \ \ \ \ r < R_{Earth} \ .
 \end{equation}
 So now we have an expression for the constant potential for $r < 
R_{Earth}$. We also know that the potential is zero for $r > R_{max} > 
R_{Earth}$, in the $\lambda_D \to 0$ approximation. All that remains is 
to find the potential in the transition region $R_{Earth} < r < 
R_{max}$. In this regime, setting the right hand side of 
\eref{poissoneqnrewritten} to zero with $\beta = 0$ 
yields
 \begin{equation}
\label{e.ftransitionsmallcharge}
\varphi_{esc}(r) =\alpha_{\hat{e}} \frac{ \alpha_{\hat{e}}\left(2r^2-R_{max}^2\right)-\sqrt{\alpha^2_{\hat{e}}R_{max}^4+4r^2(r^2-R_{max}^2)(2\alpha_\mathrm{\hat{H}}+ 3\alpha_\mathrm{\hat{H}e})  } }
{2r^2(\alpha_\mathrm{\hat{H}}+2\alpha_\mathrm{\hat{H}e})(3\alpha_\mathrm{\hat{H}}+4\alpha_\mathrm{\tilde{H}e})} \ .
\end{equation}
 From this we can find the value of $R_{max}$ by imposing continuity of 
$\varphi_{esc}$ at $r = R_{Earth}$. This leads to the potential shown as 
the solid gray curve in Fig.~\ref{fig.phi}. 

This solution has been obtained neglecting the gravitational potential 
of the Earth. Including the Earth's gravity leads to a modest correction 
to the mirror electrostatic potential which acts to compensate for the 
fact that the effect of gravity is larger on the positively charged 
mirror nuclei than on the negatively charged mirror electrons. As noted 
earlier, gravity has only a subdominant effect on our results.

Now we consider the realistic Maxwell-Boltzman distribution of speeds at 
infinity, rather than the delta function. This case is more complicated, 
and can only be solved numerically. Our approach is to solve recursively 
to find $R_{max}$ as a function of $v_{\infty}$. This is done by 
discretizing the speed distribution as $v_{\infty,k}$ and noting that $R_{max}$ goes down 
as the speed is increased. We start therefore with all $R_{max}(v_{\infty,k})$ at 
zero, except that of the lowest speed in the discrete range, which is 
taken as the reference value. We can then find $R_{max}$ for the next 
speed in the range by solving the Poisson equation numerically, imposing 
the condition Eq.~(\ref{eq:Rmax_cond}) and raising it from zero. This is 
then done recursively for the entire range of speeds. This procedure 
works because for $r>R_{max}(v_{\infty,k})$, the solution is the same as if 
$R^i_{max}$ is taken to zero and at $r=R_{max}(v_{\infty,k})$ the derivative of 
$\varphi_{esc}(r)$ is continuous. This numerical solution is shown as 
the solid black line in Fig.~\ref{fig.phi}.

To go beyond the $\lambda_D \to 0$ limit in the regime of small charge 
accumulation, we can solve \eref{poissoneqnrewritten} as a perturbation 
series in $\lambda_D$. We substitute the solutions we have obtained into 
the right hand side of \eref{poissoneqnrewritten} to determine the 
$\mathcal{O}(\lambda_D)$ correction to the solution for the potential 
and so on. The resulting solutions are shown as the dashed lines in 
Fig.~\ref{fig.phi}, demonstrating that the discontinuities in the first 
derivative of the potential are smoothed out on scales of order the 
Debye length at $r = R_{Earth}$ and $r = R_{max}$.

Our careful analysis of screening effects shows that to determine capture and evaporation rates in the small charge regime, we can make use 
of the simple estimate,
 \begin{equation}
\varphi_{esc} = \left\{ \begin{array}{lll}
\varphi_{esc}^\mathrm{Earth} & \mathrm{for} & r \leq R_{Earth} \\
0 & \mathrm{for} & r > R_{Earth}
\end{array}
\right.
 \end{equation}
 with $\varphi_{esc}^\mathrm{Earth}$ as calculated above. We neglect the 
effect of the small nonvanishing potential just above the Earth's 
surface, which does not have a significant effect on capture or 
evaporation since $(R_{max} - R_{Earth}) \ll R_{Earth}$. The mirror 
particle escape velocities near the surface are then modified as 
follows,
 \barray
 \label{e.vesccharge}
v_{esc,{\rm \hat{H}e}}=\sqrt{v_{esc}^2-\frac{2}{3}2v_{\odot,\hat{\rm H}e}^2 \varphi^{\rm Earth}_{esc} }
\,,\qquad v_{esc,{\hat{e}}}=\sqrt{v^2_{esc}+\frac{2}{3}v_{\odot,\hat{e}}^2\varphi^{\rm Earth}_{esc}}\, .
 \earray
 Here $v_{esc} \approx 11$ km/s is the gravitational escape velocity at 
the Earth's surface. This suppresses accumulation and enhances 
evaporation of mirror helium. For the purposes of evaporation below, any 
$(v_{esc,{\rm \hat{H}e}})^2 < 0$ is interpreted to mean that 
$v_{esc,{\rm \hat{H}e}} = 0$ at the surface, since any $\mathrm{\hat 
He}$ that reaches the surface immediately flies away from the Earth.

\subsubsection{Evaporation of Mirror Nuclei}\label{sec.helium.evaporation}

We now consider the evaporation rate of mirror helium. We conservatively 
assume that any $\hat{{\rm H}}{\rm e}^{++}$ captured inside the Earth 
thermalizes with the Earth's interior, which is at temperature 
$T_{\rm Earth} \sim 4000 \mathrm{K}$. Since mirror baryons typically have 
higher kinetic energies than this when they are captured, this can only 
underestimate evaporation and hence overestimate mirror helium 
accumulation. The resulting thermal velocity of mirror helium $\langle 
v_{\rm\hat{H}e,th}\rangle =\sqrt{3T_{\rm Earth}/m_{\rm\hat{H}e}}\sim 5$ 
km/s. The virial radius corresponding to $v \sim \langle 
v_{\rm\hat{H}e,th}\rangle$ is comparable to the radius of the Earth, 
$R_\mathrm{Earth}$. Therefore, to simplify the discussion, we assume 
that the captured particles are distributed homogeneously inside the 
Earth.

The mean free path of $\mathrm{\hat He}$ in the Earth for the $\epsilon 
\lsim 10^{-9}$ regime we consider can be estimated as
 \begin{equation}
\ell_{\rm \hat{H}e}  \sim \frac{1}{n_O \sigma_{\rm\hat{H}e\,O}}\approx 10^{-2} R_{\rm Earth}  \left(\frac{10^{-10}}{\epsilon}\right)^2\,.\label{e.mfp}
 \end{equation}
  Here $\sigma_{\rm\hat{H}e\,O}$ represents the cross section for 
scattering off an oxygen target, since this constitutes the most 
effective target that has a sizable abundance inside the Earth. The 
expression for $\sigma_{\rm\hat{H}e\,O}$ has the same form as in 
Eq.~(\ref{eq:sigmav}).

When $\epsilon\lsim10^{-11}$, the penetration depth of mirror helium 
inside the Earth is comparable to $R_{\rm Earth}$. Evaporation can 
therefor occur from everywhere in the Earth, with a timescale set by the 
mirror helium collisional time scale $\sim\ell_{\rm \hat{H}e}\langle 
v_{\rm\hat{H}e,th}\rangle^{-1}$ weighted by the probability that a 
mirror helium particle has enough kinetic energy to escape the earth's 
gravitational field. The evaporation rate is given by,
 \beq\label{eq.evp}
n_{{\rm \hat He}}^CR_{evp,{\rm \hat{H}e}}\sim
\frac{n_{{\rm \hat He}}^C}{R_{\rm Earth}^3}
\int_0^{R_{\rm Earth}}
dr \ 
r^2
\left[
\left. \langle v_{\rm\hat{H}e,th}\rangle \right|_{T = T_{\rm Earth}(r)}
\right]
\ell_{\rm \hat{H}e}^{-1}
P_\mathrm{ejection}(r)\,.
 \eeq
 Here we have defined an $r$-dependent ejection probability, which is 
simply the fraction of mirror helium nuclei in local thermal equilibrium 
with the Earth at temperature $T = T_{\rm Earth}(r)$ that have speeds 
greater than the escape velocity for helium $v_{esc}(r)$ at that location,
 \begin{eqnarray}
P_\mathrm{ejection}(r) &=& 
\int_{v_{esc}(r)}^{\infty} f_{MB}(v; T_{\rm Earth}(r)) dv \ ,
 \end{eqnarray}
 where $f_{MB}$ is the Maxwell-Boltzmann distribution. We can determine 
$v_{esc}(r)$ from the following differential equation,
 \begin{equation}
\label{e.vescdiffDE}
\frac{1}{2} m_{\mathrm{\hat He}}\frac{d v^2}{d \tilde r} = - \frac{dU_{\rm grav}}{d \tilde r}
 \end{equation}
 with boundary condition $v_f = v(\tilde r = R_{\rm Earth}) = 11$ km/s. 
Here $U_{\rm grav}(\tilde{r})$ is the gravitational potential energy at 
radius $\tilde{r}$ from the center of the Earth, so the term on the 
right side of the equation represents the kinetic energy lost in 
climbing out of the Earth's potential well. In the 
$\epsilon\lesssim10^{-11}$ case we are considering, the accumulated 
mirror electric charge is too small to generate enough electric force at 
the Earth's surface to eject mirror helium. This is in contrast to the 
$\epsilon\gsim10^{-10}$ case we will study in \sref{appendix10}. Since 
we know the temperature profile inside the Earth $T_{\rm Earth}(r)$, 
this allows us to determine $P_\mathrm{ejection}(r)$ as a function of 
$r$ and hence the evaporation rate.

\subsubsection{Results}

 Having determined the capture and evaporation rates, we can solve the 
following equations to obtain the mirror particle abundance,
 \barray
\frac{dn_{{\rm \hat He}}^C}{dt}&=&\left(R^{<v_{\rm  cap}}_{cap,{\rm \hat{H}e}}+R^{>v_{\rm cap}}_{cap,{\rm \hat{H}e}}\right)n_{{\rm \hat He}}^F-R_{evp,{\rm \hat{H}e}}n_{{\rm \hat He}}^C\,,\quad (v_{esc}=v_{esc,{\rm \hat He}})
\\
\label{e.nHediffeqns}
\nonumber n_Q^C&=&2n_{{\rm \hat He}}^C-n_{\hat e}^C\,.
 \earray
 We can obtain number densities in the equilibrium configuration, 
corresponding to ${dn_{{\rm \hat He}}^C}/{dt}= dn_Q^C/dt = 0$ in 
\eref{nHediffeqns}. As discussed in~\ref{sec.electron}, when $n_Q^C$ is 
small, the number density of captured mirror electrons is small and can 
be neglected. Making this assumption and setting $n_{\hat e}^C = 0$, we 
have $n_Q^C = 2n_{{\rm \hat He}}^C$. Since it is the value of $n_Q^C$ 
that determines the capture and evaporation rates of ${\rm \hat He}$, 
this allows us to self-consistently solve for the $n_{{\rm \hat He}}^C$ 
and $n_Q^C$ that satisfy the equilibrium condition.

For a Maxwell-Boltzmann distribution of velocities with $\epsilon = 
10^{-11}$, $\hat v/v = 3, \hat Y_\odot = 1$ and local mirror baryon DM 
fraction $r_\odot = 0.05$, we obtain,
 \begin{equation} 
N_\mathrm{\hat He}  \lesssim 
\left\{ \begin{array}{ll} 
5 \times 10^{23} & \mathrm{(disk)}\\ 1 
\times 10^{22} & \mathrm{(halo)} 
 \end{array} \right. 
 \end{equation} 
 and accordingly, $\varphi_{esc}= 0.15(0.004)$ for the disk (halo) cases. 
Since electrostatic shielding is only important when $\varphi_{esc} \gtrsim
\mathcal{O}(1)$, its effects on the direct detection signal are negiligible.
The escape velocity of mirror electrons due to the accumulated mirror 
helium charge is large compared to the escape velocity from 
gravitational effects alone, of order $\mathcal{O}(100 \mathrm{km/s})$ 
for both halo and disk distributions, but it is still much smaller than 
the thermal width of the Fermi surface in iron. Consequently evaporation 
of mirror electrons is highly efficient, and the number density of 
captured mirror electrons is small (see the discussion around Eq.~\eqref{eq.Ne_th}). Therefore our 
assumption that the captured population of mirror electrons can be 
neglected is self-consistent. Comparing the number of captured mirror 
helium nuclei against \eref{ignoreselfcapture}, we see that we are far 
away from the regime where self-capture becomes important.

It follows from this that for $\epsilon \lesssim 10^{-11}$, there is no 
significant effect on the mirror helium flux or velocity distribution in 
direct detection experiments, either from electrostatic or collisional 
shielding effects. Mirror electrons are accelerated towards the Earth by 
the electrostatic potential, resulting in just a slight increase in 
their velocity (compared to their already high ambient velocity in the 
plasma). This makes our sensitivity estimates for mirror electron direct 
detection mildly conservative.

Other choices of $\hat{v}/v$ or $\hat Y_{\odot}$ do not significantly 
change these results. This includes the limiting case when $\hat Y_\odot 
= 0$, when the mirror baryons are entirely composed of twin hydrogen. In 
all cases, we find that the lowering $r_\odot$ results in 
$N_\mathrm{\hat He} \propto r_\odot$ (mirror baryon fraction of local DM 
density) because the mirror electrostatic screening is negligible. Very 
small $r_{\odot}$ might result in a non-negligible Debye length, but 
this will only modify our results by $\mathcal{O}(1)$. Therefore, 
capture of mirror matter has no significant effect on the direct 
detection signal for $\epsilon \lesssim 10^{-11}$.

\subsection{{\bf $ 10^{-10} \lesssim \epsilon \lesssim 10^{-9} $}}
\label{s.appendix10}

We now discuss the case in which the kinetic mixing parameter $\epsilon 
\gtrsim 10^{-10}$. In this regime the net accumulated mirror charge 
density can become comparable to the ambient density, resulting in 
significant electrostatic effects. In addition, the potential can become 
large enough to suppress mirror electron evaporation, and therefore 
their captured fraction must be taken into account. Therefore the 
physics of mirror baryon accumulation in this regime is quite different 
from that of $\epsilon \lesssim 10^{-11}$.

\subsubsection{Equilibrium in the Large Charge Regime}
\label{s.largecharge}

For the limiting case of a delta function distribution of speeds at 
infinity, we obtained the Poisson equation \eref{poissoneqnrewritten} 
for the dimensionless potential $\varphi_{esc}(r)$. We now consider this 
equation in the regime in which the net captured charge is large, so 
that $\beta$ can no longer be assumed to be small. 
Working in the limit that $\lambda_D \to0$ we immediately run into the 
roadblock that either the the potential $\varphi(r)$ must be 
discontinuous across the surface of the Earth, or $\varphi(r)$ is not 
constant inside the Earth. This is because the solution outside the 
Earth, given by Eq.~(\ref{e.ftransitionsmallcharge}), leads to a maximal 
possible value of $\varphi \simeq 0.06$ at the surface for the 
delta function distribution of speeds at infinity. For this limiting 
value of the potential inside the Earth, we can determine the captured 
mirror helium density that sets the right hand side of the Poisson 
equation to zero. Then, by integrating over the volume of the Earth we 
obtain the net captured charge,
 \begin{equation}
\left.2N_\mathrm{\hat He}-N_\mathrm{\hat e}\right|_{\rm large~q} \approx 2\times10^{25}  \times \left( \frac{r_\odot}{0.05} \right)\ .
 \end{equation}
 It follows that in the $\lambda_D \to0$ limit, for any value of the 
captured charge larger than this, either the potential must be 
discontinuous at $R_{Earth}$, or it must have a varying profile inside 
the Earth. In general, we expect that the solution in the large charge 
regime will exhibit both these features. This greatly complicates the 
calculation of the mirror electric potential.

The discussion in the paragraph above focused on the simple case in 
which the speed distribution far away from the Earth takes the limiting 
form of a delta function. However, in the $\lambda_D \to 0$ limit, the 
solution for the potential in the case of the fully realistic 
Maxwell-Boltzmann distribution that we obtain numerically exhibits the 
same characteristic features, which we list below.
 \begin{itemize}
 \item
 For $r>R_{Earth} $ the solution of the Poisson equation sets a unique 
upper bound on the potential just outside the Earth's surface, which we 
denote as $\varphi^+_{esc}(R_{Earth})\simeq 0.06 (0.15)$, where the 
value shown is for the delta-function (Maxwell-Boltzmann) distribution.
 \item
 Just inside the Earth, there could be a sharp potential jump within a
Debye length of the surface as 
the effects of $\beta$ ``turn on". In the
$\lambda_D \to 0$ limit, we account for this possible jump by
allowing for a discontinuity in the potential at the Earth surface.
We denote as $\varphi^-_{esc}(R_{Earth})$ and $\varphi^+_{esc}(R_{Earth})$
the potentials just inside and outside the Earth, with
$\varphi^-_{esc}(R_{Earth}) > \varphi^+_{esc}(R_{Earth})$.
 \item 
 Inside the Earth, there is some continuous potential $\varphi_{esc}(r)$ 
for $0 \leq r \leq R_{Earth}$ that satisfies $\varphi_{esc}(R_{Earth}) = 
\varphi^-_{esc}(R_{Earth})$. In general the potential is expected to 
vary inside the Earth. However, if mirror electron capture is so 
efficient that their number density is much larger than the net free 
charge, the captured mirror electrons will distribute themselves inside 
the Earth in such a way that $\varphi_{esc}(r)$ will tend to a constant 
value independent of $r$.
 \end{itemize}
 The numerical values of $\varphi^+_{esc}(R_{Earth})$ and the functional 
form of $\varphi_{esc}(\beta)$ exhibit a mild dependence on 
the precise mirror helium fraction of the dark plasma which does not 
affect our conclusions. They are notably independent of $r_\odot$ and 
whether we have a halo or disk distribution, since the dark baryon 
fraction and overall velocity drop out of the right hand side of the 
Poisson equation. The mirror matter fraction $r_\odot$ does affect the 
Debye length since $\lambda_D \sim r_\odot^{1/2}$, see \eref{lambdaD}. 
If $\lambda_D$ is taken to be nonvanishing (but still much smaller than 
the size of the Earth), then the above piece-wise defined potential is 
smoothed out on scales of $\lambda_D$ near $R_{Earth}$, similar to what 
is shown in Fig.~\ref{fig.phi}. We account for effects from the 
nonvanishing Debye length in our discussion below, but it does not 
significantly change our conclusions about the general features of the 
potential.

A varying potential inside the Earth would necessarily result in a 
mirror electric force that repels the captured mirror helium particles 
and pushes them towards the surface. Then, to obtain the potential 
$\varphi_{esc}$ we would first need to solve a complicated set of 
coupled equations that determines the distribution of mirror helium in 
the Earth as a function of $\varphi$ taking into account the capture of 
mirror helium, its diffusion in the presence of the mirror electric 
field and its eventual evaporation. It follows that solving for the 
mirror electric potential of the Earth in the regime of significant 
charge accumulation is extremely challenging. However, we can make use 
of the fact that the potential $\varphi$ outside the Earth continues to 
take the same form to derive robust \emph{upper bounds} on the 
mirror potential at equilibrium, both near the surface and deep within 
the Earth's interior. As we now show, this can be used to place limits 
on the suppression of incoming mirror helium flux near the surface where 
direct detection experiments are located.

\subsubsection{Upper Bound on the Net Captured Charge}
\label{s.limit_on_charge}

At equilibrium, the flux of mirror helium exiting the Earth is equal to 
the flux entering it. By requiring that the flux entering any region of 
the Earth not exceed the flux leaving it, we can derive upper bounds on 
$\varphi_{esc}(r)$ for $0 \leq r \leq R_{Earth}$. We begin by 
considering the mirror helium flux at the surface. The \emph{outgoing} 
flux can be bounded from below under the assumption that mirror helium 
nuclei start out stationary just below the surface and are then 
accelerated by the potential jump as they leave the surface,
 \begin{eqnarray}
\label{e.Foutmin}
\mathcal{F}_{out} > \mathcal{F}_{out}^{min} &=& n^C_\mathrm{\hat{H}e} v_{out}  =
\left(\sum{q^2_i n_i^F}\right)
\beta_{\mathrm{\hat He}}(R^-_{Earth}) \sqrt{\frac{4}{3}} \sqrt{ \varphi^-_{esc}\left(R_{Earth})-\varphi^+_{esc}(R_{Earth}\right)} v_{\odot, \mathrm{\hat He}} \; , \nonumber\\
 \end{eqnarray}
 where 
 \begin{equation}
\beta_{\mathrm{\hat He}} = \frac{ 2 e n^C_\mathrm{\hat He}(r)}{\sum{q^2_j n_j^F}} 
\end{equation}
is the contribution to $\beta$ from mirror helium alone, 
and $v_{out}$ is the velocity of the outgoing particles. This is 
related to the difference between the potentials just inside and just 
outside the Earth, 
$\varphi^-_{esc}\left(R_{Earth})-\varphi^+_{esc}(R_{Earth}\right)$, by 
energy conservation,
 \begin{equation}
v_{out} = \sqrt{\frac{4}{3}} \sqrt{ \varphi^-_{esc}\left(R_{Earth})-\varphi^+_{esc}(R_{Earth}\right)} v_{\odot, \mathrm{\hat He}} \;.
 \end{equation}
 This should be considered a lower bound on the outgoing flux because 
the electrostatic drift inside the Earth may increase the velocity of 
mirror nuclei just before they cross the jump. The bound is valid as 
long as the mean free path near the surface, \eref{mfp}, is larger than 
the Debye length, \eref{lambdaD}, so that collisions have no effect 
across the potential jump. 

On the other hand, we can place an upper bound on the flux of mirror 
helium entering the Earth. To do this we note that the incoming flux, 
given by $\left<n_\mathrm{\hat{H}e}v_r\right>$, is always less than 
$\left<n_\mathrm{\hat{H}e}\left|v\right|\right>$, where the radial velocity 
has been replaced by the total speed. This allows us to set an upper
bound on the incoming flux,
 \begin{equation}
\label{e.Finmax}
\mathcal{F}_{in} < \mathcal{F}_{in}^{max} = \int{f_{\hat{He}}\left(R^+_{Earth},v\right) v dv} = 
\left(\sum{q^2_i n_i^F}\right)
\alpha_{\mathrm{\hat He}} \sqrt{\frac{2}{3\pi}}v_{\odot, \mathrm{\hat He}} e^{-2\varphi_{esc}^-(R_{Earth})} \; , 
 \end{equation}
 where $f_\mathrm{\hat{H}e}\left(R^+_{Earth},v\right)$ is given by 
Eq.~(\ref{eq:speed_dist_nuc}). Note that this upper bound takes into 
account the full Boltzmann suppression, since for large potentials only 
the fastest mirror nuclei penetrate below the surface. Since the 
outgoing flux cannot exceed the incoming flux, $\mathcal{F}_{out} \leq 
\mathcal{F}_{in}$, we obtain the relation
 \begin{equation}
\beta_{\mathrm{\hat He}}(R^-_{Earth})  \sqrt{\frac{4}{3}}\sqrt{ \varphi_{esc}^-\left(R_{Earth}\right)-\varphi_{esc}^+\left(R_{Earth}\right)} \leq \alpha_{\mathrm{\hat He}} \sqrt{\frac{2}{3\pi}} e^{-2\varphi_{esc}^-(R_{Earth})} \; .
 \end{equation} 
 We now make use of the fact $\beta_{\mathrm{\hat He}} > \beta$ (we are here neglecting the effect of mirror hydrogen for simplicity, since it evaporates efficiently and its accumulation has negligible effect), which allows us to write:
  \begin{equation}
\label{e.surfaceinequality}
\beta(R^-_{Earth})  \sqrt{\frac{4}{3}}\sqrt{ \varphi_{esc}^-\left(R_{Earth}\right)-\varphi_{esc}^+\left(R_{Earth}\right)} \leq \alpha_{\mathrm{\hat He}} \sqrt{\frac{2}{3\pi}} e^{-2\varphi_{esc}^-(R_{Earth})} \; .
 \end{equation} 
 For the limiting case in which the speed distribution at infinity is a 
delta function, \eref{poissoneqnrewritten} allows us to relate 
$\varphi^-_{esc}$ to $\beta(R_{Earth})$ in the 
$\lambda_D \to 0$ limit. This is accomplished by setting the right hand 
side of \eref{poissoneqnrewritten} to zero inside of the potential jump 
near the surface. This is a valid approximation in the $\lambda_D \to 0$ 
limit since the potential does not rapidly change on Debye length scales 
once we are past the potential jump near the Earth's surface. This 
allows us to determine $\beta(R^-_{Earth})$, just 
inside of the potential jump, as a function of 
$\varphi^-_{esc}\left(R_{Earth}\right)$. With this, 
\eref{surfaceinequality} can be translated into an upper bound on the 
potential just inside the Earth,
 \begin{equation}
\label{e.fescbound}
  \varphi^-_{esc}\left(R_{Earth}\right) < 0.6 (0.35)
   \end{equation}
 calculated using the Maxwell-Boltzmann (delta-function) distribution at 
infinity. Notice that $v_{\odot, \mathrm{\hat He}}$ dropped out of 
\eref{surfaceinequality}, which means that this upper bound applies to 
both halo and disk distributions. The resulting electrostatic 
suppression of the incoming mirror helium flux for direct detection 
experiments near the surface can be obtained by considering the radial 
flux,
 \begin{equation}
\label{eq:speed_flux}
 \mathcal{F}_i(r) = \int dv \frac{r_{\infty}^2}{r^2} 
\frac{dv_{\infty}}{dv} f_i(r_\infty,v_\infty) \int v_r 
\frac{v_{r,\infty}}{v_r} \gamma\left(v_\infty, v_{r,\infty},r_\infty ,r 
\right) \frac{dv_{r,\infty}}{v_{\infty}} \; ,
 \end{equation} 
 evaluated at $r=R_{Earth}$. This suppression factor turns out to be 
simply ${\rm min}(e^{-2\varphi^-_{esc}\left(R_{Earth}\right)},1) \simeq 
0.3 $ for the Maxwell-Boltzmann distribution of speeds at infinity. While we can neglect the effects of captured mirror hydrogen, any incoming mirror hydrogen flux would nonetheless also be affected by this electrostatic screening. The corresponding suppression is ${\rm 
min}(e^{-\varphi^-_{esc}\left(R_{Earth}\right)},1)\simeq 0.55 $ (halo). 
The effect of this suppression on the direct detection bounds for 
$\epsilon$ is marginal --  less than 50\%~(25\%) for mirror helium 
(hydrogen) searches. 
Our sensitivity projections for $\epsilon \sqrt{r_{\odot}}$ from mirror 
nuclear recoils will therefore have at most a factor of 2 uncertainty, but only if $r_{\odot} \ll 0.01 $ is so small that the sensitivity boundary lies 
near or above $\epsilon \sim 10^{-10}$. 
For $r_{\odot} \sim 0.01$, we find that future experiments probe kinetic mixings that are far below $10^{-10}$, meaning the projected sensitivities are not affected by this modest suppression.
In our discussion to this point we have neglected 
the effects of the Earth's gravity, which acts to slightly reduce in the 
extent of electrostatic shielding. The resulting correction to the 
results is modest in the disk case and negligible in the halo case.

We move to find a similar bound on $\varphi_{esc}$ anywhere inside the 
Earth. As we will see, this can be used to place a limit on the net 
number of captured particles inside the Earth, and thereby show that 
collisional shielding does not limit direct detection.
 \begin{figure}
\center{\includegraphics[width=12cm]{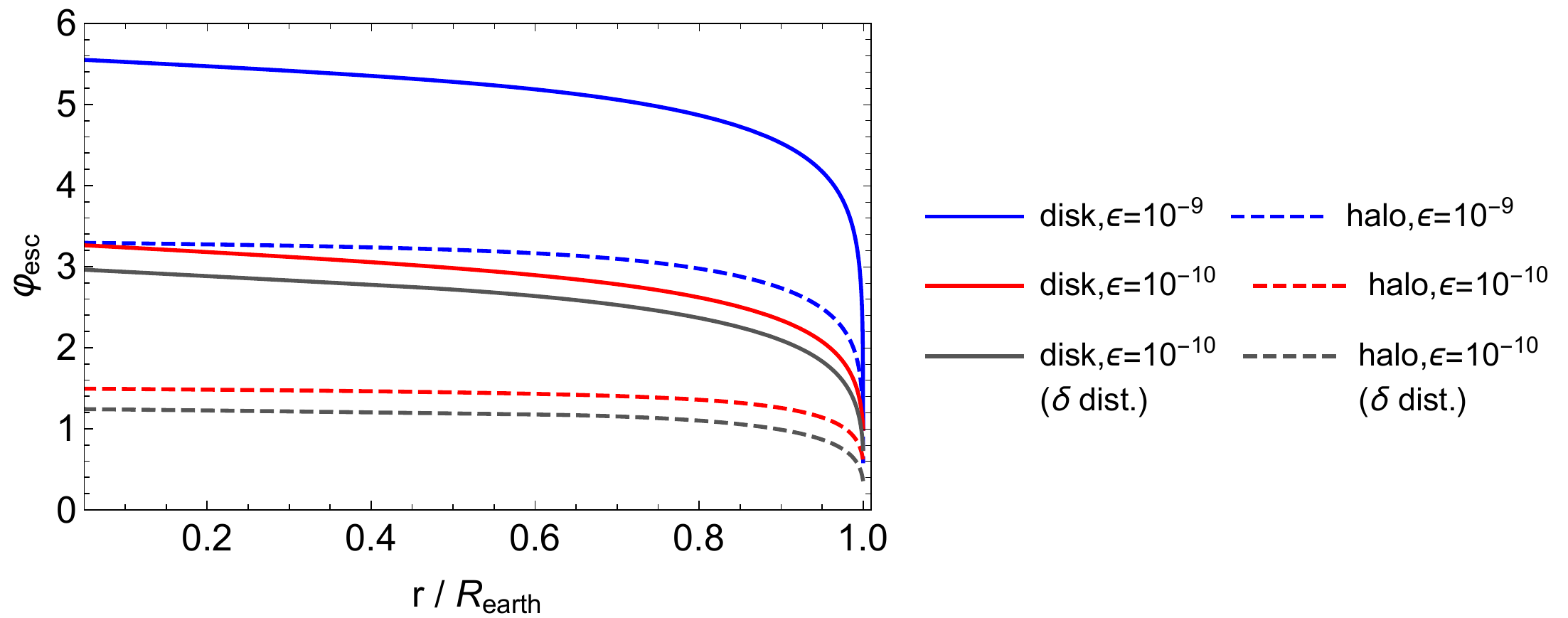}
}
 \caption{Maximum possible mirror electric potential inside the Earth, 
parametrized in terms of $\varphi_{esc}(r)$, shown for mirror particles 
in a halo (solid) and disk (dashed) distribution. Results are shown for 
a delta function distribution of speeds at infinity with $\epsilon = 
10^{-10}$ (grey) and for a Maxwell-Boltzmann distribution of speeds at 
infinity with $\epsilon = 10^{-10}$ (red) and $\epsilon = 10^{-9}$ 
(blue).
 }\label{fig.phimax}
 \end{figure}
 A non-uniform mirror-electric potential $\varphi_{esc}(r)$ inside the 
Earth initiates a \emph{drift velocity} towards the surface,
 \begin{equation}
 v_{drift} \ \sim \ a \  \Delta t_{coll} \; ,
 \end{equation}
 where $\Delta t_{coll}$ denotes the average time between collisions, 
and $a$ represents the radial acceleration of mirror helium charges due 
to the electric field. The value of $\Delta t_{coll}$ is determined by 
the mirror helium mean free path inside the Earth, \eref{mfp}, and the 
average velocity of mirror helium nuclei inside the Earth. The latter is 
dominated either by the drift velocity itself or by the thermal 
contribution that arises from the mirror nuclei coming into equilibrium 
with the Earth's interior. To approximately account for the limits where 
either contribution could dominate, we write
 \begin{equation}
\Delta t_{coll} \sim \frac{\ell_\mathrm{\hat He} }{v_{drift} + v_{\rm\hat{H}e,th}} \; ,
 \end{equation}
which yields,
 \begin{eqnarray}
v_{drift} &\sim& 
\sqrt{a \ell_\mathrm{\hat He} + \left( \frac{v_{\rm\hat{H}e,th}}{2}\right)^2} - \left( \frac{v_{\rm\hat{H}e,th}}{2}\right)
\nonumber \\
&=& 
\sqrt{
2\varphi'_{esc}(r) \ell_\mathrm{\hat He} 
\left(\frac{v_{\odot, \mathrm{\hat He}}}{\sqrt{3}}\right)^2
+ \left( \frac{v_{\rm\hat{H}e,th}}{2}\right)^2} - \left( \frac{v_{\rm\hat{H}e,th}}{2}\right) \; ,
 \end{eqnarray}
 where we have taken
 \begin{equation}
a=\frac{2}{3}\varphi'_{esc}(r)v^2_{\odot, \mathrm{\hat He}} \; .
 \end{equation}

 The outgoing radial flux must be less than the incoming radial flux at 
any arbitrary location $r < R_{Earth}$. As in \eref{Finmax}, we 
overestimate the incoming flux by replacing the radial component of the 
velocity by the total speed. From this we obtain the inequality,
 \begin{equation}
\beta(r)  v_{drift} \leq \beta_{\mathrm{\hat He}}(r)  v_{drift} \leq
 \alpha_{\mathrm{\hat He}} \sqrt{\frac{2}{3\pi}} v_{\odot, \mathrm{\hat He}} e^{-2\varphi_{esc}(r)} \; .
 \end{equation}
 In the case of a delta function distribution of velocities we can 
relate $\beta(r)$ to $\varphi_{esc}(r)$ by setting 
the right hand side of \eref{poissoneqnrewritten} to zero. We can obtain 
the corresponding relation for the case of a Maxwell-Boltzmann 
distribution from our numerical solution. Saturating the above 
inequality yields a differential equation for the maximum possible 
mirror electric potential $\varphi^{max}_{esc}(r)$. This differential 
equation can be solved numerically. The solution is shown in 
Fig.~\ref{fig.phimax} for the delta function and Maxwell-Boltzmann 
distributions of velocities, for both halo and disk profiles. The 
effects of gravity, although subdominant, have been included. From the 
corresponding $\beta(r)$, we can obtain an upper 
bound on the net captured charge by integrating over the volume of the 
Earth. For $\epsilon \lesssim 10^{-9}$ we obtain,
 \begin{equation}
\label{e.NHeupperbound}
2N_\mathrm{\hat He}-N_\mathrm{\hat e} < 0.5 \sum{q^2_i n_i^F} \frac{4}{3} \pi R^3_{Earth} \approx 
3(8) \times 10^{25} \times \left( \frac{r_\odot}{0.05}\right) \; ,
 \end{equation}
 for the Maxwell-Boltzmann (delta-function) distribution of speeds at 
infinity.  This bound is almost identical for the halo and disk 
profiles, since the lower repulsive potential in the disk case is 
compensated for by the smaller incoming flux.

If $r_\odot \lesssim 10^{-7}$ $(10^{-5})$ in disk (halo) distributions, 
$\lambda_D$ becomes comparable to or larger than the mirror helium mean 
free path $\ell_\mathrm{\hat He}$. In this parameter range the results 
we obtained above are slightly modified. The potential jump just below 
the Earth's surface is now smoothed out on scales of $\lambda_D \gtrsim 
100~\mathrm{km}$, and we have to take collisions into account when 
computing the outgoing flux near the surface. The collisions suppress 
the outgoing velocity for the same potential difference by a factor of 
${\cal O}( \ell_\mathrm{\hat He}/\lambda_D )$. However, we see that the 
potential is only logarithmically sensitive to this due to the 
exponential dependence of the incoming flux on $\varphi_{esc}$ in 
\eref{Finmax}. Therefore, although this effect increases the upper bound 
on the potential `inside the jump' in the Earth's interior by a factor 
of $\sim \log (\lambda_D / \ell_\mathrm{\hat He})$ compared to 
\eref{fescbound}, the potential at the location of direct detection 
experiments at depth $d \sim \mathcal{O}(\mathrm{km})$ below the Earth's 
surface is actually smaller by a factor of $\sim d/\lambda_D$ since the 
potential varies smoothly on distance scales of order the Debye length 
instead of sharply increasing at the surface. This means that the 
effective maximum $\varphi_{esc}$ that electrostatically shields direct 
detection experiments is reduced by a factor of $\sim (d/\lambda_D) \log 
(\lambda_D / \ell_\mathrm{\hat He})$. Lowering $r_{all}$ therefore 
further reduces the modest electrostatic suppression of the incoming 
mirror helium flux.

\subsubsection{Upper Bound on the Total Number of Captured Mirror Particles}
\label{s.electroncaptureepsilon10e-10}

The constraint calculated in Eq.~(\ref{e.NHeupperbound}) represents an 
upper bound on the net charge of the captured mirror particles, $2 
N_\mathrm{\hat{H}e} - N_{\hat{e}}$. In order to establish that the effects of 
collisional shielding are not important, we need to obtain upper bounds 
on $N_\mathrm{\hat{H}e}$ and $N_{\hat{e}}$ separately. We do this by noting 
that if the number of captured mirror electrons is much larger than the 
net captured charge, so that $N_{\hat{e}} \gg 2 N_\mathrm{\hat{H}e} - 
N_{\hat{e}}$, the mirror electrons will move freely inside the Earth to 
neutralize any gradients in the electric field. Therefore, in this limit 
the value of the potential $\varphi_{esc}$ takes on the same value at 
every point in the Earth{\footnote{This can be seen by solving the 
Debye-Huckel equation for $\varphi_{esc}$ inside the Earth.}}. This 
constant value is necessarily less than or equal to the upper bounds on 
$\varphi_{esc}^-(R_{Earth})$ shown in Eq.~(\ref{e.fescbound}) 
for the halo and disk cases.

We now show that this behavior can be used to set a bound on the neutral 
component that complements the bound on the net captured charge in 
Eq.~(\ref{e.NHeupperbound}). In Section~\ref{sec.electron} we obtained 
an upper bound on the net number of captured mirror electrons that 
depends on the potential $\varphi_{esc}$ as,
 \begin{equation}
N_{\hat{e}} \lesssim 10^{23} \exp\left(\varphi_{esc}\frac{T_{mirror}}{T_e}\right)
 \label{e.em_population}
 \end{equation}
 for the disc (halo) case. Setting $\varphi_{esc}$ in this equation to 
its upper bound in Eq.~(\ref{e.fescbound}) we find $N_{\hat{e}} \lesssim 
10^{24}$ for both the disk and halo cases. This upper bound on 
$N_{\hat{e}}$ has been obtained under the assumption that the number of 
captured mirror electrons is much larger than the net captured charge, 
$N_{\hat{e}} \gg 2 N_\mathrm{\hat{H}e} - N_{\hat{e}}$. If this assumption is 
valid, the upper bound on $N_{\hat{e}}$ also translates into an upper 
bound on the number of captured helium nuclei, $N_\mathrm{\hat{H}e} \ll 
N_{\hat{e}}$. It follows that the number of captured mirror particles is 
at most of order $10^{24}$, far too small to result in significant 
self-capture.

If the assumption that the number of captured mirror electrons is much 
larger than the net captured charge is not valid, so that $N_{\hat{e}} 
\lesssim 2 N_\mathrm{\hat{H}e} - N_{\hat{e}}$, we can nevertheless still obtain 
an upper bound on the captured population. In this case the upper bounds 
on the net captured charge in Eq.~(\ref{e.NHeupperbound}) translate into 
upper bounds on the $N_{\hat{e}}$ and $N_\mathrm{\hat{H}e}$ individually, so 
that $N_{\hat{e}} \; , N_\mathrm{\hat{H}e} \lesssim 10^{26}$. These are still 
far below the numbers required for self-capture to play a major role. We 
therefore conclude that self-capture and collisional shielding do not 
significantly affect direct detection.


\bibliography{MTH_cosmo}
\bibliographystyle{JHEP}


\end{document}